\documentclass [badsci, printsupervisor, printschool, copyrightpage, titlesmallcaps]{uqthesis}
\pdfoutput=1
\input{deps/defs_thesis.ptex}
\usepackage{hyperref}
\usepackage{url}
\makeindex
\usepackage{bm}
\usepackage{braket}
\usepackage{color}

\newcommand{\norm}[1]{\lVert#1\rVert}

\newcommand{\deriv}[2]{\frac{\mathrm{d} #1}{\mathrm{d} #2}}

\newcommand{\pl}[2]{\frac{\partial #1}{\partial #2}}

\begin{document}

\frontmatter

\logo{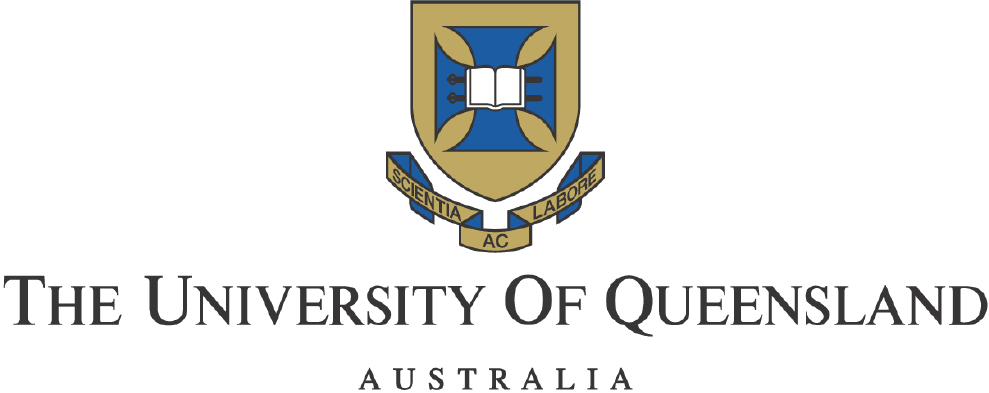}
\logoscale{1}

\title{Numerical Probe of Dynamical Properties of a Two-Dimensional
Triangular Ising Lattice with Long-Range Interactions\\ 
}
\author{Tomohiro Hashizume}
\authorqual{Degree of Physics in Bachelor of Advanced Science with Honours}
\supervisor{Dr Ian Peter McCulloch}
\department{Physics Department}
\school{Mathematics and Physics}

\titlepage

\chapter{Abstract}
\label{Abstract}
The invention of the Density Matrix Renormalization Group (DMRG) algorithm in 1992 \cite{White1992} opened a new door 
for high-precision computational condensed matter physics.
The reformulation of the algorithm with Matrix Product States (MPS) motivated DMRG-like  algorithms which allows us
to perform time evolutions and other various numerical probes.
Although the numerical probe of the ground state using DMRG can be applied on small two-dimensional lattices,
until now, there was no working algorithm that can perform a time evolution of a two-dimensional lattice with long-range interactions. 

In 2012, researchers at the National Instate of Standards and Technology performed an experimental realisation of the two-dimensional triangular spin lattice. 
Using a Penning trap of ions, this apparatus allows the realisation of a global quench terms with long-range interactions.
Motivated by this experiment, we developed a new algorithm which allows a time evolution of an MPS with a Hamiltonian with long-range interactions
based on generalised Suzuki-Trotter decompositions with Krylov subspace expansion steps.

The algorithm provided a good agreement on a benchmark calculation of real and imaginary time simulations, 
which performed on models which have long-range interactions.
In benchmark tests, it predicted the correct local magnetisation behaviour within $5$ decimal places up to $t\sim 8$,
it provided the ground state energy of Haldane-Shastry (HS) model correctly up to 4 decimal places for evolution with $64$ sites;
and it gave a global magnetisation behaviour which agrees with Time Dependent Variational Principle algorithm.
The algorithm is applied to a global quench on the triangular Ising lattice on the surface of an infinite cylinder.
The algorithm successfully showed the thermalisation of the initial state after a deep quench proving that the algorithm works on time evolutions in
a quasi-two-dimensional system.

This is believed to be the world's first calculation on the time evolution of a MPS in dimensions higher than one with long-range interactions.
It is hoped that in the future the algorithm is used to perform beyond what is discussed in this thesis.
The generation of entangled states, search of thermalisation-protected quenching parameters, and extending the algorithm to perform
local quench are recommended as future research. Application to quantum computing and modeling of quantum simulators 
are also hoped to be achieved.

\input{front/publications.ptex}
\chapter{Acknowledgements}
\label{Acknowledgements}
First and foremost, I would like to thank my supervisor Dr. Ian McCulloch.
I cannot thank him enough for all the opportunities that he has given me.
In fact, there are so many things that I want to thank that I don't even know where to start.
I think thanking him for the development of the Matrix Product Toolkit is the good place to start.
All of the code that I wrote upon completion this thesis were based on. 
He was always helpful and led me in the right direction whenever I was struggling with the sea of functions in the toolkit.
Without his help and the user friendly, simple, and easy to use toolkit it was impossible to finish coding in a suitable 
timescale to conduct actual numerical simulations. 
Constructing an algorithm was like finding the laws of physics written by the developer and 
I really enjoyed that a lot. 
I also had a great experience at the Engineered Quantum System (EQUS)  workshop held in the Hunter Valley and a subsequent trip to the 
University of Sydney cannot have happened without him.
The deeper understanding of the field of strongly correlated quantum systems and the skills to perform a numerical simulation have opened the door to 
various post-graduate opportunities, and that is all because he gave me this project. 

Upon completing this thesis, firstly I would like to thank Dr. Nariman Saadatmand for the fruitful discussion on the quantum phase diagram of 
the two-dimensional triangular Ising lattice. 
I would like to thank Jason Pillay for the helpful advice on the presentations and the scientific communications. 
I would like to thank Simon Deeley for putting soul into finding and correcting the grammatical mistakes in the drafts 
of this thesis. Finally, I would like to acknowledge Eliot Sandberg for creating a great Web Applet which 
enables me to jump to the PDF version of journal articles with a click of a button. 
Without you four, this thesis would have never been completed. 

I must thank cohorts in the physics degree for constituting my wonderful years of the undergraduate study in the University of Queensland. 
The hiking in the outback, all the discussion we had on various topics in physics,
and the International Physics Tournament we attended in Paris are a few of the many unforgettable memories. 
Also, I would not have made it through without your help on the various aspects of the university life.
Especially, I would like to thank Oliver Sandberg and his family for all the help that they have provided when I was in a struggle.

The last but not least, I would like to thank my parents, my sisters and my grand parents. 
They supported me throughout the undergraduate years,
Without their help I would not have finished the degree. 
They also always welcome me whenever I go back to Japan.
I must have caused a lot of trouble for moving to a foreign country before even finishing high school.
Thank you for always trusting, supporting, and being patient on my life decisions.

\tableofcontents

\listoffigures
\listoftables
\chapter{List of Symbols}


The following list is neither exhaustive nor exclusive, but may be helpful.
\begin{list}{}{%
\setlength{\labelwidth}{24mm}
\setlength{\leftmargin}{35mm}}
\item[$\otimes$] Tensor product
\item[$\ket{\uparrow},\ket{\downarrow}$] Positive and negative eigenstates of the spin-1/2 $\hat{S}^z$ operator
\item[$A^{\bm{\sigma}}$,$A^{s_n}_{i,j}$] Local matrix of a matrix product state
\item[$A^{\bm{\sigma}}_{\bm{\sigma}'}$,$W^{s_n}_{s'_n}$] Local matrix of a matrix product state
\item[$a_i$] Bond biasing coefficient of $\hat{H}_A$
\item[$\alpha$] Parameter governing the strength of Long-range interaction
\item[$\alpha_i$] Diagonal element of the tridiagonal matrix obtained through Lanczos steps
\item[$b_i$] Bond biasing coefficient of $\hat{H}_B$
\item[$\beta_i$] Off diagonal element of the tridiagonal matrix obtained through Lanczos steps
\item[$D$] Diagonal matrix
\item[$\delta_{i,j}$] Kronecker's delta
\item[$\bm{\gamma}$] Decay rate
\item[$\bm{\Gamma}$] Magnetic Field vector
\item[$\Gamma$] Transverse magnetic field strength
\item[$\hat{H}$,$\hat{H}_A$,$\hat{H}_B$] Hamiltonian
\item[$h$] Transverse magnetic field strength
\item[$I$,$\hat{I}$] Identity (operator)
\item[$J$] Spin coupling strength
\item[$J_0$] zeroth Bessel function of the first kind
\item[$L$] Left orthogonal matrix ($L^\dag L = I$)
\item[$M$] A matrix
\item[$m$] Number of states (the dimensions of the Auxiliary dimension)
\item[$m_0$] Magnetisation ($m_0=\bra{\Psi}S^{z}(i)\ket{\Psi}$ or $\sum \bra{\Psi}S^{z}(i)\ket{\Psi}/N$)
\item[$N$] Number of sites in a system
\item[$\mathcal{N}$] Number of Lanczos vectors
\item[$N_R$] Number of sites around a ring of a cylinder.
\item[$N_e$] Number of exponential series used to fit algebraic decay. 
\item[$\mathcal{O}$] Order of
\item[$R$] Right orthogonal matrix ($RR^\dag = I$)
\item[$\bm{r}_i$] Position vector to site $i$
\item[$s$] spin quantum number
\item[$\hat{S}^x_i$,$\hat{S}^y_i$,$\hat{S}^z_i$] Local spin operators
\item[$s_i$] configuration of a spin at site $i$
\item[$\bm{\sigma}$] configurations of all the spins in a system
\item[$T$] Temperature 
\item[$T_c$] Critical temperature
\item[$t$] Real Time
\item[$\tilde{t}$] Some time scale
\item[$\tau$] Imaginary time ($1/k_bT$).
\item[$V$] Potential
\item[$\ket{V_i}$] $i$\textsuperscript{th} Lanczos vector.
\item[$W(i)$] Bond biasing function
\end{list}


\mainmatter

\chapter{Introduction}
\label{Introduction}
\section{Quantum Spin Lattice Models}
\label{Quantumspinlattice}
The existence of ferromagnetic materials has been known as early as ancient Greece.
However, the theoretical explanation of its origin had to wait until 
the establishment of quantum mechanics and the discovery of the `spin'\index{Spin},
in the early 1900s \cite{Stern1922,Pauli1925,Heisenberg1928,Schrodinger1926}.
Spin is an intrinsic angular momentum of the fundamental particles. 
An electron has a magnetic dipole moment arising from the rotating electric charge due to its spin.
The ferromagnetism can be modelled and explained with the localised electrons in the sites of a crystal lattice
interacting via their magnetic dipole moments.

The Heisenberg model \cite{Heisenberg1928,Dirac1929}\index{Heisenberg model} is the simplest model which 
modelled the dipole interactions of electrons in a lattice.
The Hamiltonian of the model is given by
\begin{align}
    \hat{H} &= \sum_{i \neq j} J_{i,j} \bm{\hat{S}}_i\cdot\bm{\hat{S}}_j\label{HeisenbergSpinChain},
\end{align}
where $\hat{S}_i$ is a spin-1/2 operator at $i$\textsuperscript{th} site, 
and $J$ is a parameter which defines an interaction strength.
Heisenberg \cite{Heisenberg1928} predicted that, 
the ground state is ferromagnetic for negative values of $J$ and anti-ferromagnetic otherwise. 

The difficulty in solving for the exact ground state of the model lies on the exponentially increasing 
dimensions of the Hilbert space with respect to the number of particle in the system ($N$). 
Not only is the calculation of the exact energy spectrum of the model by hand impossible, numerical exact diagonalisation fails 
when $N$ is greater than $\sim30$\footnote{For $N=30$, exact diagonalisation must be done on about a billion by a billion matrix!}
Since an ordinary magnet contains a number of particles comparable to a mole ($10^{23}$),
finding the ground state energy and energy spectrum of this model was thought to be impossible.

Thankfully, Bethe \cite{Bethe1931} found the exact ground state for the positive values of $J$, using a technique 
now called the Bethe ansatz\index{Bethe ansatz}.
Against our classical intuition, the ground state was not antiferromagnetic. 
It is found that the local expectation value of the $S^{z}$ operator is $0$, but it has a long-range correlation which decays algebraically \cite{Koma1993}. 
However, the applicability of the technique is very limited.
Adding a small anisotropy, changing the magnitude of the spin of the operators, additional interactions, or changing the geometry of
the model make it generally impossible to obtain exact ground states.
Only the ground states of a few special cases of the modified Heisenberg model (quantum spin lattice \index{Quantum spin lattice})
were solved exactly \cite{Kasteleijn1952,Bonner1964,Yang1966,Takahashi1971,Affleck1987,Haldane1988,Shastry1988,Bernard1995}.
In fact, up to this date, there is no known solution to the model in two and higher dimensions. 

Most of the last century was devoted to finding the exact solutions and the mean-field studies of the
static (equilibrium) properties of the quantum spin lattice models.
Most of the studies focused on the low-temperature regime where quantum fluctuations dominate over thermal fluctuations.
Such studies showed that quantum spin lattice systems not only explain the magnetic properties of materials, 
but also showed that strongly interacting quantum spin lattices can exhibit exotic phases.
Such exotic phases include:
Anderson localisation\index{Anderson localisation} \cite{Anderson1958}, topologically ordered states\index{Topologically ordered states}
\cite{Haldane1983a,Affleck1989}, the spin glass states\index{Spin glass states} \cite{Physik1986} 
and the spin liquid states\index{Spin liquid state} \cite{Anderson1973,Zhou2017}. 
Due to a simple but exotic nature of the model, studies of model's variants form an important branch of modern physics.
Today such exotic phases are attracting attention due to their applicability in 
high-temperature superconductivity and the physical realisation of quantum computers \cite{Qi2011,halcrow2013spin,Wolf2001}.

\subsection{Ising Model and Quantum Phase Transition}
A key toy model of the spin lattice models studied is the Ising model \cite{Ising1925}\index{Ising model}.
The model simplifies the Heisenberg model by only considering spin-spin interactions in one direction. 
The Quantum Hamiltonian of the model is
\begin{align}
    \hat{H} &= \sum_{i} \hat{S}^{z}_i\hat{S}^{z}_{i+1} + \sum_{i}\Gamma_i \cdot \hat{\bm{S}}_{i}
\end{align}
It is known that a $d$-dimensional model of this Hamiltonian 
can be mapped onto a $(d+1)$ lattice with classical spins \cite{Suzuki1976}.
Classical Ising lattices in two dimensions are well studied using the Transfer Matrix method \cite{schultz1964a,Baxter1971},
and it is one of the first model which was solved exactly \cite{Onsager1944}.
Quantum Ising lattices in one and two dimensions are also well studied.
In particular, the natures of the short-range interactions and square lattice models are well known
\cite{BaxterRodneyJ1982Esmi,Brush1967,Wu1971,Bortz1975}.

\begin{figure}
    \begin{minipage}{0.5\textwidth}
        \includegraphics[width=\textwidth]{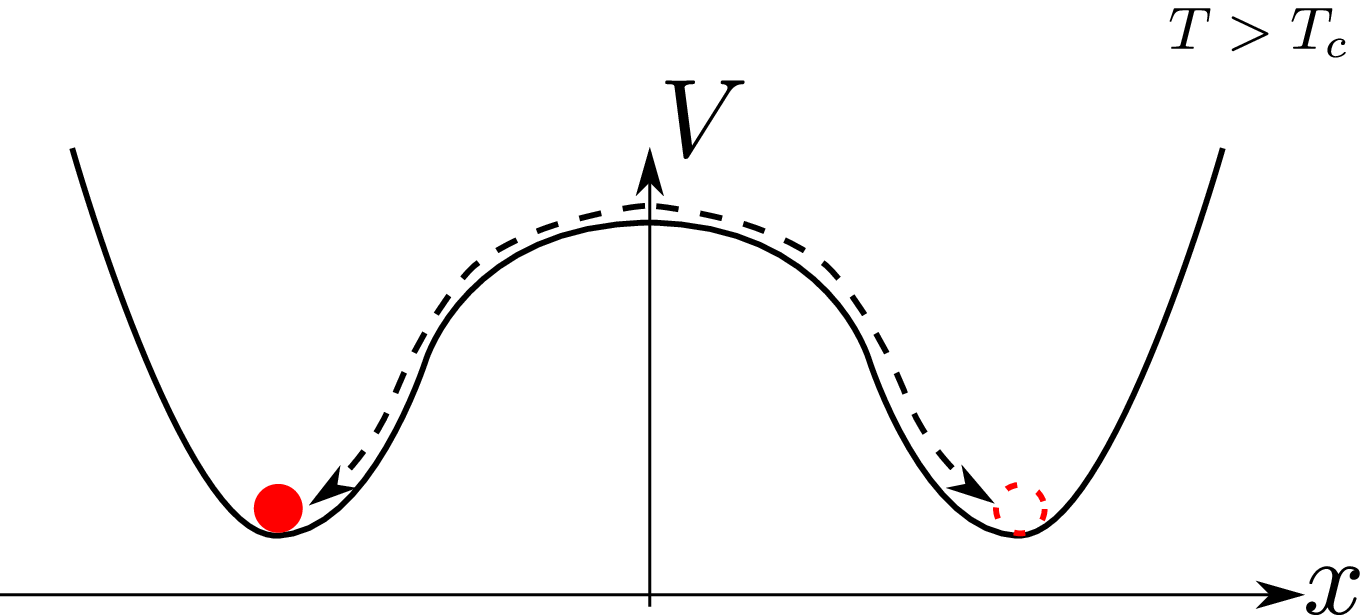}
    \end{minipage}\hfill
    \begin{minipage}{0.5\textwidth}
        \includegraphics[width=\textwidth]{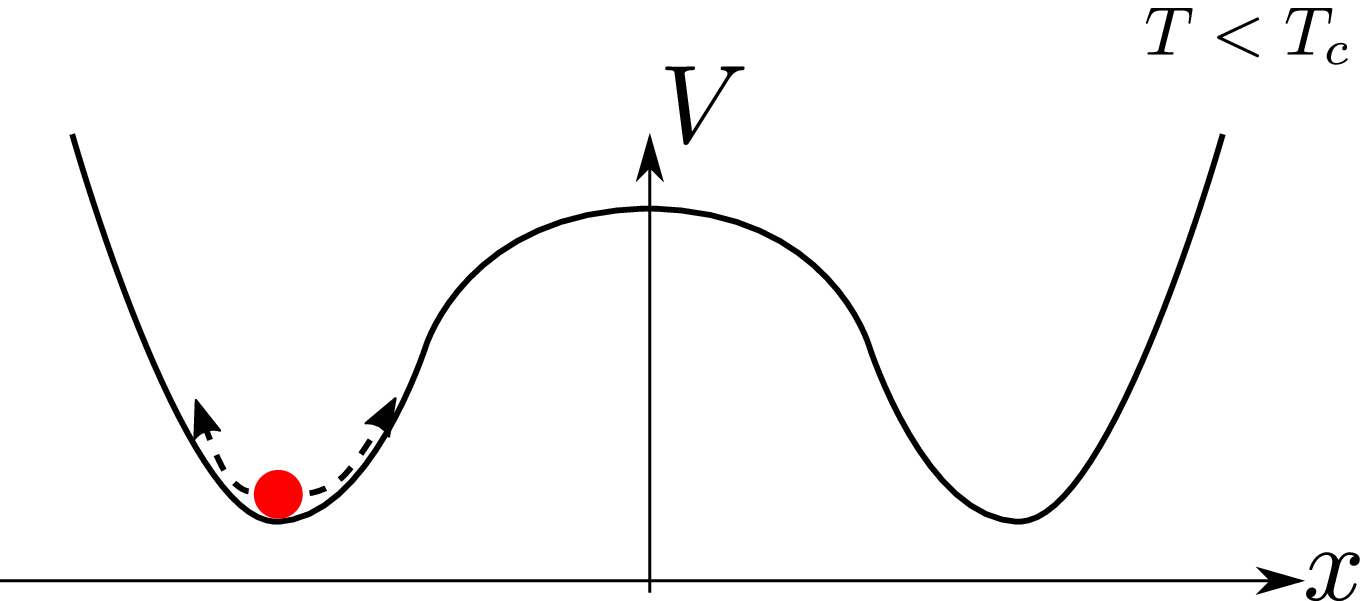}
    \end{minipage}
    \caption{A simple example which relates a phase transition to spontaneous symmetry breaking}
    \legend{At high temperature, a particle (red ball) can move freely because it has enough energy to overcome the potential barrier at 
        the centre (left). 
        Therefore the parity symmetry always exist. However, in the low temperature regime, a particle gets trapped in either of the well, which breaks
        the parity symmetry. 
    }
    \label{simpleSSBex}
\end{figure}

The Ising model, in spite of its simplicity, can be used to model quantum phase transitions\index{Quantum phase transition}\index{Quantum phase}.
A quantum phase transition is a phase transition which occurs on a ground state due to the change in the phases of the ground state (quantum phase) of a system. 
In the theory of an ordinary phase transition\index{Phase transition} the phase transition occurs because of the spontaneous symmetry breaking
which can be explained by Landau's famous symmetry breaking theory \cite{Landau1969}.
For example, magnetisation breaks a rotational symmetry due to a spontaneous alignment of the electrons in a metal; freezing (liquid-solid transition) 
breaks the translational symmetry due to the spontaneous alignment of water molecules into a periodic crystalline structure.
The broken symmetry, which occurs when the temperature drops below the critical temperature\index{Critical temperature} ($T_c$), is 
known to be triggered by the loss of thermal fluctuations in a system (see \fig{simpleSSBex} and its caption for an example).
It is thought that because all the thermal fluctuations in a system disappear, the phase at absolute zero temperature must be a symmetry broken phase.

\begin{figure}[tb]
    \centering
    \includegraphics[width=0.5\textwidth]{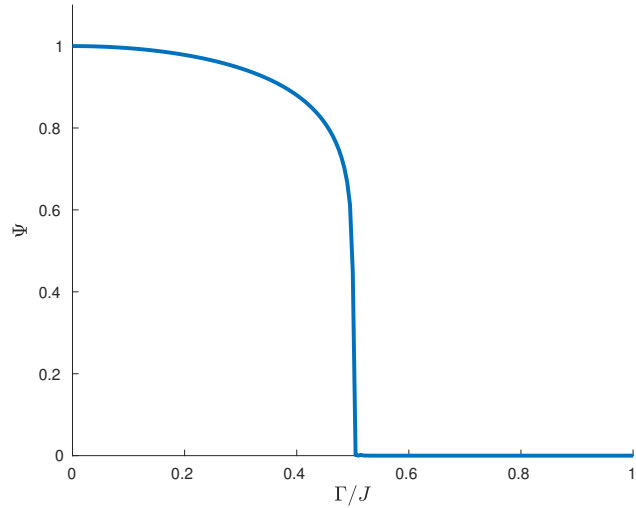}
    \caption{Quantum Phase Transition in the spin-1/2 nearest-neighbour transverse Ising model.
    There is a sharp drop off of the order parameter at $\Gamma/J = 1/2$.}\label{OrderParamIsingQPT}
\end{figure}

However, thermalisation are not the only source of the fluctuations.
In quantum systems there always some quantum fluctuations due to the uncertainty principle. 
In fact, a behaviour analogous to finite-temperature symmetry breaking can happen in quantum phases 
due to the change in the amount of quantum fluctuations.
Consider a spin-1/2 nearest-neighbour transverse Ising lattice with the Hamiltonian
\begin{align}
    \hat{H} &= J\sum_{j} \hat{S}^{z}_{j}\hat{S}^{z}_{j+1} + \Gamma \sum_{j}S^{x}_{j}
\end{align}
If there is no external magnetic field, then the ground state of the model 
is either ferromagnetic or antiferromagnetic.
They are two fold degenerate because the local state of a site can be spin up or spin down, but choosing a spin will 
immediately defines the local spin states of rest of the system. 
The order parameters ($\Psi$) of the each of the phases are defined as follows:
\begin{align}
    \Psi &=\frac{2}{N} \times 
    \begin{cases}
        \left|\sum_{j} \bra{\psi}S^{z}_j\ket{\psi}  \right| &(\text{ferromagnetic}) \\
        \left|\sum_{j} i^{j}\bra{\psi}S^{z}_j\ket{\psi} \right| &(\text{anti-ferromagnetic})
    \end{cases}
\end{align}
where $i$ is the imaginary unit with $i\times i = -1$. 
Starting from the ground state at $\Gamma = 0$ and quasi-statically changing $\Gamma$ towards $\infty$,
the state undergoes a phase transition into the disordered state (no order in $z$ direction).
This is because increasing $\Gamma$ causes the eigenstates of $\hat{S}^{z}$ to no longer be the eigenstates of the Hamiltonian.
Like in finite temperature classical systems, the order parameter goes to $0$ precisely at $|\Gamma/J|=1/2$ (\fig{OrderParamIsingQPT}).
It is worth noting that this is exactly the same (mathematically equivalent) process 
as the thermal phase transition which occurs in the two-dimensional Ising model. 

It is later found that quantum fluctuation can recover the broken symmetry, leading to a
symmetry protected ground state (Haldane phase\index{Haldane phase} \cite{Wen2012,Haldane1983a,Affleck1987,Affleck1989,Wen1989a}). 
The mechanism behind this type of phase transition is fundamentally different from 
the translations which can be derived from the symmetry breaking theory.
Such a phase transition can be realised in some of the two-dimensional Ising models \cite{Jalal2016},
however the exact nature of the realisation of such a phase in the Ising model is still unknown.
In particular, effects of long-range interactions and geometrical structure of the lattice (in particular the triangular lattice) 
on the quantum phases of a system, even in the Ising model, are not very well known.
\section{The Density Matrix Renormalization Group Algorithm and Matrix Product States}
\label{Thedensitymatrix}
At the end of the last century, a numerical scheme called the \textit{density matrix renormalization group} 
(DMRG\index{Density Matrix Renormalization Group}\index{DMRG}) was invented by Steven White \cite{White1992,White1992a}.
It is an algorithm which searches for the ground state of a spin chain by iteratively truncating out the high-energy sectors of 
the Hilbert space.
The invention revolutionised the studies of quantum lattice systems.
On the very first numerical calculation using this technique, White showed that the algorithm gives a $1000$ times 
improvement in precision in finding the ground state of a one-dimensional spin chain compared to previously known methods 
(e.g. quantum Monte Carlo\index{Quantum Monte Carlo}). 
Due to the drastic technological advancement, today the algorithm allows to determine the ground state energy of a spin chain\index{Spin chain}
(one-dimensional spin lattice) with the precision that is comparable to the machine epsilon precision\index{Machine epsilon} \cite{Schollwock2005}.

\begin{figure}[tbp]
    \begin{center}
    \includegraphics[width=0.5\textwidth]{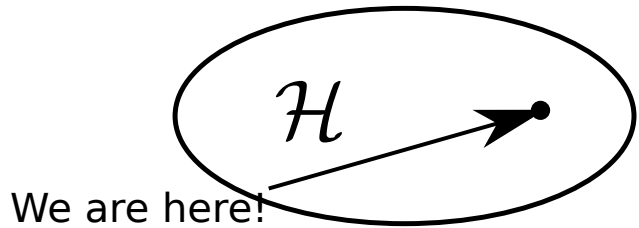}
    \caption{Physically realisable states in a Hilbert space}
    \label{HSpaceTruncation}
    \legend{Although the dimensions of Hilbert Space tend to be very large, most of it is physically unreachable.
        In fact, all the region in white are physically impossible to be realised in a reachable time scale.
        By imposing a physical constraints such as the macroscopic behaviour of a many-body system is dominated by the dominant states of the density matrix
        a system, one can efficiently find a relevant state which can be physically realised (e.g. the ground state)}
    \end{center}
\end{figure}

The core of DMRG lies on truncation of Hilbert space.
Consider a lattice  with a mole ($10^{23}$) of spin-1/2 particles. 
Even considering only the spin degree of freedom, the system spans Hilbert space with dimensions $2^{10^{23}}$.
It is obvious that performing the exact diagonalisation of this system and finding the desired state is impossible.
However, if one thinks deep enough, then it can be realised that physical realisation of all the states in the system is also impossible.
Let's say that the most dominant state in the system changes in a time scale of $\tilde{t}$.
Within the age of the universe, the number of physically realisable states since the beginning of the universe is roughly $10^{17}/\tilde{t}$.
Even setting the Planck time as the time scale of $\tau$, the number is nowhere close to $2^{10^{23}}$.
This simple analysis tells us that not the entire Hilbert Space is necessary to describe a physical system.
In fact, the region of interest for physicists is just a dot in a Hilbert Space (\fig{HSpaceTruncation}).
DMRG, in essence, relies on the fact that the macroscopic behaviour of a physical state is determined by the dominant state in the density matrix of a system. 
This acts as a physical constraint when truncating the Hilbert space starting from two particle states
(or any small number of particles which are computationally manageable).

The algorithm was later integrated with a new representation of a state called a Matrix Product State\index{Matrix Product State} (MPS). 
It was formulated originally by Klumper et al. \cite{Klumper,Klumper1992,Fannes1992},
to represent a variational ansatz for anisotropic case of the AKLT model \cite{Affleck1987}.
The MPS is represented by products of matrices (c.f. \Chap{Anewtimeevolution} and reviews \cite{Perez-Garcia2007,Schollwock2011}).

In quantum mechanics, the states of an isolated system can be represented as a superposition of basis states. 
For example, if a state ($\ket{\Psi}$) in a spin-1/2 chain of length $N=4$ 
is in a superposition of ferromagnetic states with all spins pointing $+z$ direction ($\uparrow$)
and all spins pointing in the $-z$ direction ($\downarrow$), then the state can be written as a sum of basis states:
\begin{align}
    \ket{\Psi} &= \frac{1}{\sqrt{2}}\left( \ket{\uparrow\uparrow\uparrow\uparrow} + \ket{\downarrow\downarrow\downarrow\downarrow} \right)
    \label{normalstate}
\end{align}
This state can be then written in terms of a matrix product, where each matrix consists of the local basis.
Then the state can be written as:
\begin{align}
    \equiv \frac{1}{\sqrt{2}}
    \begin{pmatrix}
        \ket{\uparrow}_1 & \ket{\downarrow}_1
    \end{pmatrix}
    \begin{pmatrix}
        \ket{\uparrow}_2 & 0\\
        0 & \ket{\downarrow}_2
    \end{pmatrix}
    \begin{pmatrix}
        \ket{\uparrow}_3 & 0\\
        0 & \ket{\downarrow}_3
    \end{pmatrix}
    \begin{pmatrix}
        \ket{\uparrow}_4 \\ \ket{\downarrow}_4
    \end{pmatrix}
\end{align}
This representation of a state is the Matrix Product State. 
It can be expanded to represent a translationally invariant state in a thermodynamical limit\index{Thermodynamical limit}
(the limit where $N=\infty$). 
For example, the thermodynamical limit of the state in \eqn{normalstate} can be written as:
\begin{align}
    \ket{\Psi} 
    &=
    \cdots \begin{pmatrix}
        \ket{\uparrow} & 0\\
        0 & \ket{\downarrow} 
    \end{pmatrix} \cdots
\end{align}
The detailed theoretical framework of the Matrix Product States and allowed mathematical operations are covered in \Chap{Anewtimeevolution}.

In this representation, it can clearly be seen that the entanglement between the nearest neighbour sites lives in between the matrices (bonds\index{Bonds}). 
The long range entanglement is then realised through the entanglements propagating through the nearest neighbour bonds of the matrices. 
Increasing the dimension of the matrices ($m$) increases the number of basis states which can be used to represent $\ket{\Psi}$. 
Therefore, for sufficiently large $m$ ($\sim100$), it can efficiently represent states with locally confined correlations
(i.e. high entanglement between the nearest neighbours but very low entanglement between sites that are far apart).
This is exactly one of the properties of the ground state of most of the spin chains \cite{Schollwock2011,sachdev2007quantum,Latorre2003}. 

It was proven that DMRG is equivalent to a variational minimisation algorithm of a Matrix Product State with respect 
to the Hamiltonian of both infinite and finite systems \cite{Rommer1995,Rommer1997,Takasaki1999,McCulloch2007b,Pirvu2010}.
This realisation motivated physicists to reformulate the DMRG in terms of MPS. 
Such a reformulation gave rise to the formulation of a matrix product representation of non-local 
operators (such as Hamiltonian), which are now known as Matrix Product Operators \index{Matrix Product Operators}\index{MPO}(MPO)
\cite{Verstraete2004a,Vidal2004,McCulloch2007b}.

\begin{figure}[tb]
        \centering
        \includegraphics[width=0.4\linewidth]{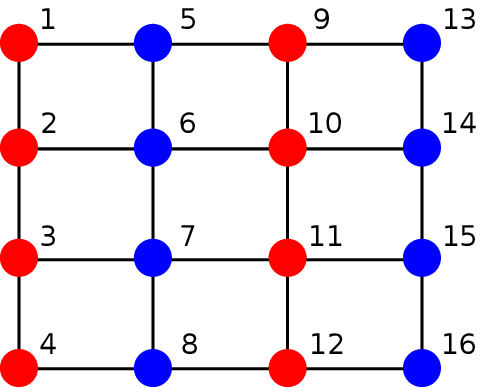}
        \includegraphics[width=0.8\linewidth]{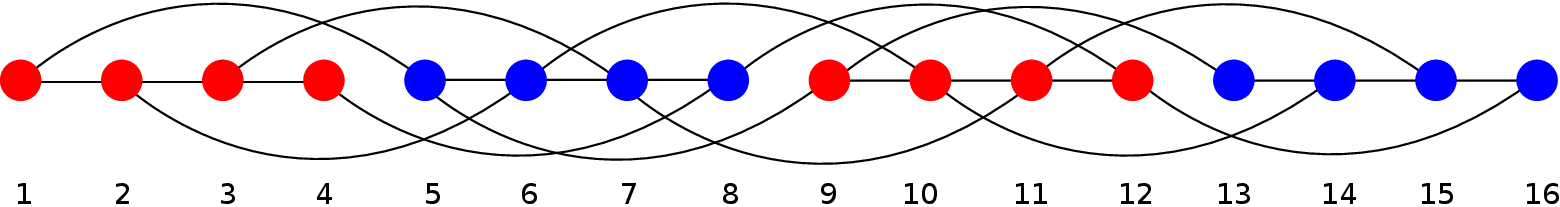}
    \caption{Projection of a two-dimensional system onto a one-dimensional spin chain}
    \legend{
        (top) The system is described by the Hamiltonian 
        \begin{minipage}{\linewidth}
        \begin{equation}
            \protect\hat{H} = -\sum_{<i,j>} \hat{\bm{S}_i}\cdot\hat{\bm{S}_j} 
        \end{equation}
        \end{minipage}
        where $<i,j>$ denotes a nearest neighbour sites. The system is translationally invariant and has only nearest neighbour interactions (above).
        The solid circles correspond to the spin vector operator at each site.
        Different colours are used only to distinguish the different columns for visualisation purpose.
        Black lines represent the bonds\index{Bonds} in the Hamiltonian.
        (bottom) The projection onto a 1D chain distorts the translational symmetry and introduces long-range interactions.
        }\label{projection}
\end{figure}

These new representations for states and operators gave new ways of performing classical simulations of quantum lattice systems. 
In particular, the behaviour of a quantum spin chain system at finite-temperature, and out of equilibrium dynamics attracted 
the interests of many areas; to name a few, condensed matter, ultracold atomic gases, and quantum information.
This motivated the development of sophisticated time evolution algorithms for MPS.
There are two major algorithms which can be used to evolve MPS: time-evolving block decimation \cite{Orus2007} (TEBD) 
and time-developing variation principle \cite{Haegeman2011} (TDVP). 

However, the applicability of such numerical methods is limited.
TEBD can only be readily applied to a Hamiltonian with interactions shorter than nearest neighbour \cite{Pirvu2010}, and 
TDVP requires a translationally invariant initial state \cite{Haegeman2011}.
This is problematic because not all of the systems of interest can be described in terms of a Hamiltonian with 
translational invariance and/or short-range interactions.
This is especially problematic when a small two-dimensional lattice is projected onto a one-dimensional spin-chain to make use of an MPS representation.
Even for a two-dimensional spin lattice with an initial state with translational invariance and a Hamiltonian with only nearest neighbour interactions, 
projection distorts the symmetry making it difficult to apply those algorithms (\fig{projection}). 
Therefore there are no known results of numerical simulations or any proposals of algorithms for the time evolution of 
two-dimensional quantum spin lattice systems. 

\newpage
\section{Quantum Simulators and Two-Dimensional Quantum Spin Lattice System}
\label{Quantumsimulatorsand}
In the latter half of the last century, due to the breakthrough in cooling technologies \cite{Adams1997,Phillips1998}
there has been active research on materials which possess magnetic properties of the Heisenberg type models in the low-temperature regime
\cite{Stinchcombe1973,halcrow2013spin,Wolf2001}.
Such probes showed that Heisenberg-type models can explain magnetic properties very well with experimentally determined coupling constants
(e.g $J_{ij}$ in \eqn{HeisenbergSpinChain}).
Experiments conducted on such materials cannot control bonding strengths or individual spins in a lattice.
As a consequence, predict-and-test type of investigation had been impossible, especially for the models with two and higher dimensions. 
Therefore, to probe of the models, there has been a strong interest in developing fully controllable artificial spin lattice systems
(i.e. quantum simulators\index{Quantum simulator}). 

\begin{figure}[tb]
    \centering
    \begin{minipage}{0.48\textwidth}
        \centering
    \includegraphics[width=0.95\textwidth]{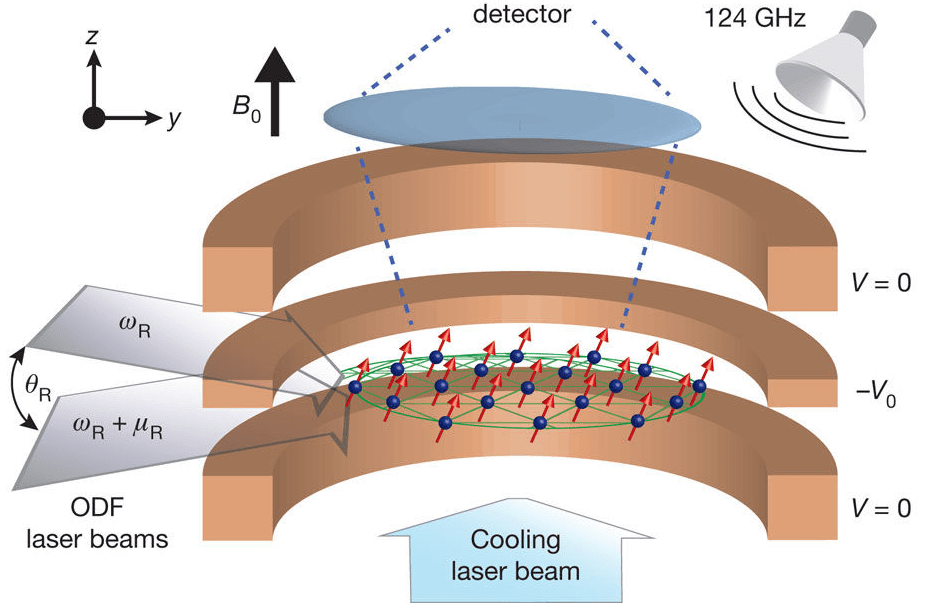}
    \end{minipage}\hfill
    \begin{minipage}{0.48\textwidth}
        \centering
    \includegraphics[width=0.80\textwidth]{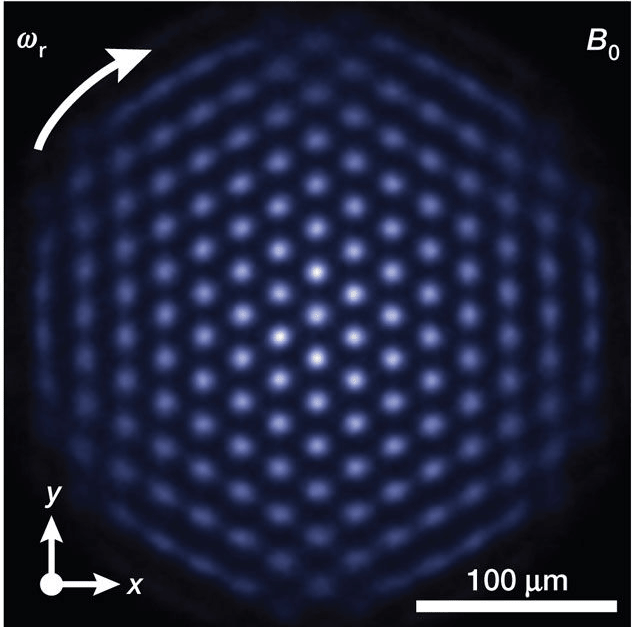}
    \end{minipage}
    \caption{The experimentally realised long-range Ising lattice \cite{Britton2012}}
    \legend{(left) An illustration of an apparatus at NIST which realises a transverse Ising lattice model with long range interaction (left).
     Each ion acts as a spin-1/2 particle. Due to the electric property of the ions, the application of a strong uniform magnetic field and 
     quadrupolar electromagnetic wave traps ion in a two-dimensional plane (Penning trap).
    The long range interactions between individual atoms are then realised by the optical dipole force between atoms induced by transverse laser lights 
    with the right frequencies. The detected spin alignment in the apparatus with a uniform magnetic field with no spin coupling interactions. 
    (right) Bright spots are spin which points upwards ($\ket{\uparrow}$) 
    and dark spots are spin pointing downwards ($\ket{\downarrow}$).}
    \label{2012NIST}
\end{figure}

In 2012, an experimental apparatus which allowed such parameters to be controlled 
was finally realised by researchers at the NIST (National Institute of Standards and Technology)\index{NIST} and their international collaborators
\cite{Biercuk2009,Britton2012}.
The apparatus can simulate a two dimensional triangular quantum Ising lattice (\fig{2012NIST}) with long-range interactions.
The Hamiltonian of the lattice is given as
\begin{align}
    \hat{H} &= \sum_{i\neq j}\frac{\hat{S}_{i}^z\hat{S}_{j}^z}{|\bm{r}_{i}-\bm{r}_j|^{\alpha}} + \sum_{i}\bm{\Gamma}\cdot\bm{\hat{S}}_i
    \label{LLHamil}
\end{align}
With tunable parameters $\alpha$ (controls the strength of long-range interactions) and $\bm{\Gamma}$ 
(controls the strength and the direction of the applied magnetic field).

It is known that this type of system can be used as a quantum simulator \cite{Buluta2009}.
Quantum simulators are a highly controllable artificial quantum system, which can be used to simulate other quantum systems 
more efficiently than classical computers. The applications are not limited to the field of physics.
It is suggested that quantum simulators can be used for solving computationally hard optimisation problems 
and performing quantum chemistry simulations \cite{Lucas2013,Ronnow2014}.
The quantum simulators are attracting attention as they could be the first generation general purpose 
\footnote{although the usage is limited compare to the classical computers, it is general in a sense that its use is not limited to only one type of problem}
quantum computer.
It is also hoped that the classical MPS simulations and the experimentally realised quantum simulators can complement each other in \textit{ab initio}
\index{\textit{Ab initio}} (formulation of the properties of a built matter from the fundamental laws) studies in condensed matter physics. 

\begin{figure}[tb]
    \begin{minipage}{0.5\textwidth}
    \centering
    \includegraphics[width=0.5\textwidth]{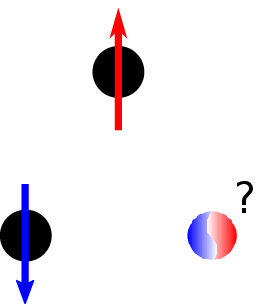}
    \caption{An example of spin frustration in a triangular lattice}
    \legend{Although two of the spins in the sites are determined, the third spin can point in either direction and
    in either of the cases the energy is the same.}
    \label{degentrilat}
    \end{minipage}\hfill
    \begin{minipage}{0.5\textwidth}

    \fboxsep=0pt
    \fboxrule=1pt
    \begin{minipage}{0.48\textwidth}
        \fbox{\includegraphics[width=\textwidth]{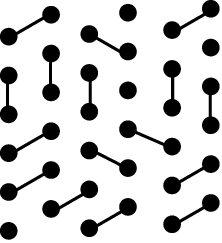}}
    \end{minipage}\hfill
    \begin{minipage}{0.48\textwidth}
        \fbox{\includegraphics[width=\textwidth]{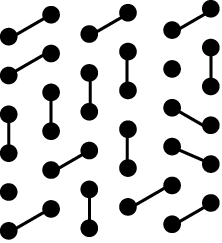}}
    \end{minipage}

    \caption{A schematic drawing of two of the degenerate RVB states}
    \legend{Each of the solid lines connecting two nearest neighbour sites is a valence bond between the sites, which is described by a 
        state in \eqn{nnstate}.
    The valence bond configuration in the left and the valence bond configuration in the right gives the same ground state energy.}
    \label{RVBstate}
    \end{minipage}
\end{figure}

Among other lattice geometries, the triangular lattice is at the center of attention for use in spin lattice quantum simulators (c.f. \fig{2012NIST}).
As it can be seen from the \fig{2012NIST}, the triangular lattice is at the center of attention among other lattice geometries for the geometry of spin lattice
quantum simulators. 
This is because of the interesting nature of a quantum phase called a spin liquid phase.
When there is a triangular lattice with isotropic interactions, the states in the system become massively degenerate. 
As shown in \fig{degentrilat}, when the Hamiltonian of a triangular Ising lattice 
only contains nearest neighbour interactions which favour anti-parallel alignment,
spin frustration occurs.
When two of the sites in a lattice are in the anti-parallel alignment, 
then the spin at the third site can point in either direction without changing the total energy of the lattice.
This causes a massive degeneracy in the energy spectrum. 

A new approach was first proposed by Anderson in 1973 \cite{Anderson1973} as an extension of the work by Pauling \cite{Pauling1949}.
The original intention was to determine whether a material with electrons that can move around freely is always a metal.
Pauling, in 1949, gave an answer to this question with a physical intuition based on the argument of why Mott insulators 
(materials which are insulator due to the strongly interacting valence electrons) 
do not conduct electric current.
Anderson made a more quantitative argument by representing the ground state of the triangular Heisenberg lattice
as a tensor product of the nearest neighbour sites.
The nearest neighbour bonds can be written as:
\begin{align}
    \ket{\Psi (i,j)} &=\frac{1}{\sqrt{2}}\left(\ket{\uparrow_i\downarrow_j} - \ket{\downarrow_i\uparrow_j}\right)\label{nnstate}
\end{align}
where $i$ and $j$ are the nearest neighbour sites in a triangular lattice. 
This representation hides away the geometric frustration, 
but introduces a massive degeneracy on the ground state, which causes a very large energy gap to the first excited state (\fig{RVBstate}).
This massively degenerate state caused by the different arrangement of nearest neighbour bonds 
(physically, this bond can be realised through valence bonds between the nearest neighbour atoms) is called resonating valence bond (RVB)
\index{Resonating valence bond}\index{RVB}. 
It is thought that RVB theory can explain high-temperature superconductivity \cite{Baskaran1993,Anderson2013}; 
the experimental realisation and the development of a more fundamental theoretical framework have been a long-sought goal in condensed matter physics. 

\begin{figure}[tb]
    \begin{minipage}{0.48\textwidth}
    \centering
    \includegraphics[width=1.0\textwidth]{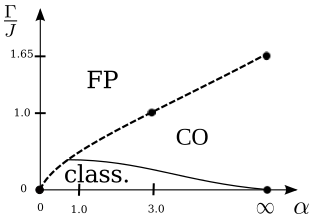}
    \caption{The ground state phase diagram of the two-dimensional triangular Ising lattice.}
    \legend{ The Hamiltonian of the system is given by \eqn{LLHamil2}. 
    The phase diagram is constructed from mean-field and quantum Monte Carlo study by Humeniuk\cite{Humeniuk2016}.
    \textit{FP} is a ferromagnetic phase; \textit{class} is a classical phase; and \textit{CO} is a clock ordered phase.}
    \label{MFQMC}
    \end{minipage}
    \begin{minipage}{0.48\textwidth}
    \centering
    \includegraphics[width=1.0\textwidth]{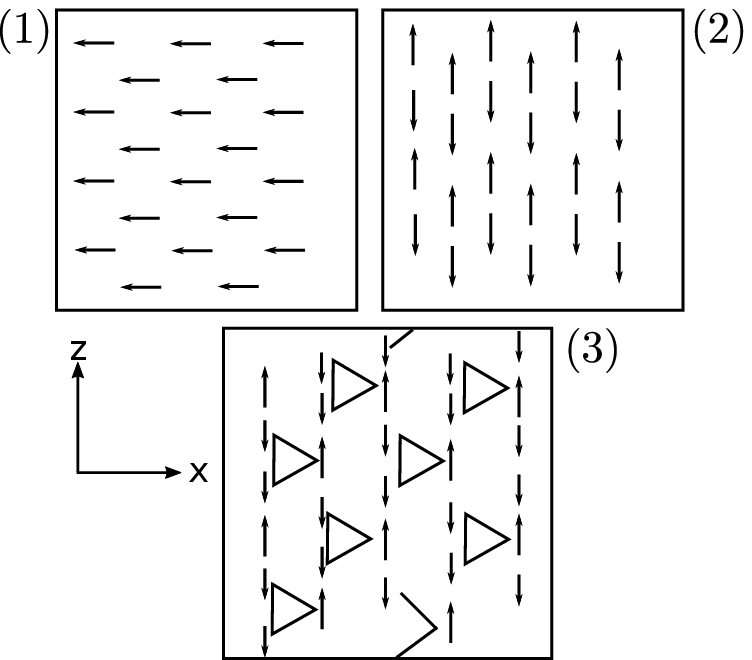}
    \caption{The schematic diagram of a ferromagnetic (1), classical (2), and clock (3) phase}
    \legend{Arrows are pointing in the direction of the spin at each site.}
    \label{VisualPhase}
    \end{minipage}
\end{figure}

The exact quantum phases that exist in the experimental setup of NIST \etal \cite{Britton2012} (described by the \eqn{LLHamil}) is still not known.
Whether it has a spin liquid phase is a question that still remains.
The mean-field and quantum Monte Carlo calculation done by Humeniuk \cite{Humeniuk2016} provided a very brief 
ground state phase diagram (\fig{MFQMC})) of the two-dimensional triangular Ising lattice with a Hamiltonian
\begin{align}
    \hat{H} &= J\sum_{i\neq j}\frac{\hat{S}_{i}^z\hat{S}_{j}^z}{|\bm{r}_{i}-\bm{r_j}|^{\alpha}} + \Gamma\sum_{i}\hat{S}_i^x
    \label{LLHamil2} &(J>0, \Gamma>0)
\end{align}
with parameters $\alpha$ that controls an algebraically decaying long-range interactions and $\Gamma$ which controls a transverse magnetic field. 

According to the study, the system contains three distinct region in the ground state phase diagram which are visualised in \fig{VisualPhase}.
The first region is a trivial ferromagnet.
As the effect from the $\Gamma$ term increases, the spins try to align themselves against the direction of the magnetic field, and the ground state becomes
ferromagnetic in the negative $x$-direction. The second region is a classical (columnar) order, which is an antiferromagnetic phase, where 
the directions of the nearest neighbour spins are antiparallel.
Finally, the last region is a clock ordered phase.
It is an order in which the magnetisation in the direction of the $z$-axis is $(1,-1/2,-1/2)$.

The full nature of the model is still unknown. 
Humeniuk's study did not reveal the quantum behaviour of the ground state at the thermodynamical limit or the existence of a spin liquid phase.
Most of the nature of the two-dimensional triangular Ising lattice is unknown, especially with a Hamiltonian with long-range interactions.
It is hoped that an \textit{ab initio} simulation and further theoretical studies will reveal a new type of spin material
and fundamental laws of condensed matter physics beyond the equilibrium regime and the semi-classical approximations. 
\section{The Dynamical Phase Transitions in A Quantum System}
Compared to the static phase transitions, the nature of the dynamical phase transitions are not known in detail. 
In fact, due to the time scale of the thermalisation, and the theoretical and computational difficulty, 
the dynamical behaviours of closed quantum many-body systems with strong interactions in general are not known very well either. 
As a consequence there are not many references in the literature that are available on the topic of dynamical phase transitions.

However, there have been a few attempts, mainly from the theoretical sector, to study the dynamical behaviours of exactly solvable spin chains
\cite{Schutz1999,Igloi2000,Barouch1970,Barouch1971a,Barouch1971}.
Due to the experimentally realised, highly controllable quantum simulators (\S\ref{Quantumsimulatorsand}), 
attempts to cultivate the last frontier of condensed matter physics | the fundamental laws that governs the dynamical behaviour of 
the strongly correlated quantum many-body systems | are beginning to be made. 

A type of non-equilibrium behaviour which seems to be promising is a procedure called \textit{global quenching}\index{Global quench}.
Global quenching is done as follows. First the ground state is prepared in some Hamiltonian $\hat{H}_i$.
Then at $t=0$, the Hamiltonian of the system is changed by tuning a parameter 
and evolving the state according to Schr\"{o}dinger's equation with the new Hamiltonian.
This type of dynamics can be realised in quantum simulators.
It is also easier to implement than \textit{local quenching}\index{Local quench}, a type of quench where a state is deformed locally
and followed by a measurement of how the deformation propagates through the state \cite{Diehl2010,Eisert2014}.

It is known that dynamical phase transitions naturally occur in system in which a global quenching is performed \cite{Diehl2010}.
A key paper that is relevant to this thesis on global quenching is a study done by Halimeh \etal \cite{Halimeh2016}. 
It is a study of the dynamical properties of a transverse Ising chain with long-range interactions.
It is described by the Hamiltonian:
\begin{align}
    \hat{H} &= -\sum_{i\neq j} \frac{\hat{S}_i^z \hat{S}_j^z}{4|\bm{r}_i-\bm{r}_j|^{\alpha}}-\frac{h}{2}\sum_{i}\hat{S}^x_i
\end{align}
Their numerical study using a time-dependent Density Matrix Renormalization Group algorithm \cite{Daley2004}
showed signs of order-to-disorder phase transition due to the quantum decoherence for the parameters $\alpha=3$ and $h=0.99$.
Also, it showed that there is a regime where the order is protected ($\alpha=3$ and $h=0.28$). 
Although the sign of the long-range interaction term in the Hamiltonian is different to \eqn{LLHamil} and it is one-dimensional, 
giving \textit{ab initio} evidence of order protected dynamical phase gave new insight into the possibility of such a dynamical
phase transition existing in different systems with higher dimensions and non-trivial interactions.
\section{Motivation and Aim}
\label{Motivationandaim}
As reviewed, there are still many mysteries concerning the nature of the quantum spin lattice models.
In particular the dynamical and non-equilibrium natures of the models are not known at all.
Questions such as 
\begin{itemize}
    \item Do ground state phase structure affect finite temperature dynamics?
    \item Is there fundamental laws (analogous to the laws of thermodynamics) which are governing dynamics of quantum many-body systems?
    \item Is there any difference between the quench dynamics of a symmetry broken and topologically ordered ground state?
    \item What are the effects of long-range interactions on many-body dynamics?
\end{itemize}
are still unanswered.
It seems that the key to answer these questions lies in the study of quantum spin lattice systems.
It has a rich ground phase which includes topologically ordered ground states and it has experimental apparatus which can be used 
to probe the dynamics \cite{Bohnet2016}.
However, it seems that the experimental realisation is not yet possible.
Therefore the current interest is in performing the numerical simulation of the system.

The aim of this project is to develop a new algorithm for evolving the Matrix Product States in a spin chain with long-range interactions.
Since, some of the two-dimensional spin lattices can be mapped on to a spin chain with long-range interactions,
the algorithm is hoped to simulate two-dimensional spin lattice systems. 
The model at the focus of the thesis lies on the two-dimensional Ising lattice described by the Hamiltonian in \eqn{LLHamil}, as this model is 
experimentally realised and has some study regarding to its equilibrium behaviours.
\section{Structure of Thesis}
In \Chap{Anewtimeevolution} a new algorithm for simulating the time evolution of Matrix Product States (MPS) is introduced. 
The chapter begin by introducing MPS and Matrix Product Operators (MPO).
Then the theory of the arithmetics of MPS and MPO are explained.
Then the theory of numerical implementation of the calculation of matrix exponential is explained.
The new algorithm is then introduced.
The chapter ends by providing the results of the benchmark tests of the algorithm on the exactly solvable cases 
(Transverse Ising model\cite{Search2010}, Haldane-Shastry Model\cite{Shastry1988}), and a previously studied case\cite{Halimeh2016}.
In \Chap{dynamicalphasetransitions}, the algorithm is applied on the two-dimensional Ising lattice with long-range interactions.
The chapter begin with explaining how the geometry of the lattice is simplified and mapped onto a spin-chain.
Then the ground state phase diagram of the lattice is introduced. 
The result of the simulation of global quenching with the new algorithm 
for different sets of initial and final Hamiltonians are then presented with a discussion on their physical implications.
A conclusion, which summarises the thesis and the proposes the future research prospects, are provided in the last chapter (\Chap{Conclusion}).

\chapter{A New Time Evolution Algorithm for Matrix Product States with Long Range Interactions}
\label{Anewtimeevolution}
\section{Matrix Product States and Matrix Product Operators}\label{Matrixproductstates}
\subsection{Matrix Product States}
As introduced in \Chap{Thedensitymatrix}, a state with a locally confined entanglement profile can be represented using the Matrix Product State (MPS)
\index{Matrix Product States}.
In this section, the theoretical background on MPS and its operator generalisation (Matrix Product Operator, MPO)\index{Matrix Product Operators} are discussed.
This section is based on the three great review papers \cite{Daley2004,Schollwock2011,McCulloch2007}.
Although mathematical rigor is important, the focus of this section is on filling in between the lines on what is covered in those reviews in a context
of the new algorithm that is introduced in the next section (\S\ref{Anewtime}) in a self-consistent manner.

In quantum mechanics, the Hilbert space of a spin lattice system is spanned by the tensor product of local basis. 
In a quantum spin lattice system with $N$ sites, the basis states of the system ($\ket{\bm{\sigma}})$) 
can be written in terms of the tensor product of local spin states.
\begin{align}
    \ket{\bm{\sigma}} = \ket{s_1}\otimes\ket{s_2}\otimes \cdots \otimes \ket{s_{N-1}}\otimes\ket{s_{N}}
\end{align}
With this notation, any state of the system can be written as a superposition of the basis states as follows
\begin{align}
    \ket{\Psi} &= \sum_{i}^{(2s+1)^{N}}A_{i}\ket{\bm{\sigma}_i}
    & \sum_{i}^{(2s+1)^N} |A_i|^2 = 1
\end{align}
where $\bm{\sigma}_i$ is the $i$\textsuperscript{th} basis state, $s$ is the magnitude of the spin of the particles at each site (for spin-1/2 particle $s=1/2$), 
and $s_i$ is a local configuration of a spin.
There are $(2s+1)^{N}$ possible configurations of $\bm{\sigma}_i$; this is the dimension of the Hilbert space spanned by the lattice. 
Crudely, one needs $(2s+1)^{N}$ numbers to describe a state in the system using vector notation.

However, like continuous wavefunction can be approximated with a sum of a small number of eigenfunctions, 
well behaved states on a quantum lattice can be approximated with a small number of basis states.
Let $m$ be the number of basis states used to approximate the state $\psi$.
Consider a spin lattice of $N=4$, $s=1/2$ and $m=2$, 
if the state ($\ket{\psi}$ of the lattice is in a superposition of states where one is all the spins pointing upwards, 
and the other is all the spins pointing downwards, then the state can be written as
\begin{align}
    \ket{\Psi} &= \frac{1}{\sqrt{2}} \left( \ket{\uparrow\uparrow\uparrow\uparrow} + \ket{\downarrow\downarrow\downarrow\downarrow} \right)
\end{align}
As it is shown in \Chap{Thedensitymatrix}, this state can be written in terms of the product of matrices with local basis states:
\begin{align}
    \ket{\Psi} = \frac{1}{\sqrt{2}}
    \begin{pmatrix}
        \ket{\uparrow}_1 & \ket{\downarrow}_1
    \end{pmatrix}
    \begin{pmatrix}
        \ket{\uparrow}_2 & 0\\
        0 & \ket{\downarrow}_2
    \end{pmatrix}
    \begin{pmatrix}
        \ket{\uparrow}_3 & 0\\
        0 & \ket{\downarrow}_3
    \end{pmatrix}
    \begin{pmatrix}
        \ket{\uparrow}_4 \\ \ket{\downarrow}_4
    \end{pmatrix}
\end{align}
By keeping track of the direction of the local basis in the third index, a probability amplitude of some basis state $\ket{\sigma}$ can be written 
as a product of the matrices. 
\begin{align}
    \braket{\Psi|\bm{\sigma}} &= A^{s_1,j}A_{j}^{s_1,k} A_{k}^{s_3,l} A_{l}^{s_4}
\end{align}
where $i,j,k,l$ are the indices of a matrix; the Einstein summation convention\index{Einstein summation convention}
\footnote{indices with same letter in the different tensors are summed from $0$ to $m-1$.}
is used to represent the contractions\index{Contraction}.
The matrix $A^{s_n}$ with a fixed local basis configuration to $\bar{s}_n$ is a matrix at site $n$ with the coefficients of 
the components with the local basis $\bar{s}_n$. 
For example, the tensor $A^{s_2}$ in the above example can be written as
\begin{align}
    A^{\uparrow_2} &=
    \begin{pmatrix}
        1 & 0  \\
        0 & 0 
    \end{pmatrix}\\
    A^{\downarrow_2} &=
    \begin{pmatrix}
        0 & 0  \\
        0 & 1 
    \end{pmatrix}
\end{align}

The probability amplitude of some basis state can be computed by calculating the product of the local matrices with the corresponding local spin configuration. 
For example, the amplitude of a basis state $\ket{\uparrow\uparrow\uparrow\uparrow}$ of the state from the above example is
\begin{align}
    \braket{\Psi|\uparrow\uparrow\uparrow\uparrow} = \frac{1}{\sqrt{2}}
    \begin{pmatrix}
        1 & 0
    \end{pmatrix}
    \begin{pmatrix}
        1 & 0  \\
        0 & 0 
    \end{pmatrix}
    \begin{pmatrix}
        1 & 0  \\
        0 & 0 
    \end{pmatrix}
    \begin{pmatrix}
        1 \\
        0 
    \end{pmatrix}
    = \frac{1}{\sqrt{2}}
\end{align}
and similarly, the probability amplitude of a basis state $\ket{\uparrow\downarrow\uparrow\downarrow}$ of the state is
\begin{align}
    \braket{\Psi|\uparrow\downarrow\uparrow\downarrow} = \frac{1}{\sqrt{2}}
    \begin{pmatrix}
        1 & 0
    \end{pmatrix}
    \begin{pmatrix}
        0 & 0  \\
        0 & 1 
    \end{pmatrix}
    \begin{pmatrix}
        1 & 0  \\
        0 & 0 
    \end{pmatrix}
    \begin{pmatrix}
        0 \\
        1 
    \end{pmatrix}
    = 0
\end{align}

This can be generalised into any state $\ket{\Psi}$ which lives in the Hilbert space.
A state $\ket{\Psi}$ in the Hilbert space spanned by $N$ spin-$s$ particles can be written, crudely, as follows
\begin{align}
    \ket{\Psi} &= \sum_{s_1,s_2,s_3,\ldots,s_{N-2},s_{N-1},s_{N}}C^{s_1.s_2,s_3,\ldots,s_{N-2},s_{N-1},s_{N}} \ket{s_1,s_2,s_3,\cdots,s_{N-2},s_{N-1}.s_{N}}
\end{align}
where $C$ is a tensor of rank-$N$.  Of course this tensor cannot be stored into a memory of a computer because it requires $(2s+1)^N$ numbers.
However, since we know that there are only $m$ states that gives non-zero amplitudes, we can factorise the tensor ($C^{s_1,\ldots,s_N}$) by factoring out the
system into two parts; one with only site 1 and the other with the rest of the system. 
Such a factorisation is achievable by a process called Schmidt decomposition. 
It allows a big tensor $C$ to be rewritten as a vector product of $M$, 
which describes the local state of site $1$, and a new tensor $A$ with one less spin index, 
which describes the configuration of the rest of the system:
\begin{align}
    C^{s_1,\ldots,s_N} &= \sum_{i=0}^{m-1}A^{s_1}_{i} C^{s_2\ldots,s_{N}}_{i}
\end{align}
Here, $i$ is some indices (usually referred as \textit{auxiliary dimension} \index{Auxiliary dimension}in the literatures), which 
determines the number of states to keep.
This process can be continued $N-3$ times to get $N-2$ matrices and two vectors (the boundary condition).
\begin{align}
    C^{s_1,\ldots,s_N} &= A^{s_1}_{i} C^{i,s_2\ldots,s_{N}}\nonumber\\
    &= A^{s_1}_{i} A^{i,s_2}_{j}C^{s_3\ldots,s_{N},j} \nonumber\\
    & \cdots \nonumber\\
    &= A^{s_1}_{i} A^{i,s_2}_{j}\cdots A^{k,s_{N-1}}_{l}A^{l,s_{N}}
\end{align}
For a finite system, a state can be described by $2(2s+1)$ vectors at each end and $(N-2)(2s+1)$ matrices.
By using this formulation, a state can be represented with $\sim Nm^2$ numbers in the Matrix Product State formalism;
the amount of numbers needed to represent a state increases linearly with the number of sites not exponentially.
\newpage
\subsection{Matrix Product Operators}\label{Matrixproductoperators}
MPS can be generalised to operators. Crudely, an operator $\hat{O}$ can be written in some basis as:
\begin{align}
    \hat{O} &= \sum_{\bm{\sigma}, \bm{\sigma'}} A^{\bm{\sigma}}_{\bm{\sigma'}}\ket{\bm{\sigma}}\bra{\bm{\sigma'}}
\end{align}
where $\bm{\sigma}=(s_1,\cdots,s_N)$ and $\bm{\sigma'} = (s_1',\cdots,s_N)$ are the basis states 
represented by the tensor product of some configurations of the local basis states ($s_n$);
and the tensor $A^{\bm{\sigma}}_{\bm{\sigma'}}$ contains the values of the matrix components.
Like the formulation of MPS, A rank-$2N$ tensor $A$, which holds $(2s+1)^{2N}$ numbers, 
can be broken into a product of $N$ matrices ($W^{s_i}_{s'_i}$) where each matrix consists of local operators as follows:
\begin{align}
    A^{\bm{\sigma}}_{\bm{\sigma'}}
    &= W^{s_1}_{s'_1}A'^{s_2,\ldots,s_N}_{s'_2,\ldots,s'_N}\\
    &\cdots\\
    &= W^{s_1}_{s'_1,i_1}W^{s_2,i_1}_{s'_2,i_2} \cdots W^{s_2,i_{N-2}}_{s'_2,i_{N-1}}W^{s_2,i_{N-1}}_{s'_2}
\end{align}
This decomposed form of the matrix component of an operator ($A^{\bm{\sigma}}_{\bm{\sigma}'}$) is the Matrix Product Operator (MPO)\index{Matrix Product Operator}.
Like MPS, the value of the matrix component $\ket{\bm{\sigma}}\bra{\bm{\sigma'}}$ can be 
obtained by the product of the matrices with the corresponding configuration of the local basis states. 

For example, the Hamiltonian of a spin-1/2 Ising model with only nearest neighbour interactions can be written as:
\begin{align}
    \hat{H} &= J\sum_{i=0}^{N-2}\hat{S}_{i}^z\hat{S}_{i+1}^z
\end{align}
where $\hat{S}^z_n$ is a spin $z$ operator at site $n$; and $J$ is some coupling constant.
This can be written in terms of a product of matrices with local operators as:
\begin{align}
    \hat{H} &= J\left( \hat{I}_1,\hat{S}^z_1,0 \right) 
    \begin{pmatrix}
        \hat{I}_2&\hat{S}^z_2&0\\
        0&0&\hat{S}^z_2\\
        0&0&\hat{I}_2
    \end{pmatrix}
    \begin{pmatrix}
        0\\
        \hat{S}^z_3\\
        \hat{I}_3
    \end{pmatrix}
\end{align}
For an arbitrary number of sites:
\begin{align}
    \hat{H} &= J\left( \hat{I}_1,\hat{S}^z_1,0 \right) 
    \begin{pmatrix}
        \hat{I}_2&\hat{S}^z_2&0\\
        0&0&\hat{S}^z_2\\
        0&0&\hat{I}_2
    \end{pmatrix}
    \cdots
    \begin{pmatrix}
        \hat{I}_{N-1}&\hat{S}^z_{N-1}&0\\
        0&0&\hat{S}^z_{N-1}\\
        0&0&\hat{I}_{N-1}
    \end{pmatrix}
    \begin{pmatrix}
        0\\
        \hat{S}^z_N\\
        \hat{I}_N
    \end{pmatrix}
\end{align}
$\hat{H}$ in the form of an MPO is then
\begin{align}
    \hat{H} &= \sum_{\bm{\sigma},\bm{\sigma}'} W^{s_1}_{s'_1}W^{s_2}_{s'_2} \cdots W^{s_{N-1}}_{s'_{N-1}}W^{s_N}_{s'_N}\ket{\bm{\sigma}}\bra{\bm{\sigma}'}
\end{align}
where $W^{s_n}_{s_n'}$ is a matrix with components corresponding to the matrix component of a local operator $\ket{s_n}\bra{s'_n}$.
For example, using the eigenstates of $\hat{S}^z$ as the local basis states, $W^{\uparrow_1}_{\downarrow_1}$ is
\begin{align}
W^{\uparrow_1}_{\downarrow_1} &= 
    \begin{pmatrix}
        0 & 0 & 0 & 0 & 0 & 0 \\
        0 & 0 & 0 & 0 & 0 & 0  
    \end{pmatrix}
\end{align}
and similarly $W^{\uparrow_2}_{\uparrow_2}$ is
\begin{align}
W^{\uparrow_2}_{\uparrow_2} &= 
    \begin{pmatrix}
        1 & 0 & \frac{1}{2} & 0 & 0 & 0\\
        0 & 0 & 0 & 0 & 0 & 0\\
        0 & 0 & 0 & 0 & \frac{1}{2} & 0\\
        0 & 0 & 0 & 0 & 0 & 0\\
        0 & 0 & 0 & 0 & 1 & 0\\
        0 & 0 & 0 & 0 & 0 & 0\\
    \end{pmatrix}
\end{align}
etc\ldots

MPOs of nearest neighbour interactions can be generalised to long-range interactions. 
However as the interaction gets longer the dimensions of the MPO gets larger.
Of course the mixture of the interactions with different interaction lengths make the dimensions go crazy. 
The exponentially decaying interaction is a lucky exception. 
They can be expressed by putting the decay rate ($\gamma$, ($0<\gamma<1$)) at the center of all the matrices in nearest neighbour MPOs.

For example, the Hamiltonian with all-to-all exponentially decaying $\hat{S}^z$ 
interactions in the spin-1/2 Ising model with $4$ sites can be written as:
\begin{align}
    \hat{H} &= \sum_{i=1}^{3}\sum_{j=i+1}^{4}\gamma^{|j-i|}S^z_{i}S^z_{j}
\end{align}
$\hat{H}$ can then be written in the form of an MPO:
\begin{align}
    \hat{H} &=\gamma \left( \hat{I}_1,\hat{S}^z_1,0 \right) 
    \begin{pmatrix}
        \hat{I}_2&\hat{S}^z_2&0\\
        0&\gamma\hat{I}_2&\hat{S}^z_2\\
        0&0&\hat{I}_2
    \end{pmatrix}
    \begin{pmatrix}
        \hat{I}_3&\hat{S}^z_3&0\\
        0&\gamma\hat{I}_3&\hat{S}^z_3\\
        0&0&\hat{I}_3
    \end{pmatrix}
    \begin{pmatrix}
        0\\
        \hat{S}^z_4\\
        \hat{I}_4
    \end{pmatrix} \label{expdecayH}
\end{align}
By explicitly expanding the matrix product, one may confirm that
\begin{align}
    \hat{H} &=
    \gamma \hat{S}^z_3\hat{S}^z_4
    +\gamma^2 \hat{S}^z_2 \hat{S}^z_4
    +\gamma \hat{S}^z_2\hat{S}^z_3
    +\gamma^3 \hat{S}^z_1 \hat{S}^z_4
    +\gamma^2\hat{S}^z_1 \hat{S}^z_3 +\gamma \hat{S}^z_1 \hat{S}^z_2\nonumber\\
    &=\gamma \hat{S}^z_1 \hat{S}^z_2
    +\gamma^2\hat{S}^z_1 \hat{S}^z_3 
    +\gamma^3 \hat{S}^z_1 \hat{S}^z_4
    +\gamma \hat{S}^z_2\hat{S}^z_3
    +\gamma^2 \hat{S}^z_2 \hat{S}^z_4 
    +\gamma \hat{S}^z_3\hat{S}^z_4 
\end{align}
Therefore, if a monotonically decreasing coupling function governs the long-range interactions, 
it can be efficiently expressed as an MPO by approximating it in terms of a sum of exponentially decaying MPOs. 

\subsection{Tensor Network Diagrams}\label{Tensornetworkdiagrams}
\begin{figure}[t]
    \begin{minipage}{0.48\textwidth}
        \centering
        \includegraphics[width=0.4\textwidth]{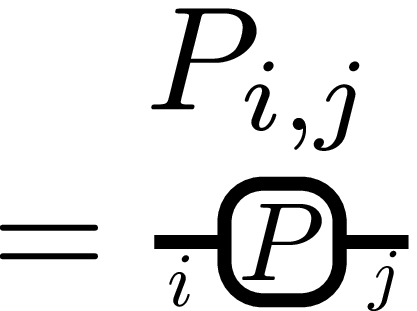}
    \caption{TND of a Matrix}\label{TNDmatrix}
    \end{minipage}\hfill
    \begin{minipage}{0.48\textwidth}
        \centering
        \includegraphics[width=0.6\textwidth]{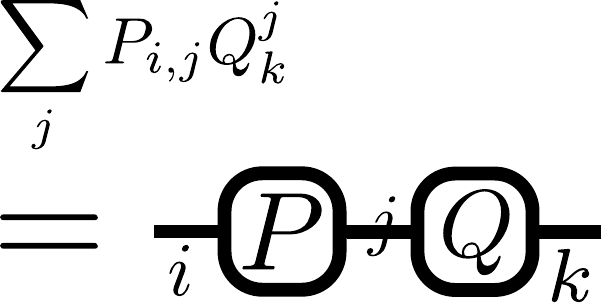}
    \caption{TND of a contraction of the matrices $P$ and $Q$}\label{TNDmatrixcontract}
    \end{minipage}
\end{figure}
\begin{figure}[b]
    \begin{minipage}{0.32\textwidth}
        \centering
        \includegraphics[width=0.8\textwidth]{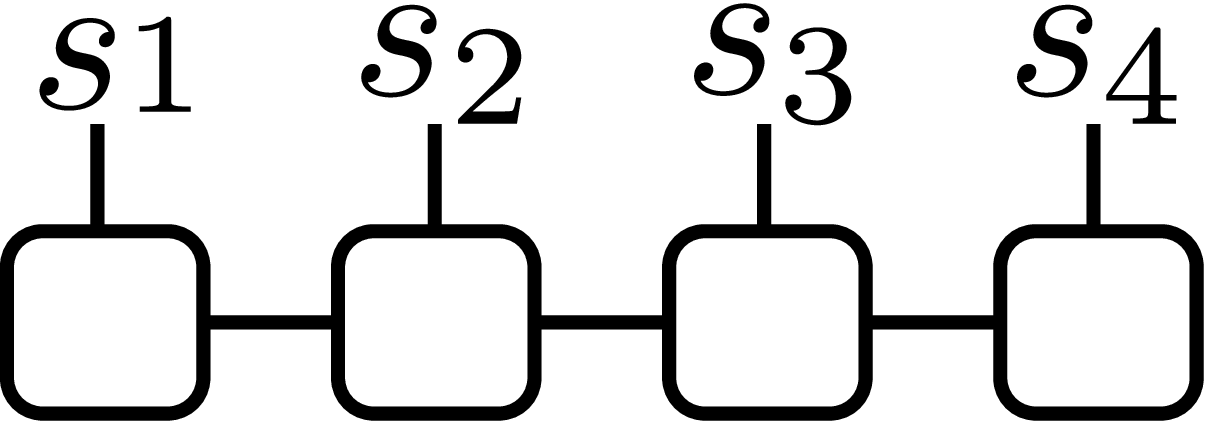}
    \caption{TND representation of MPS ($\ket{\Psi}$).}\label{TNDpsi}
    \end{minipage}\hfill
    \begin{minipage}{0.32\textwidth}
        \centering
        \includegraphics[width=0.8\textwidth]{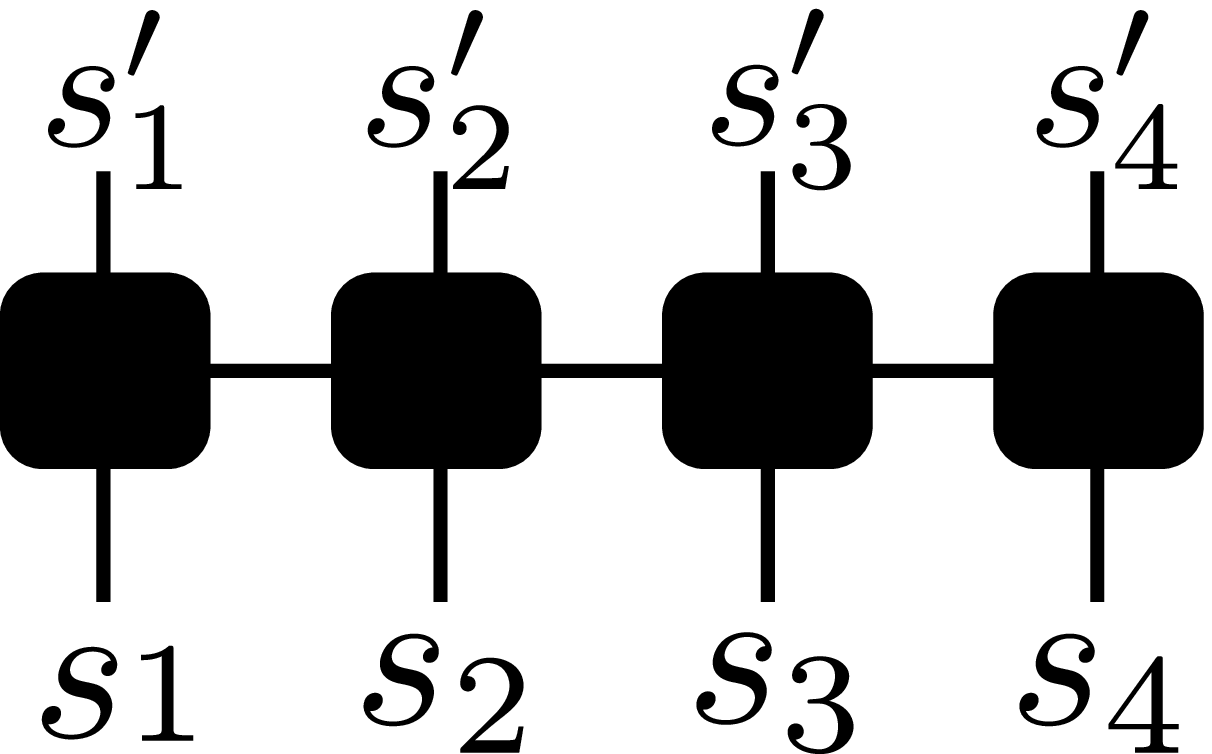}
    \caption{TND representation of MPO ($\hat{O}$).}\label{TNDo}
    \end{minipage}\hfill
    \begin{minipage}{0.32\textwidth}
        \includegraphics[width=0.8\textwidth]{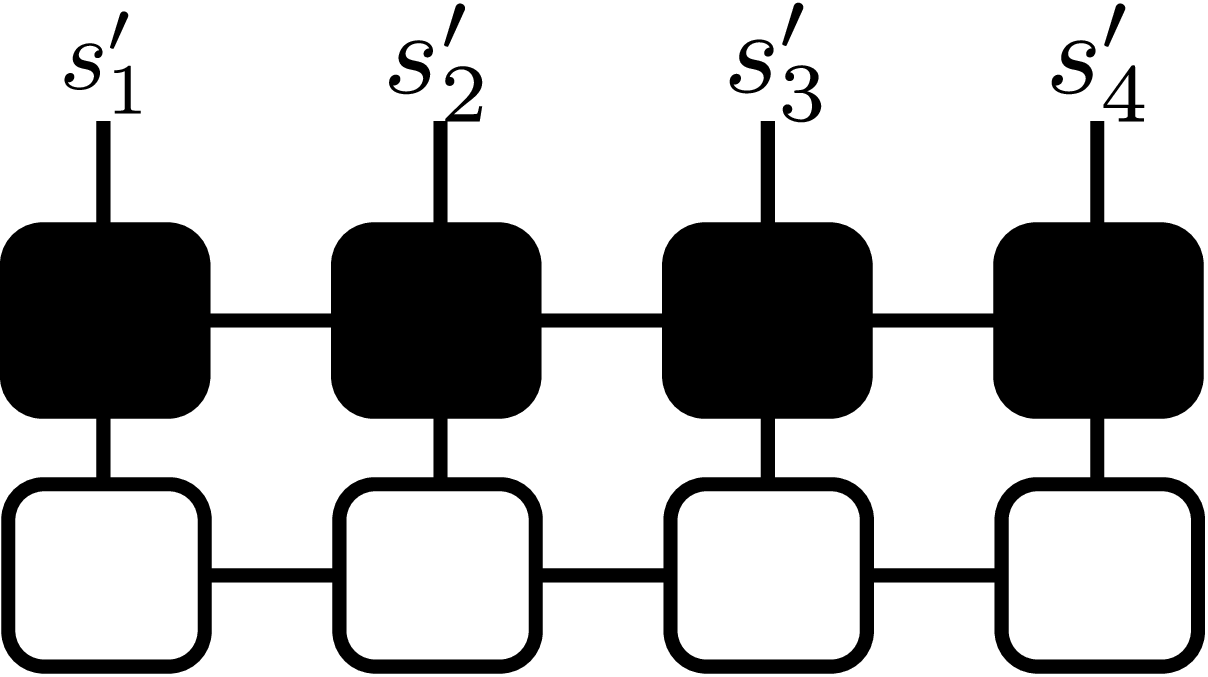}
        \caption{TND representation of an operator acting on a wavefunction ($\hat{O}\ket{\Psi}$).}\label{TNDopsi}
    \end{minipage}
\end{figure}

Like Dirac's Bra-Ket notation in matrix mechanics, there is a useful graphical representation called a tensor network diagram\index{Tensor network diagram} (TND)
that can be used to describe operations of MPS and MPO.
TND uses boxes and legs to describe different tensors and their indices.
Using the diagram, the complex algebra of tensors can be graphically represented.  
There are only two rules upon constructing TND:
\begin{itemize}
    \item A box represents a tensor and legs coming out from a box are the indices of the tensor. 
        For example, a matrix can be expressed as two legs coming out of a box (\fig{TNDmatrix}).
    \item Two tensors connected by a line shows a contraction of indices. 
        For example, a matrix product of $P$ and $Q$ can be expressed as two boxes connected by a node (\fig{TNDmatrixcontract}).
\end{itemize}
With this notation, each of the matrices in an MPS can be drawn as a box with three legs. Two legs corresponds to two auxiliary dimensions, and 
the third bond is a physical index which corresponds to a basis state.
Since a wavefunction is a tensor with $N$ physical indices, 
to represent a full wavefunction, the indices of auxiliary dimensions must be contracted (\fig{TNDpsi}).
Similarly, an MPO can be represented by contracting the indices of auxiliary dimensions (\fig{TNDo}).
An operator applied on a wavefunction can be represented by contracting the physical indices on one of the sides of the operator with the physical
indices of the wavefunction (\fig{TNDopsi}).
\newpage
\subsection{Algebra of MPS and MPO}
With the properties of MPS and MPO, the ordinary quantum mechanics can be recovered.
In this section how the normalisation condition, overlaps, expectation values, and projections 
are efficiently computed are discussed. 
\subsubsection{Orthogonality}
MPS have gauge\index{Gauge} freedom\index{Gauge freedom} since identities can be insured between any places where one of the indices of the matrix is contracted.
For example, a matrix product of $A_1$ and $A_2$ can be written as
\begin{align}
    A_1 A_3 &= A_1 G G^{-1} A_ = (A_1 G) (G^{-1} A_2 ) = M_1 M_2
\end{align}
and one obtains a new matrix product of $M_1(=A_1G)$ and $M_2=(G^{-1}A_2)$.
Using this hidden freedom, one can perform \textit{Singular value decomposition}\index{Singular value decomposition} (SVD)
to the matrices in MPS to canonicalise it in a left or right orthogonal form.

\begin{figure}[tb]
    \begin{minipage}{0.48\textwidth}
        \centering
        \includegraphics[width=0.70\textwidth]{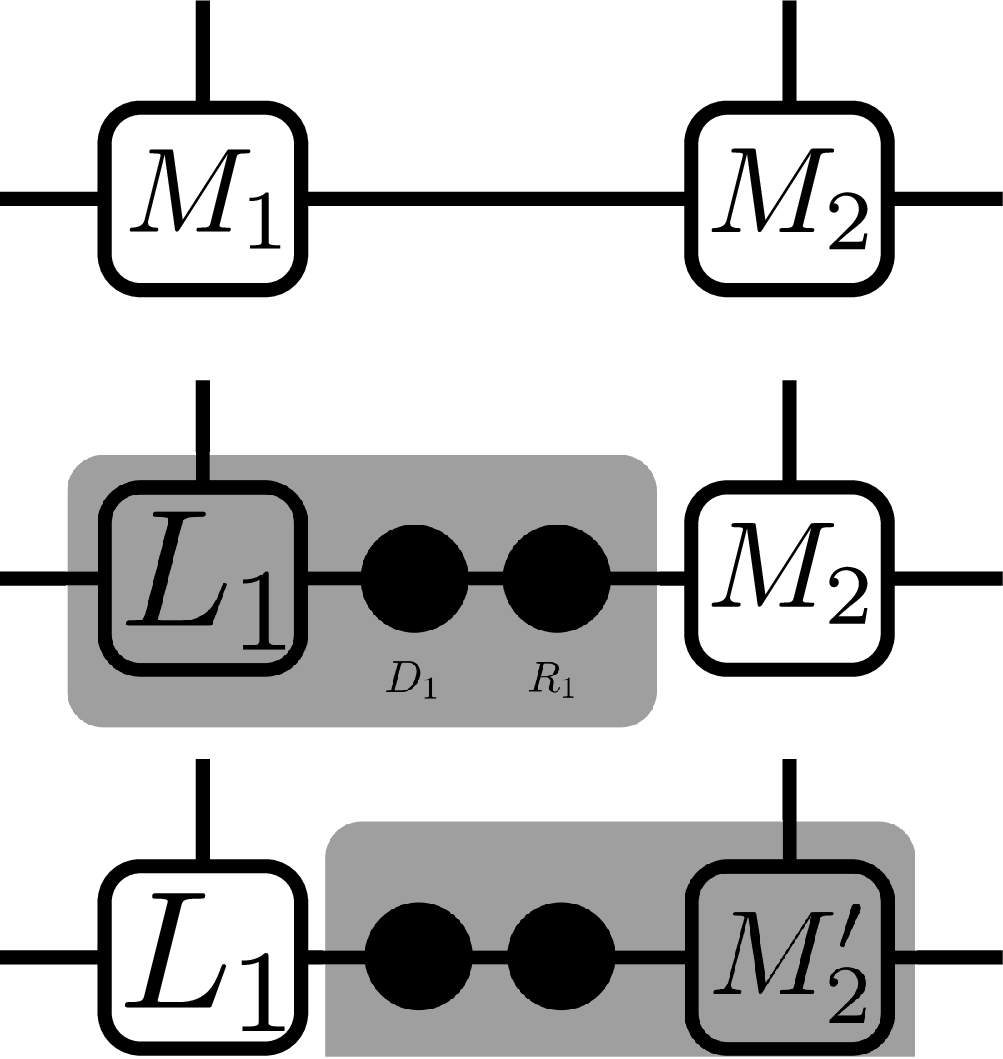}
        \caption{TND of SVD and left-canonicalisation of a wavefunction.}
        \label{TNDSVD}
    \end{minipage}\hfill
    \begin{minipage}{0.48\textwidth}
        \centering
        \includegraphics[width=0.70\textwidth]{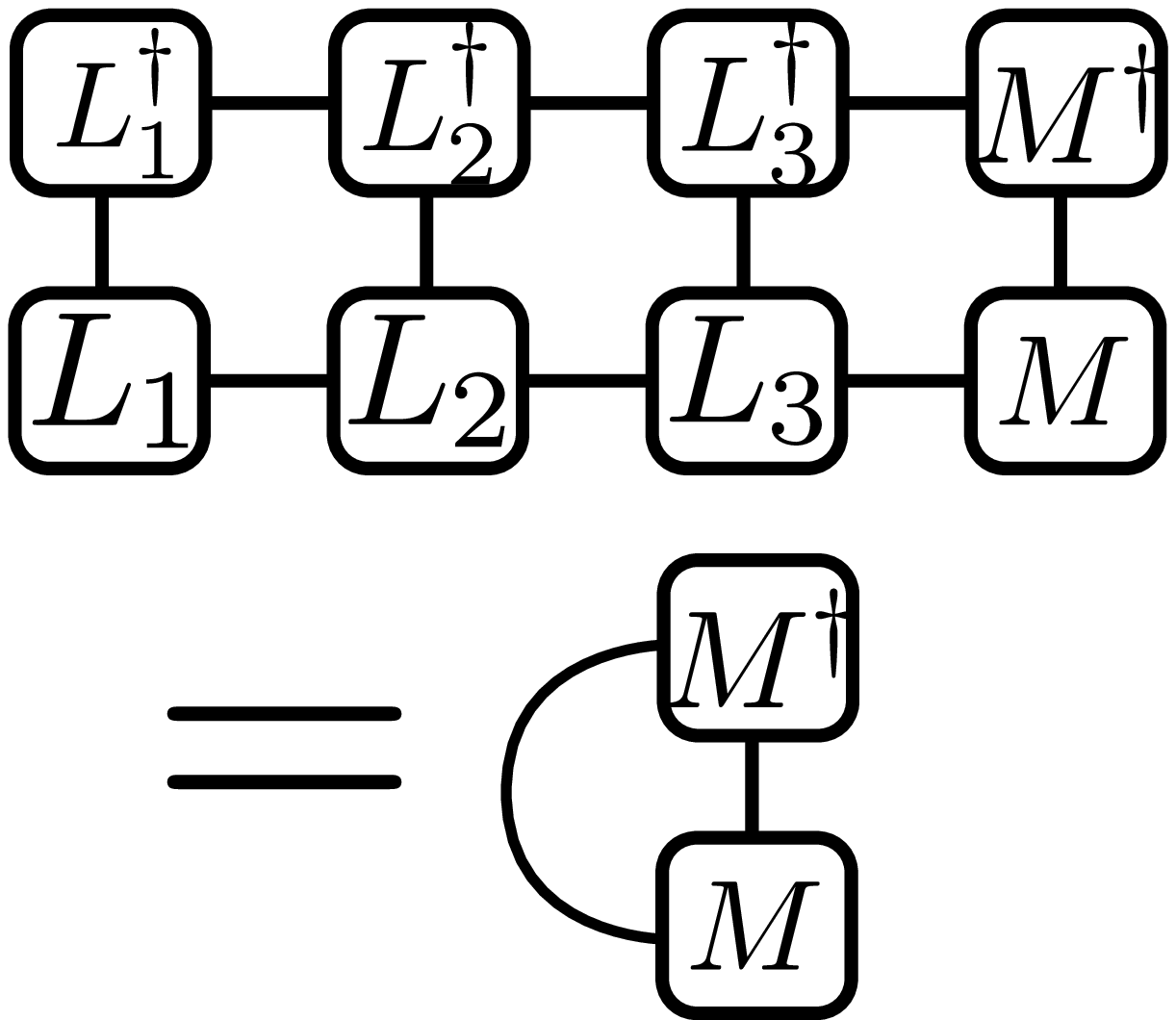}
        \caption{TND of a computation of $\braket{\Psi|\Psi}$}
        \label{TNDnormal}
    \end{minipage}
\end{figure}

SVD decomposes a matrix ($M$) into a product of three matrices ($L$,$D$,$R$) with 
the following properties:
\begin{align}
    M &= LDR^\dag \ \text{or}  \ M = L^\dag D R\\ 
    \text{s.t.} &\begin{cases}
        D_{i,j} = \lambda_{i}\delta_{i,j} \\
        L^{\dag i}_{j}L^{j}_{k} =  I^{i}_{k}\\
        R^{i}_{j}R^{\dag,j}_{k} = I^{i}_{k}
    \end{cases}
\end{align}
where $\lambda$ is an eigenvalue of $M$; $\delta_{i,j}$ is a Kronecker's delta; a matrix $L$, which has the property above is called
a \textit{left orthogonal matrix}\index{Left orthogonal matrix} or unitary matrix\index{unitary};
and a matrix $R$ which has the above property is called a \textit{right orthogonal matrix}\index{Right orthogonal matrix}.
One might think that the matrices $L$ and $R$ are  unitary as well, but this is not necessary the case because $M$ is not restricted to square matrices 
and hence $D$ may be square but $L$ and $R$ might not be
left orthogonalised form can be achieved by performing SVD on each matrix from the left. 
Keeping $L$ as the local matrix and merge $D$ and $R^\dag$ to the matrix of the next site left orthogonalises a site (\fig{TNDSVD}).
Starting from the left most site, and continue until it reaches the right most site, left orthogonalise the wavefunction. 
Achieving the right orthogonalised form is the exact opposite.

If the MPS ($\bm{\Psi}$) is in a canonicalised form, then it can be written as:
\begin{align}
    \bm{\Psi}^{s_1,s_2,\ldots,s_{N-1},s_{N}} &= L^{s_1}L^{s_2}\cdots L^{s_{N-1}}M^{s_{N}}
\end{align}
where $L$ is a left orthogonal matrix and $M$ is some matrix.
The normalisation condition is then:
\begin{align}
    \braket{\Psi|\Psi} &= 
    \sum_{\bm{\sigma}} \bra{\bm{\sigma}} \bm{\Psi}^{\dag,s_1,s_2,\ldots,s_{N-1},s_{N}} 
    \bm{\Psi}^{s_1,s_2,\ldots,s_{N-1},s_{N}}\ket{\bm{\sigma}}\\
    &= \sum_{\bm{\sigma}} \bra{\bm{\sigma}} M^{\dag,s_{N}} L^{\dag,s_{N}}L^{\dag,s_{N-1}}\cdots L^{\dag,s_2}L^{\dag,s_1} 
    L^{s_1}L^{s_2}\cdots L^{s_{N-1}}M^{s_{N}} \ket{\bm{\sigma}} \\
    & = \sum_{s_N} M^{\dag,s_{N}}M^{s_{N}} = 1\label{MdagMeq1}
\end{align}
This process can be presented in terms of TND as shown in \fig{TNDnormal}. 
\subsubsection{Projection and Minimisation}
Using the left and right canonical forms effectively, one can perform 
a projection of an MPS with large $m$ on to an MPS with smaller $m$ through a minimisation.
Projection is especially useful for time evolution since applying an MPO (such as $\hat{H}$) on a wavefunction increases the auxiliary dimensions.
A projection of $\ket{\Psi}$ onto a MPS  with smaller $m$ ($\ket{\phi}$) is the same as minimising $\ket{r}$ which is defined as:
\begin{align}
    \ket{r} &= \norm{\ket{\Psi} - \ket{\phi}}
\end{align}
The magnitude of $\ket{r}$ is
\begin{align}
    \braket{r|r} &= \braket{\Psi|\Psi} - \braket{\psi|\phi} - \braket{\phi|\psi} + \braket{\phi|\phi}
\end{align}

\begin{figure}
    \begin{minipage}{0.48\textwidth}
        \centering
        \caption{TND of the minimisation of $\braket{r|r}$ at site $n$}\label{TDNminimizeM}
        \includegraphics[width=0.8\textwidth]{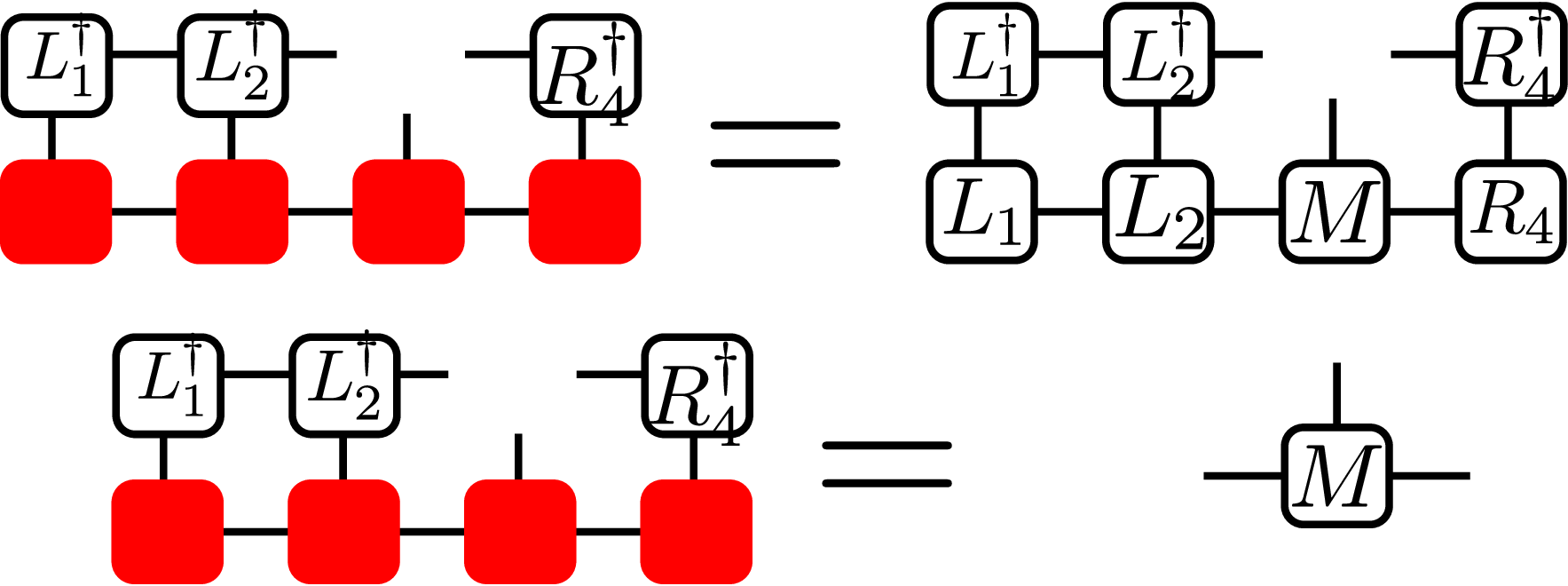}
        \legend{$\ket{\phi}$ is a MPS represented by white boxes and $\ket{\Psi}$ is a MPS represented by red.
        The form of $\ket{\Psi}$ does not play any rule in a minimisation process.}
    \end{minipage}\hfill
    \begin{minipage}{0.48\textwidth}
        \centering
        \caption{TND of the minimisation of $\braket{r'|r'}$}\label{TDNminimizeHpsi}
        \includegraphics[width=0.75\textwidth]{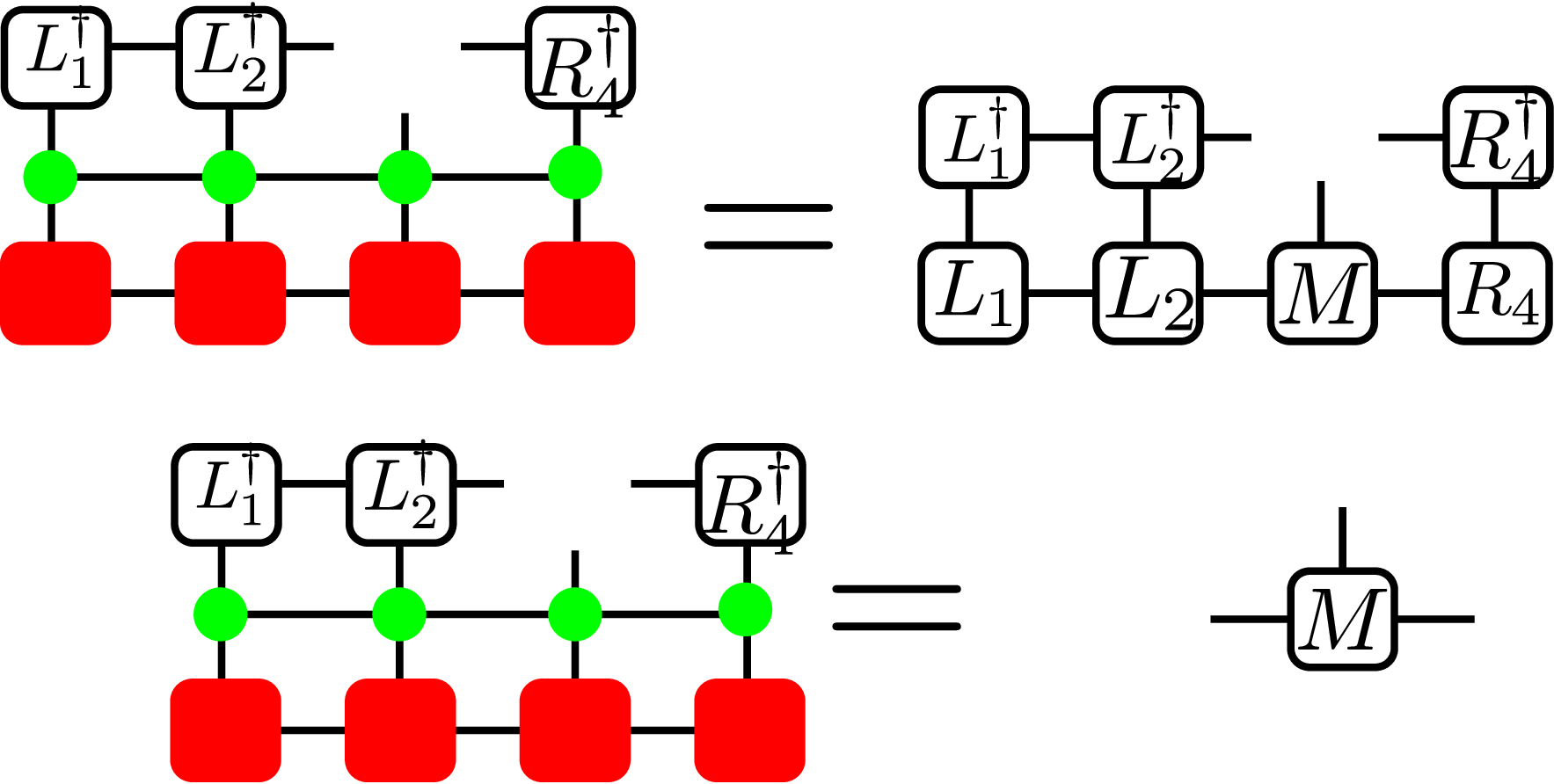}
        \legend{$\ket{\phi}$ is a MPS represented by the white boxes, $\ket{\Psi}$ is a MPS represented by the red boxes, and MPO of some operator $\hat{O}$ 
        is represented by the circles.
        Again, the form of $\ket{\Psi}$ does not play any rule in a minimisation process.}
    \end{minipage}
\end{figure}

The minimisation can be achieved through a process called
\textit{sweeping}\index{sweeping}\index{sweep}.
A scalar quantity $\braket{r|r}$ is a function of the matrices in $\ket{\phi}$, and it should only have one local minimum.
Starting from the $\phi$ in the left canonical form and taking the partial derivative with respect to the right-most matrix ($M^{s_N}$) gives 
\begin{align}
    \ket{\psi} &= \sum_{\bm{\sigma}}\bm{\psi}^{s_1,s_2,\ldots,s_{N-1},s_{N}}\ket{\bm{\sigma}}\\
    \ket{\phi} &= \sum_{\bm{\sigma}}L^{s_1}L^{s_2}\cdots L^{s_{N-1}}M^{s_N} \ket{\bm{\sigma}}\\
    \pl{}{M^{s_N}}\left( \braket{r|r} \right) &=\pl{}{M^{s_N}} \left(\braket{\Psi|\Psi} - \braket{\psi|\phi} - \braket{\phi|\psi} + \braket{\phi|\phi}\right)\\
    &= -\left( \sum_{\bm{\sigma} \neq s_N}\bra{\bm{\sigma} } \bm{\psi}^{\dag,s_1,s_2,\ldots,s_{N-1},s_{N}}L^{s_1}L^{s_2}\cdots L^{s_{N-1}}\ket{\bm{\sigma}}
    \right) 
    + M^{\dag,s_{N}}
\end{align}
At a local minimum,
\begin{align}
    \pl{}{M^{s_N}}\left(\braket{r|r}  \right) &= 0
\end{align}
Therefore
\begin{align}
    \min \{ M^{s_N} \} &= 
    \sum_{\bm{\sigma} \neq s_N} \bra{\bm{\sigma}} L^{\dag,s_{N-1}}\ldots L^{\dag,s_2}L^{\dag,s_1} \bm{\psi}^{s_1,s_2,\ldots,s_{N-1},s_{N}}\ket{\bm{\sigma}}
    \label{minimizedM}
\end{align}
This is the minimisation condition for matrix $M$.
Once $M$ is determined, perform SVD to make it a right-canonicalised form, reject the matrices  $L$ and $D$ and move to the next site to the left.
The TND of this operation is shown on \fig{TDNminimizeM}.
As it can be seen from the figure, this works because the sites to the left of the site that is being minimised are always left-orthogonalised and hence
they all get contracted to give identity; and all the sites to the right as right-orthogonalised and hence they all get contracted to give the identity. 
As a consequence, after the minimisation performed at sites $n$, the $\pl{}{M^{s_n}}\left(\braket{\phi|\phi}\right)$ always gives $M^{\dag,s_n}$.

The same method can be used to calculate the projection of $\hat{H}\ket{\psi}$ onto $\ket{\phi}$ without calculating $\hat{H}\ket{\psi}$ directly
by minimising
\begin{align}
    \ket{r'} &= \hat{O}\ket{\Psi} - \ket{\phi}
\end{align}
The direct computation of $\hat{O}\ket{\Psi}$ requires a computational cost of 
$\mathcal{O}(N(2s+1)m^3m_{o}^3)$, where $m_{o}$ is the dimension of the matrices of the MPO.
This can be reduced to $\mathcal{O}(N(2s+1)m_{o}^3)$ by the implicit projection shown on \fig{TDNminimizeHpsi}.
\subsection{Infinite Matrix Product States}
An MPS can represent a state in a thermodynamical limit by omitting a vector at each end of the finite MPS.
This is the infinite Matrix Product State (iMPS\index{Infinite Matrix Product States}\index{iMPS}).
The infinite system is defined by a unit cell of $N$ sites. In the thermodynamical limit, $N$ sites are spanned repeatedly over an infinitely long chain.
The storing cost is same as a wavefunction on a chain of length $N$, yet due to the infinite boundary condition (two legs at each end of the MPS in TND form),
the right most site and the left most site of a unit cell gains a correlation. 
As McCulloch\cite{McCulloch2008} wrote, the advantage of this representation is that extrapolation to the thermodynamical limit by increasing the size of 
the system is not required. The size of the unit cell is the smallest size which the symmetry of the state requires. 
For example, in a spin chain, a translationally symmetric ferromagnetic state requires a unit cell size of $1$, while 
an antiferromagnetic state (staggered magnetisation state) requires a unit cell size of $2$ 
because it is translationally invariant over an even number of sites.

\subsection{Numerical Implementation: \texttt{Matrix Product Toolkit}}
The numerical implementations of most of the basic operations are provided in the \texttt{C++} library \texttt{Matrix Product Toolkit}
\footnote{It is available at \url{https://people.smp.uq.edu.au/IanMcCulloch/mptoolkit/}} \cite{McCullochMPT,McCulloch2007}.
As its main feature lies on the calculation of the ground state of an infinite system through infinite-DMRG algorithm, 
it has a rich array of \texttt{classes}, \texttt{structs}, and \texttt{functions} for constructing algorithms for manipulating 
MPS which represents a physical wavefunction. 
\section{A New Time Evolution Algorithm for MPS with Long-Range Interactions}
\label{Anewtime}
The purpose of the algorithm is to time-evolve an infinite system described by iMPS with an inhomogeneous 
Hamiltonian with long-range interactions.
To perform this evolution, the algorithm for evolution is divided into two parts|a local time evolution and a global time evolution.
This can be understood as the generalised version of the Time-Evolving Block Decimation \cite{Vidal2004,Vidal2007}.
Throughout this thesis, this new algorithm is referred to as generalised-TEBD or gTEBD. 

A local time evolution can be done by applying the \textit{Lanczos method}\index{Lanczos method}.
It is one method which utilises \textit{Krylov subspace expansion}\index{Krylov subspace expansion}. 
Krylov subspace is a space spanned by a set of vectors:
\begin{align}
    \{\ket{\Psi},\hat{H}\ket{\Psi},\hat{H}^2\ket{\Psi},\hat{H}^3\ket{\Psi},\ldots\}
\end{align}
The Lanczos method finds an approximated tridiagonalised version of a hermitian operator ($\hat{H}$) as:
\begin{align}
    \hat{H}&\approx
    \begin{pmatrix}
        \alpha_0 & \beta_0 & &  \\ 
        \beta_0 & \alpha_1 & \beta_1 &  \\ 
         & \beta_1 & \alpha_2 &  \beta_2 &  & 0 \\
        & &\ddots & \ddots & \ddots &  & \\
        & && \ddots & \ddots &\ddots   & \\
        & & 0 & &\beta_{\mathcal{N}-3}   &\alpha_{\mathcal{N}-2}  & \beta_{\mathcal{N}-2}\\
        & &  & & &\beta_{\mathcal{N}-2}&\alpha_{\mathcal{N}-1}
    \end{pmatrix}
\end{align}
where $\mathcal{N}$ is a number of Lanczos step performed. 
The basis vectors can be found iteratively by solving the following recursive relationships, setting $\ket{\Psi}=V_0$ as the starting vector:
\begin{align}
    \hat{H}\ket{\Psi}  &= \alpha_0\ket{V_0} + \beta_0\ket{V_1}\\
    \hat{H}\ket{V_1}  &= \beta_0 \ket{\Psi} + \alpha_1\ket{V_1} + \beta_1\ket{V_2}\\
    \vdots & \nonumber\\
    \hat{H}\ket{V_{k}} &= \beta_{k-1} \ket{V_{k-1}} + \alpha_{k}\ket{V_k} + \beta_{k}\ket{V_{k+1}} \label{genLancozreltn}\\
    \vdots & \nonumber\\
    \hat{H}\ket{V_{\mathcal{N}-1}}  &= \beta_{\mathcal{N}-2} \ket{V_{\mathcal{N}-2} } + \alpha_{\mathcal{N}-1} \ket{V_{\mathcal{N}-1}} 
\end{align}
Since the basis vectors must be orthonormal:
\begin{align}
    \braket{V_i|V_j} &= \delta_{i,j}
\end{align}
where $\delta_{i,j}$ is a Kronecker's delta,
$\alpha_k$ can be found by acting $\bra{V_k}$ from the left on \eqn{genLancozreltn}.
\begin{align}
    \alpha_k &= \bra{V_k}\hat{H}\ket{V_k}
\end{align}
From the normalisation of $\ket{V_{k}}$, $\beta_{k}$ can be found as
\begin{align}
    \beta_{k} &= \norm{(\hat{H}-\alpha_k)\ket{V_{k}} - \beta_{k-1}\ket{V_{k-1}} }
\end{align}
and hence $\ket{V_{k+1}}$ is 
\begin{align}
    \ket{V_{k+1}} &= \frac{1}{\beta_k}\left( (\hat{H}-\alpha_k)\ket{V_{k}} - \beta_{k-1}\ket{V_{k-1}}\right)
\end{align}
The Lanczos step can be continued until the desired number of steps ($\mathcal{N}$) or a sufficiently small $\beta_k$ is achieved. 
The tridiagonal matrix  can then be exponentiated using the favourite method (such as full diagonalisation or Pade approximation).
With the $\ket{\Psi}$ as an initial vector, the time evolved vector is just
\begin{align}
    \ket{\Psi(t + \Delta t)} &= \sum_{i=0}^{\mathcal{N}-1} [\exp(-i\Delta t\hat{H})]_{i,0} \ket{V_i}\label{lastsum}
\end{align}
where $[\exp(-i\Delta t\hat{H})]_{i,j}$ is an $(i,j)$\textsuperscript{th} component of the exponentiated tridiagonal matrix $\exp(-i\Delta t\hat{H})$
\footnote{It is worth noting that the algorithm provided by Manmana \cite{Manmana2016} is incorrect}.

A global time evolution can be achieved through Suzuki-Trotter decomposition \cite{Suzuki1985}. 
The theorem states that given a symmetric operator ($\hat{H}$) in terms of a sum of 
two symmetric operators ($\hat{H}_A$ and $\hat{H}_B$), an operator exponent of $\hat{H}$ has the following limiting behaviour:
\begin{align}
    e^{\hat{H}} &= \lim_{\delta \to 0} \left( e^{\delta \hat{H}_A} e^{\delta \hat{H}_B} \right)^{\frac{1}{\delta}}\label{ForderSuzuki}
\end{align}
Therefore for small $\delta$,
\begin{align}
    e^{\delta(\hat{H}_A+\hat{H}_B)} \approx e^{\delta \hat{H}_A} e^{\delta \hat{H}_B}  + \mathcal{O}(\delta^2)\label{SorderSuzuki}
\end{align}
This approximation can be taken to a higher order. The implementation of the  second order Suzuki-Trotter decomposition is almost the same 
as the first order implementation above but with one extra step at the end:
\begin{align}
    e^{\delta(\hat{H}_A+\hat{H}_B)} \approx e^{\frac{\delta}{2} \hat{H}_A} e^{\delta \hat{H}_B} e^{\frac{\delta}{2} \hat{H}_A}
    +\mathcal{O}(\delta^3)
\end{align}

\subsection{Scheme}
\begin{figure}[tb]
    \centering
    \includegraphics[width=0.5\textwidth]{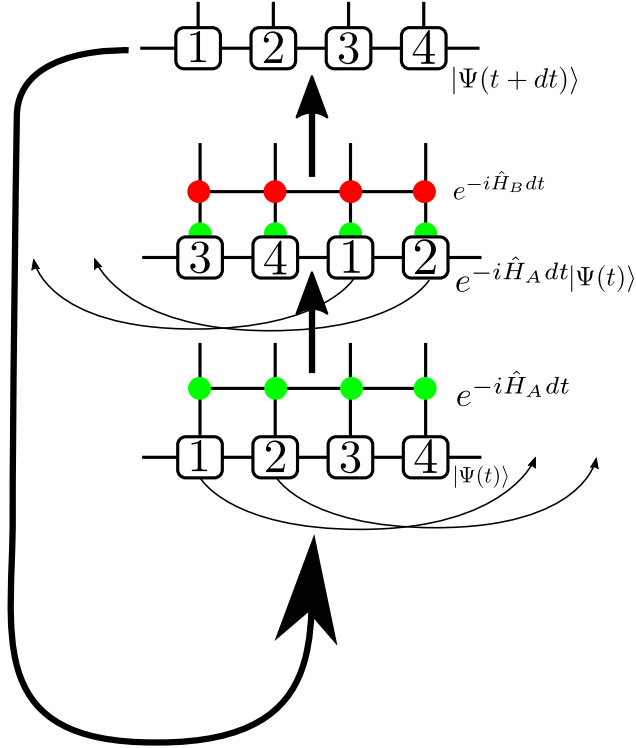}
    \caption{TND of the algorithm}
    \legend{iMPS is evolved using $\hat{H}_A$ and $\hat{H}_B$ iteratively until the desired time is reached.
    Green circles are the unitary MPOs of $\hat{H}_A$ and Red circles are the unitary MPOs of $\hat{H}_B$}\label{TNDAlgorithm}
\end{figure}

As an infinite system is described by a repeated unit-cell, take a unit cell ($U_A$) of size $N$.
Define $\hat{H}_A$ to be component of the Hamiltonian with all interactions within the unit cell A.
Then shift a unit cell by $N/2$ sites. Let us call this half-shifted unit cell $U_B$. 
Define a Hamiltonian $\hat{H}_B$ which contains all the interactions within the unit cell such that $\hat{H} = \hat{H}_A+\hat{H}_B$.
In this setup, all the interactions which cross the unit cell boundary of $U_A$ is contained within $\hat{H}_B$ and 
all the interactions which cross the unit cell boundary of $U_B$ are contained in $\hat{H}_A$.
With this setup, the time evolution can be achieved by combining Lanczos method and Suzuki-Trotter decomposition as follows:
\begin{enumerate}
    \item Evolve a unit cell iMPS by $\Delta t/2$ with a Hamiltonian $\hat{H}_A$ by approximating its operator exponent with the Lanczos method. 
    \item Shift the iMPS so that it matches the unit cell of $\hat{H}_B$. 
        Evolve it  by $\Delta t$ with a Hamiltonian $\hat{H}_B$ by approximating its operator exponent with the Lanczos method. \label{evolHA}
    \item Shift the iMPS again so that it matches the unit cell of $\hat{H}_A$ 
        and evolve it  by $\Delta t$ with a Hamiltonian $\hat{H}_A$ by approximating its operator exponent with the Lanczos method. \label{evolHB}
    \item Go back to step $2$ if $t < T$ where $t$ is the current time step and $T$ is the final time.
    \item Evolve a unit cell iMPS by $\Delta t/2$ with a Hamiltonian $\hat{H}_A$ by approximating its operator exponent with the Lanczos method.
\end{enumerate}
Note that steps \ref{evolHA} and \ref{evolHB} is a Suzuki-Trotter decomposition (\eqn{SorderSuzuki}. The extra steps
at the beginning and end makes it a second order method (\eqn{SorderSuzuki}). 
The TND of this scheme is shown in \fig{TNDAlgorithm}.

As true long-range interactions cannot be implemented in this scheme, 
the interactions must be truncated at some interaction length.
The maximum interaction to the right and left must be equal. 
As shown in \fig{TNDAlgorithm},
the left (right) most site can only have an interaction going over $N/2$ sites
to the left (right), 
the half of a unit cell size is the maximum interaction length that can be simulated using the algorithm. 

To speed up the computation, three projections are implemented in the Lanczos steps.
Firstly, to calculate $\hat{H}\ket{V_{k-1}}$, $\ket{\phi}$ with user specified auxiliary dimensions that minimises 
\begin{align}
    \braket{r|r} &= \norm{\hat{H}\ket{V_{k-1}}-\ket{\phi}}
\end{align}
is calculated. This is faster than directly applying $\hat{H}$ onto the MPS.
Then a vector $\beta_{k}\ket{V_{k}} = \left(\hat{H}-\alpha_{k}\right)\ket{V_{k-1}} - \alpha_{k-1}\ket{V_{k-1}}$
is projected down by finding $\ket{\phi}$ which minimises
\begin{align}
    \braket{r|r} &= \norm{\left(\hat{H}-\alpha_{k}\right)\ket{V_{k-1}} - \alpha_{k-1}\ket{V_{k-1}}-\ket{\phi}}
\end{align}
Finally the sum from \eqn{lastsum} can be efficiently computed by finding $\ket{\phi}$ that minimises
\begin{align}
    \braket{r|r} &= \norm{\sum_{i} [\exp(-i\Delta t\hat{H})]_{i,1}\ket{V_i} - \ket{\phi}}\label{sumporj}
\end{align}
This can be done like an ordinary state projection except there is an extra addition to be performed. 
The algorithm has $4$ parameters that must be specified:
\begin{list}{}{
        \setlength{\labelwidth}{24mm}
    }
    \item [$N$] Number of sites in a unit cell.
    \item [$m$] Maximum number of state to keep for the dimensions of a local matrix. 
    \item [$\Delta t$] Time interval for one iteration. 
    \item [$\mathcal{N}$] Number of Lanczos steps performed.
\end{list}

In terms of the computational cost, choosing a right pair of parameters $m$ and $N$ is important for efficient computation.
The number of sites in a unit cell governs the maximum length of interaction that can be simulated.
Full all-to-all long-range interaction can be achieved in the limit of $N\to\infty$.
The parameter $m$ governs the number of basis states that are used to approximate the wavefunction in the evolution.
The computational time required for calculating a matrix product is proportional to $m^3$.
Therefore the time it takes to complete a minimisation process is proportional to $Nm^3$.
One must choose $m$ and $N$ by balancing out the long-range interactions and accuracy of entanglement behaviour of the system.
A good measure is $m/N$. The condition $m/N >> 1$ gives sufficient amount of states to describe local and entanglement behaviours of a system.

A pair of parameters $\Delta t$ and $\mathcal{N}$ are also important in upon optimisation of the computation efficiency.
The coefficients of the matrix exponential of the tridiagonal matrix created via Lanczos vector gives
\begin{align}
    \exp(-\Delta t \hat{H}) \approx \sum_{i=0}^{\mathcal{N}-1} [\exp(-i\Delta t \hat{H}) ]_{i.0} \hat{H}^{i} 
\end{align}
which looks similar to a Taylor expansion, but it provides a better estimate on the coefficient than naively talking $\Delta t^i/i!$ 
as the coefficients of an expansion. 
Since $\Delta t$ is a parameter in the Suzuki-Trotter expansion shown on \eqn{SorderSuzuki}, $\Delta t$ must be small in order to minimise the error from the
commutation $(\Delta t)^2[\hat{H}_A.\hat{H}_B]$. Since the number of overlapping terms faces a combinatorial explosion 
with respect to the number of sites in the unit cell, for a sufficiently long chain, $\Delta t$ must be kept small ($\sim 10^{-3}$).
The choice of small $\Delta t$ implies that $\mathcal{N}$ should not be large.
Even with $\mathcal{N}=3$, this provides a better approximation than the second order naive Taylor expansion.
\subsection{Bond Optimisation}
\label{Bondoptimisation}
\subsubsection{Bias Function}
Generally the choice of $\hat{H}_A$ and $\hat{H}_B$ is not unique.
They have overlapping interaction terms which must be split between $\hat{H}_A$ and $\hat{H}_B$ to recover the relation $\hat{H} = \hat{H}_A + \hat{H}_B$.
This choice can be done by supplying sets of coefficients ($\{a_i\}$ and $\left\{ b_i \right\}$), 
which are multiplied with the overlapping interaction operators in the Hamiltonian of each unit cell.
The quantity to be minimised is the commutator $[\hat{H}_A,\hat{H}_B]$ with a constraint:
\begin{align}
    \begin{cases}
    a_i + b_{i+N/2} = 1 &(i < N/2)\\
    a_i + b_{i-N/2} = 1 &(i \geq  N/2)
    \end{cases}
\end{align}
Although this is the best way, it is not efficient as the coefficients $a_i,b_i$ must be calculated each time for different Hamiltonians. 

For homogeneous systems (a Hamiltonian where its interaction profile at a site is identical to the other sites),
the bond biasing function must be 
periodic with the unit cell size.
This also applies to $\hat{H}_A$ and $\hat{H}_B$.
Also, it must be symmetric over the center of a unit cell. 
As the sites near the middle have more interaction terms compared to sites near the edges, the function must have a maxima (or minima) at the center.
One choice of function that has such properties can be achieved by using error function which is parametrised by 
$\alpha$ and $\beta$ as follows: 
\begin{align}
    a_i = b_i = W(i) &= 
    \begin{cases}
        \frac{1}{2}+\beta\mathrm{erf}\left(\frac{4i+2 - N}{\alpha(N-2)}\right) &(i < N/2)\\
        \frac{1}{2}-\beta\mathrm{erf}\left( \frac{ 4i+2- 3N } {\alpha(N-2)}\right) &(i \geq N/2)
    \end{cases}
\end{align}
where $i$ is the index of a site ($ 0 < i < N-1$).
\subsubsection{Optimisation}
\label{Optimisation}
\begin{figure}[tb]
    \begin{minipage}{0.32\textwidth}
        \centering
        \includegraphics[width=0.98\textwidth]{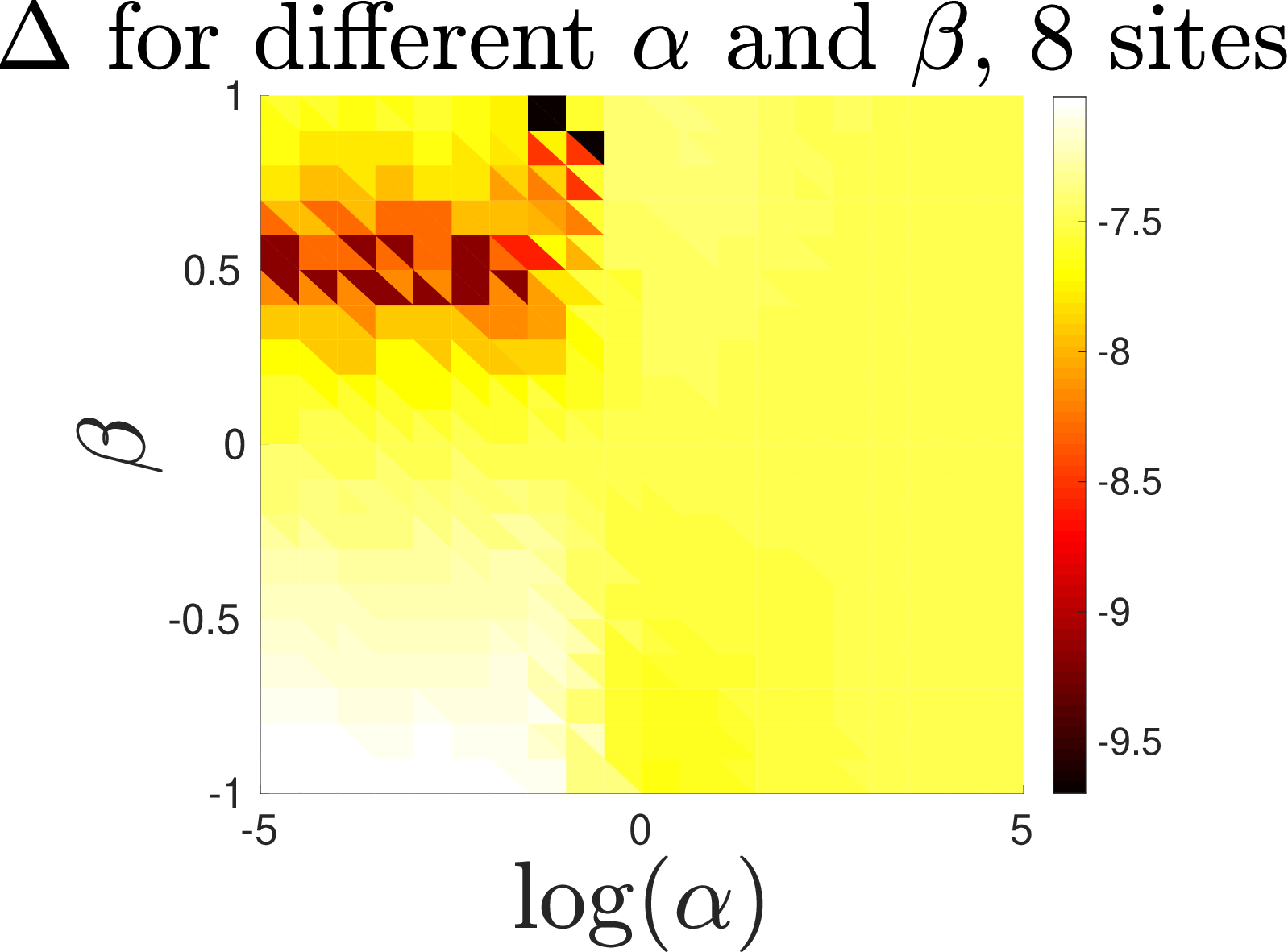}
        \caption{Plotted $\log_{10} \Delta_m $ at $t=0.1$ for $N=8$}\label{optn8}
    \end{minipage}\hfill
    \begin{minipage}{0.32\textwidth}
        \centering
        \includegraphics[width=0.98\textwidth]{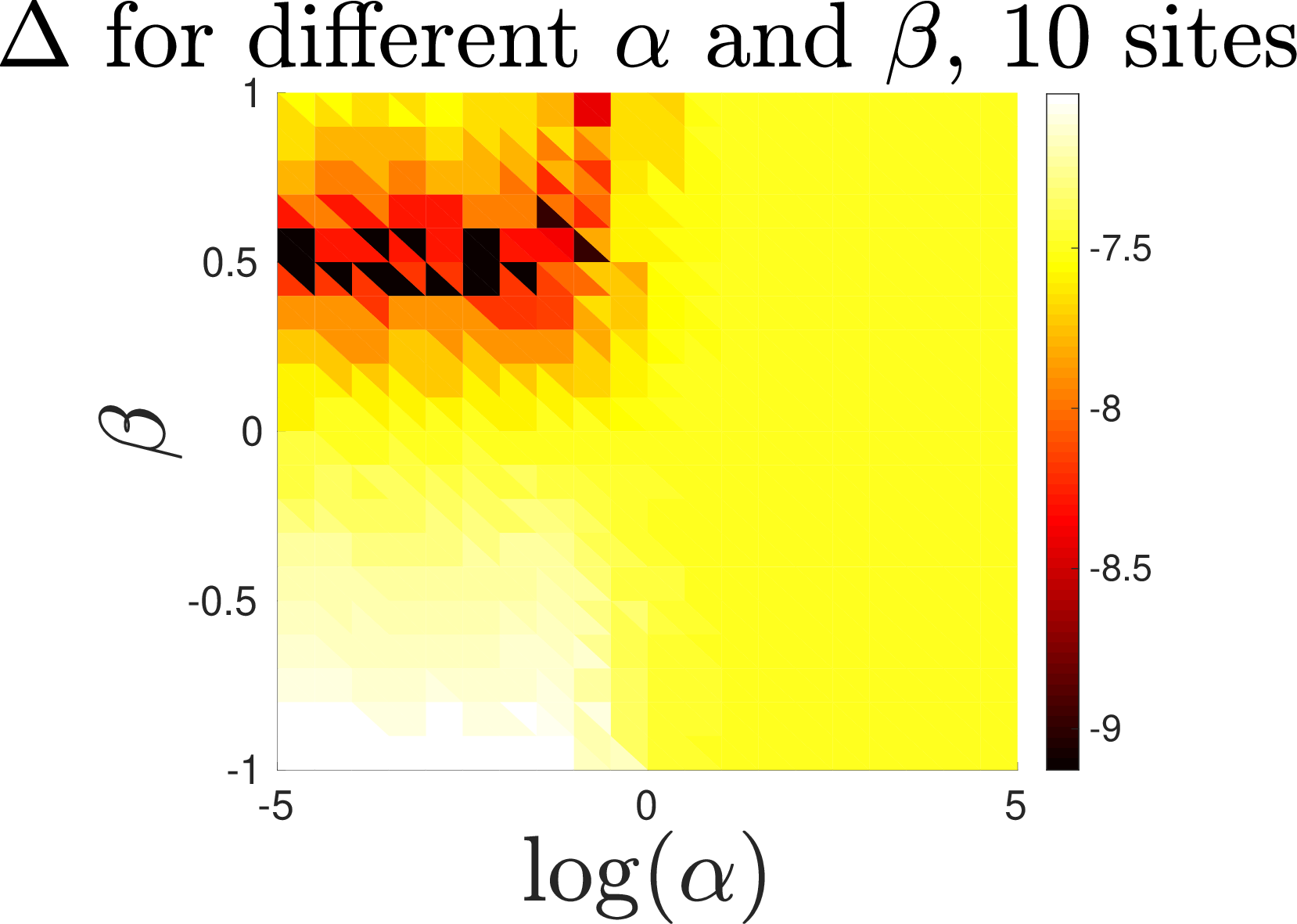}
        \caption{Plotted $\log_{10} \Delta_m $ at $t=0.1$ for $N=10$}\label{optn10}
    \end{minipage}\hfill
    \begin{minipage}{0.32\textwidth}
        \centering
        \includegraphics[width=0.98\textwidth]{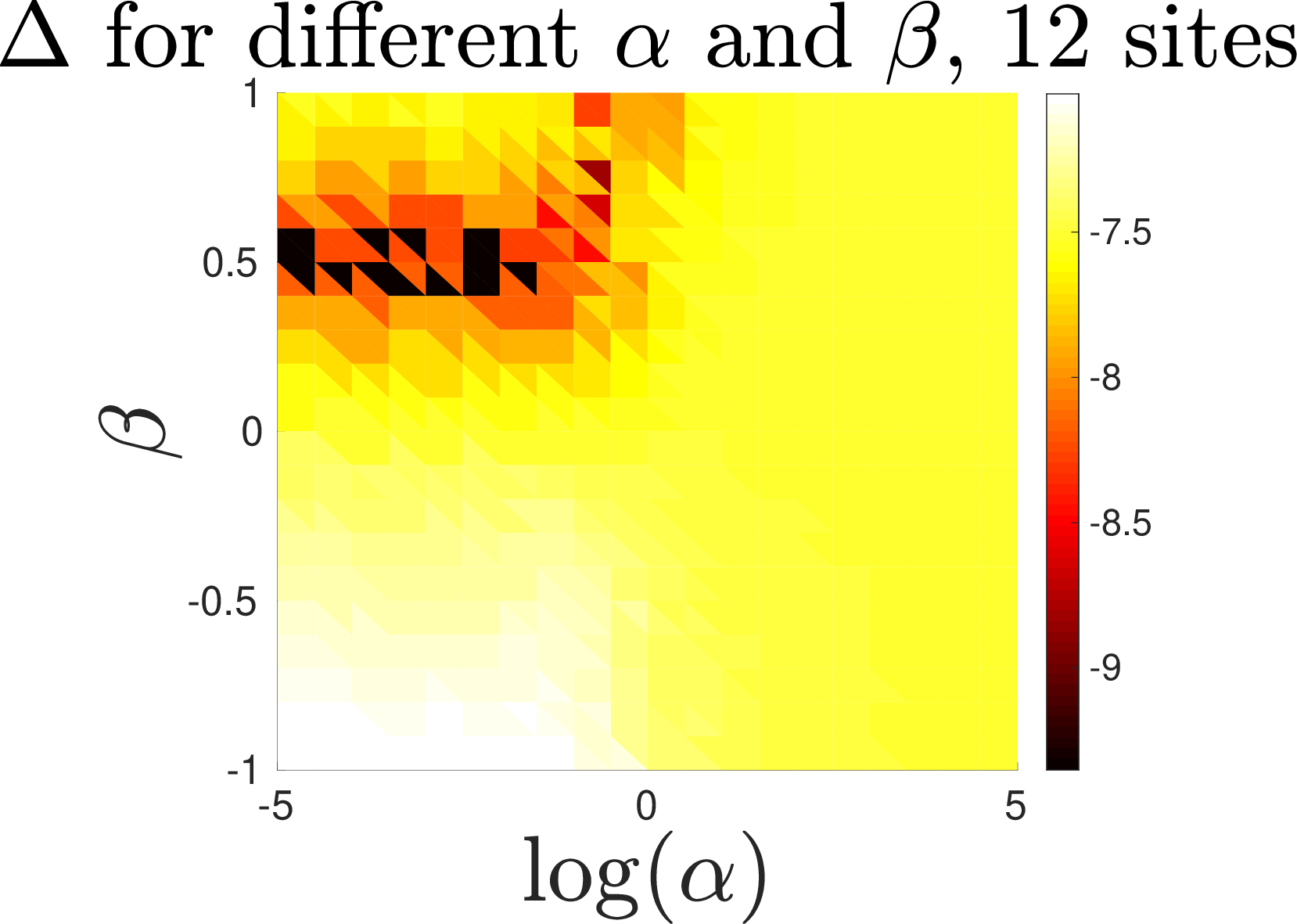}
        \caption{Plotted $\log_{10} \Delta_m $ at $t=0.1$ for $N=12$}\label{optn12}
    \end{minipage}
\end{figure}
To determine the optimal valances of the parameter $\alpha$ and $\beta$, 
The algorithm was applied to a spin-1/2 Neel state:
\begin{align}
    \hat{\Psi_N} &= \ket{\cdots\uparrow \downarrow\uparrow\downarrow\cdots}
\end{align}
and evolved with a spin-1/2 isotropic Heisenberg XY model given by the Hamiltonian
\begin{align}
    \hat{H} &= \sum_{i} \hat{S}^x_{i}\hat{S}^x_{i+1} + \hat{S}^y_{i}\hat{S}^y_{i+1}
\end{align}
up to $t=1$ with $\Delta t=0.001$ and $\mathcal{N}$ (number of Lanczos vectors per iteration) $=3$.
To find the best form of a bias function, a local magnetisation defined by
\begin{align}
    m_0(t) = 2\bra{\psi(t)}\hat{S}^z_0\ket{\psi(t)}
\end{align}
where $\hat{S}^{z}_0$ is a spin-1/2 operator at site $0$, is measured.
In the thermodynamical limit $m(t)$ has an analytical expression\cite{Search2010}:
\begin{align}
    m_z &= J_0(2t)
\end{align}
were $J_0$ is a $0$\textsuperscript{th} Bessel function of the first kind. 
The deviation from the analytical result is used as a measure for finding the best values of the parameters. 

To find the optimal set of parameters, a parameter space of $-1 < \beta <1$ and $-5 < \log(\alpha) < 5$ is searched.
The difference between the analytical value and the numerically obtained values of the magnetisation ($\Delta_m$) at $t=1$ are plotted.
Although there is a very dark region, implying very small differences, this region is numerically unstable.
This occurs because in this region the bias coefficients are very small at some sites, and such a property causes a biased MPO construction to be 
numerically unstable.
In particular, the region around $0.45<\beta<0.55$ and $-4<\log(\alpha)<0$ is so unstable that it is unusable. 
It seemed that a region at the upper-left edge is promising for the optimal values of 
$\alpha$ and $\beta$. 
The optimal pair of the parameters are found to be $\log(\alpha_o)=-5$ and $\beta_o=0.5$.
They provide a numerical stability and optimal biases which results in the smallest error per iteration for a general Hamiltonian. 
\subsection{Benchmark}\label{Benchmark}
The benchmark of the algorithm was done on the following three models.
\begin{enumerate}
    \item Spin-1/2 Heisenberg XY Model. 
    \item Imaginary time evolution of Haldane-Shastry model.
    \item Time evolution in a long-range Ising chain as studied by Halimeh \etal\cite{Halimeh2016}. 
\end{enumerate}
The first model is chosen to confirm the effectiveness of the algorithm in providing a local behaviour by comparing it to the analytical result. 
The second model is one of the few examples of a spin chain with long-range interactions that has an analytical expression for the ground state energy.
Therefore the imaginary time evolution is performed using the algorithm to confirm that the algorithm does in fact simulate
the long-range interactions effectively. 
Finally, to apply the algorithm to time evolution of a model with long-range interactions,
the algorithm is compared to the Time Dependent Variational Principle algorithm\cite{Haegeman2011} 
by performing a time evolution in a one-dimensional long-range transverse Ising chain.
The results from TDVP on a study done by Halimeh \etal\cite{Halimeh2016} where provided by the authors.
The decay time, the period and the amplitudes of oscillations are compared for two types of quenches.

\subsubsection{Spin-1/2 Heisenberg XY Model}
As discussed in the optimisation section (\S\ref{Optimisation}), the Hamiltonian of the spin-1/2 Heisenberg XY model is given as
\begin{align}
    \hat{H} &= \sum_{i}\hat{S}^{x}_{i}\hat{S}^{x}_{i+1}+\hat{S}^{y}_{i}\hat{S}^{y}_{i+1}
\end{align}
Starting from the Neel state, which can be described by MPS of dimension $1$, the magnetisation at a site evolves as
\begin{align}
    m_z &= \frac{1}{2}J_0(2t)
\end{align}
where $m_z$ is the local magnetisation defined as $\bra{\psi}\hat{S}^{z}\ket{\psi}$; and $J_0$ is the zeroth Bessel function of the first kind.

\begin{figure}[tb]
    \centering
    \caption{Time evolved Neel state in XY model}\label{figneelXY}
    \includegraphics[width=1.0\textwidth]{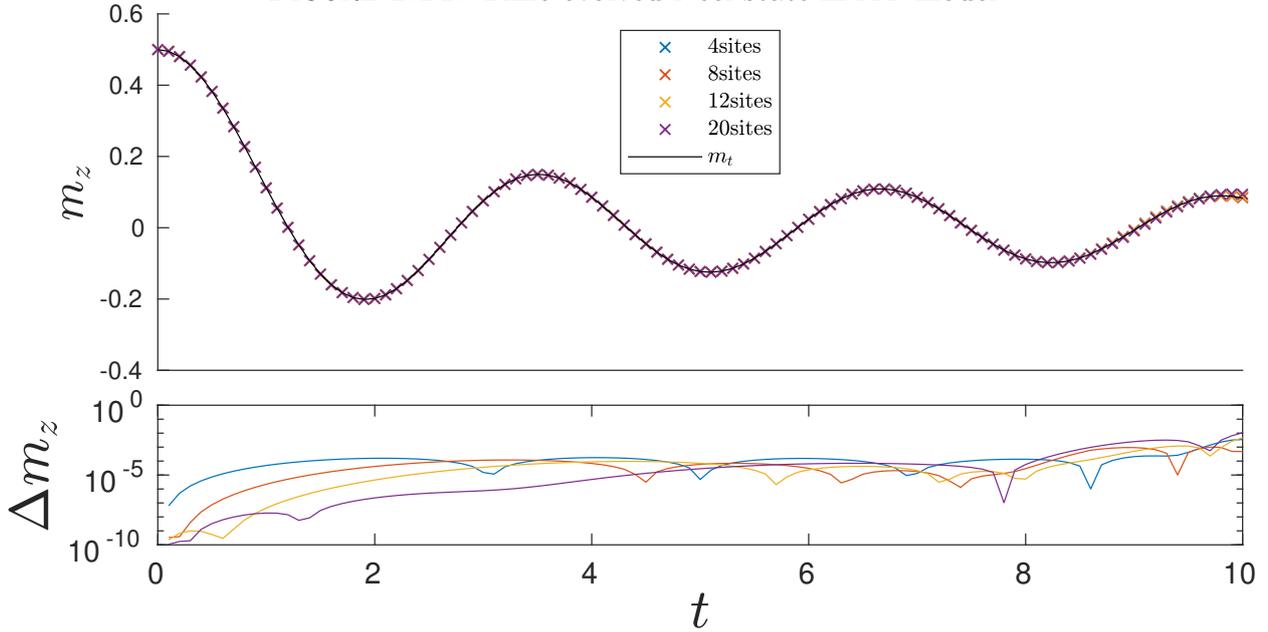}
    \legend{Magnetisation $m_z$ over time for different unit cell sizes. $m_t$ is the analytical value of the local magnetisation.}
\end{figure}

The Neel state was evolved with the algorithm using the parameters $m_{\max}=150$, $N=4$, $8$, $12$, $20$, $\Delta t=0.001$, and $\mathcal{N}=3$.
The numerical simulation showed agreement with the analytical solutions as shown in \fig{figneelXY}.
For the first few instances, the error from $N=20$ is measured to be smaller than $N=4$.
This is caused by the numerical instability of Lanczos method.
The method is unstable when $\mathcal{N}$ is large compare to the dimension of Hilbert space.
Since the Hilbert space spanned by $N=4$ spin-1/2 particles only has the dimensions of $16$  by $16$,
it is possible that the numerical error from the Lanczos step were dominant compared to the numerical error from the 
Suzuki-Trotter steps in the first instances. 
At later times, the error from the Suzuki-Trotter steps accumulates and the simulation loses its stability around $t=8$.

\subsubsection{Imaginary Time Evolution on Haldane Shastry Model}
To test the algorithm on a Hamiltonian with interactions longer than nearest neighbour interactions, an imaginary time evolution on 
Haldane-Shastry model was conducted.
Although the Lanczos and Suzuki-Trotter requires a symmetric Hamiltonian, as suggested in \cite{Orus2007,Verstraete2004a,Haegeman2011}
imaginary time evolution must be one of the exceptional case for this algorithm as well.

The Haldane-Shastry (HS) model\cite{Haldane1988,Shastry1988} is a spin-1/2 chain given by the Hamiltonian
\begin{align}
    \hat{H} &= \sum_{i < j} \frac{\bm{\hat{S}}_i\cdot\bm{\hat{S}}_j}{|\bm{r}_{i}-\bm{r}_j|^2}\label{HamilHS}
\end{align}
It has an analytical expression for the ground state energy\cite{Haldane1988}:
\begin{align}
    E_0(N) &= -\frac{1}{24}\left( \frac{2\pi}{N} \right)^2(N^2-1)\left( \frac{1}{4}N\right)
\end{align}
Where $N$ is a number of sites of the entire chain. In the thermodynamical limit, the energy per site ($E_{\infty}$) is 
\begin{align}
    E_\infty &= \lim_{N\to\infty}E_0(N) = -\frac{\pi^2}{24}
\end{align}
The Hamiltonian is implemented numerically by approximating the decaying long-range interactions with a series of exponentials
\begin{align}
    \frac{1}{|\bm{r}_i - \bm{r}_j|} \approx \sum_{n=0}^{5} \alpha_0 e^{-|\bm{r}_i-\bm{r}_j|/\xi}
\end{align}
The list of coefficients $\alpha_n$ and $\xi_n$ are given on \tab{HScoeffs}
\begin{table}[h]
\centering
\caption{Coefficients used for the exponential fit of $1/|\Delta r|^2$}
\label{HScoeffs}
\begin{tabular}{c|lllll}
$n$      & 0          & 1          & 2          & 3          & 4        \\\hline
$\alpha_n$ & 7.17784806 & 0.77706345 & 0.09122669 & 0.00737114 & 0.00030172 \\
$1/\xi_n$  & 2.49585612 & 0.85120844 & 0.27230538 & 0.06965503 & 0.01242821
\end{tabular}
\end{table}

\begin{figure}[bht]
    \caption{Imaginary time evolution on HS model}\label{figimagtimeHS}
    \begin{minipage}{0.48\textwidth}
        \centering
        \includegraphics[width=1.0\textwidth]{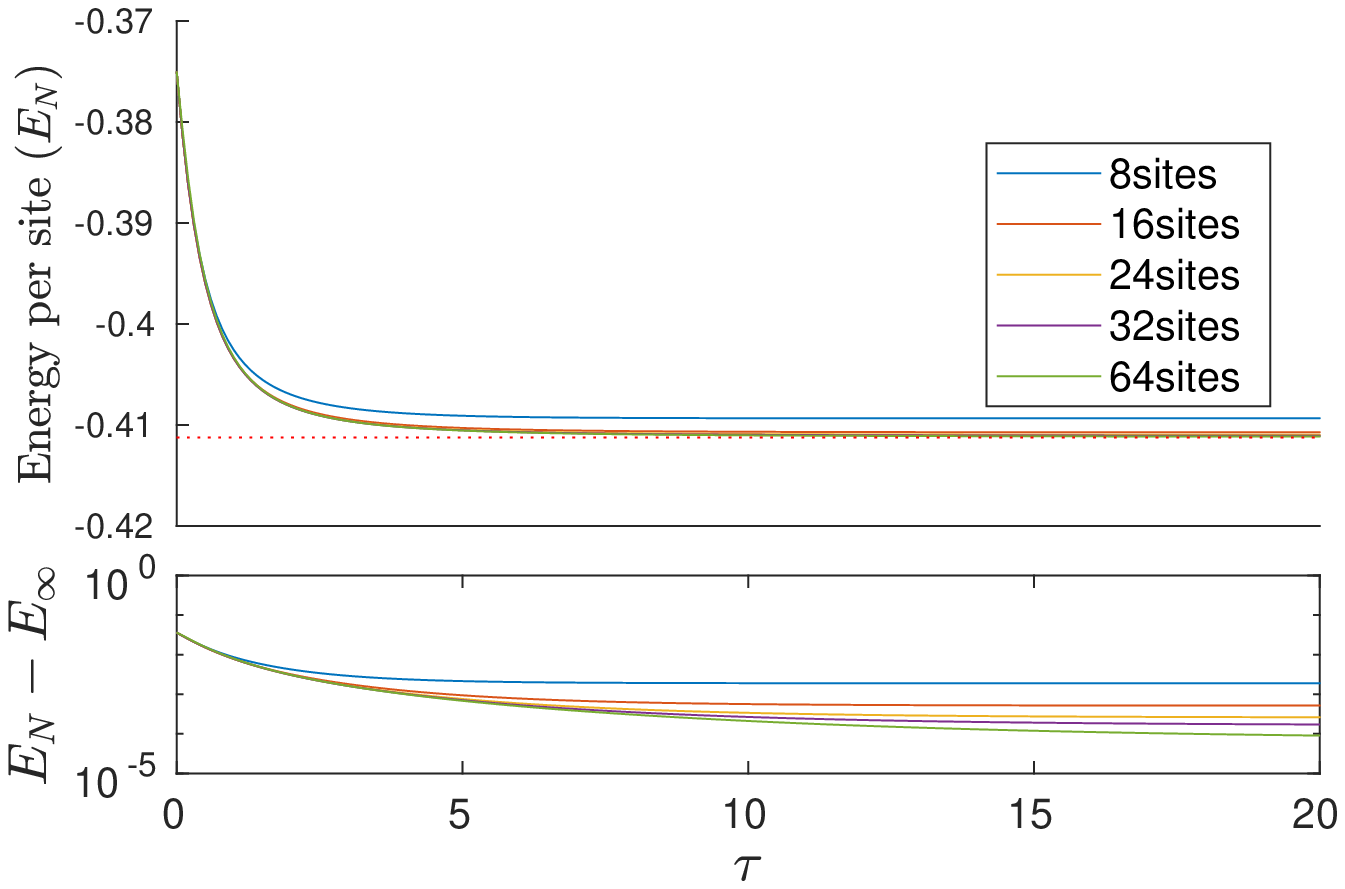}
    \end{minipage}\hfill
    \begin{minipage}{0.48\textwidth}
        \centering
        \includegraphics[width=1.0\textwidth]{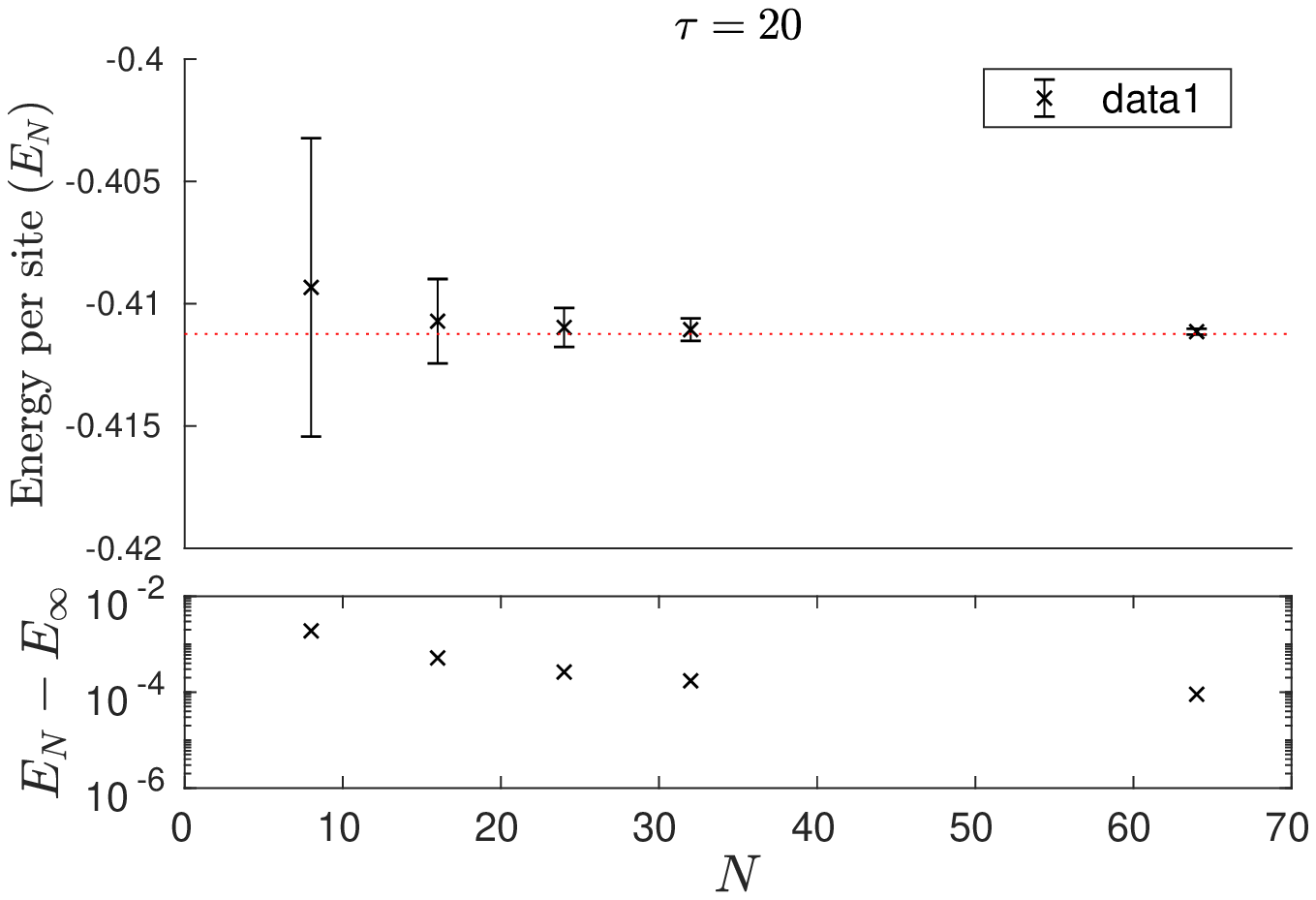}
    \end{minipage}
    \legend{(left) The evolved states asymptotically approaches towards some constant value.
        (right) At $\tau=20$, the ground state energy converges towards the correct energy as the number of sites in a unit cell increases.
        The red dotted lines on the top figures are the true ground state energy
        ($E_{\infty} = -\pi^2/24$).}
\end{figure}

An initial random $SU(2)$ symmetric state is evolved in the Hamiltonian with parameters $m_{\max}=150$, $N=8,16,24,32$, $dt=0.001$, and $\mathcal{N}=5$.
As shown in \fig{figimagtimeHS} (left), convergence toward the ground state energy of the truncated Hamiltonian is observed. 
The truncated Hamiltonian is an effective Hamiltonian which contains the interaction terms shorter than $N/2-1$
in \eqn{HamilHS}. 
The maximum error in energy at a site caused by the truncation of the interaction terms longer than $N/2$ in this model is
\begin{align}
    \mathcal{E}_{\max} &= \sum_{n=N/2}^{\infty} \frac{\xi(n)}{n^2}\label{worsterror}
\end{align}
where $n$ is the number of sites away from the site, and $\xi(n)$ is a correlation function ($\xi(n) = \braket{\bm{\hat{S}}_i\cdot\bm{\hat{S}_{i+n}}}$). 
For a non-critical quantum phase in a spin chain, $\xi(n)$ \emph{at worst} decays algebraically with $1/n$, 
and has order of $s^2$ when $n$ is close to $1$.
Therefore, the $\xi(n)$ that gives the worst error per interaction per site is:
\begin{align}
    \xi(n) &= \frac{1}{4n}
\end{align}
This function when substituted in \eqn{worsterror}, is used to estimate the error bars on \fig{figimagtimeHS} (right).
As it can be seen, the true energy lies well within the error bars, proving that algorithm is capable of performing imaginary time evolution 
with long-range interactions.
\subsubsection{Long-Range Ising Model}
Lastly, global quenching on a long-range Ising Hamiltonian 
\begin{align}
    \hat{H}(h) &= -4\sum_{i<j} \frac{\hat{S}^z_i\hat{S}^z_j}{|\bm{r}_i - \bm{r}_j|^3} -2h\sum_{i}\hat{S}^{x}_i
\end{align}
were performed (factors of $2$ and $4$ comes from the conversion between the spin-1/2 operators and the Pauli matrices).
The model and the quenching are studied in detail by Halimeh \etal \cite{Halimeh2016} using the Time Dependent Variational Principle \cite{Haegeman2011}.
The literature investigates global quenching on the ground state of the Hamiltonian at $h=0$.
Trivially, the ground state is a product state (MPS with $m=1$), where all the spins lines up in the positive $z$-direction.
At $t=0$, the value of $h$ is instantaneously changed to $h_f=0.28$ for a shallow quenching and $h_f=0.99$ for a deep quenching.
The introduction of a finite transverse field causes the state to no longer be in a stationary state, to some excited states of the quenched Hamiltonian
are observed.
The literature showed numerically the existence of a pre-thermal superposition of phase characterised by small amplitude oscillations with finite magnetisation 
($m_z$): and thermalisation of an initial state after a deep quenching characterised by the rapidly decaying global magnetisation.

To reproduce the results from this paper, numerical simulations were conducted with the same quenching parameters.
Like the previous model, the algebraically decaying long-range interactions are approximated with a sum of exponential functions
\footnote{Initial guess is made by following the scheme by Kaufmann \cite{Kaufmann2003} and further fit is done using a commercial software.}.

\begin{figure}[tb]
    \begin{minipage}{0.48\textwidth}
    \centering
    \caption{The results of time evolution simulated with gTEBD
    for the global quenching with $h_f=0.99$ (main plot) and $h_f=0.28$ (inset).}
    \label{newalgomagFigure}
    \includegraphics[width=0.98\textwidth]{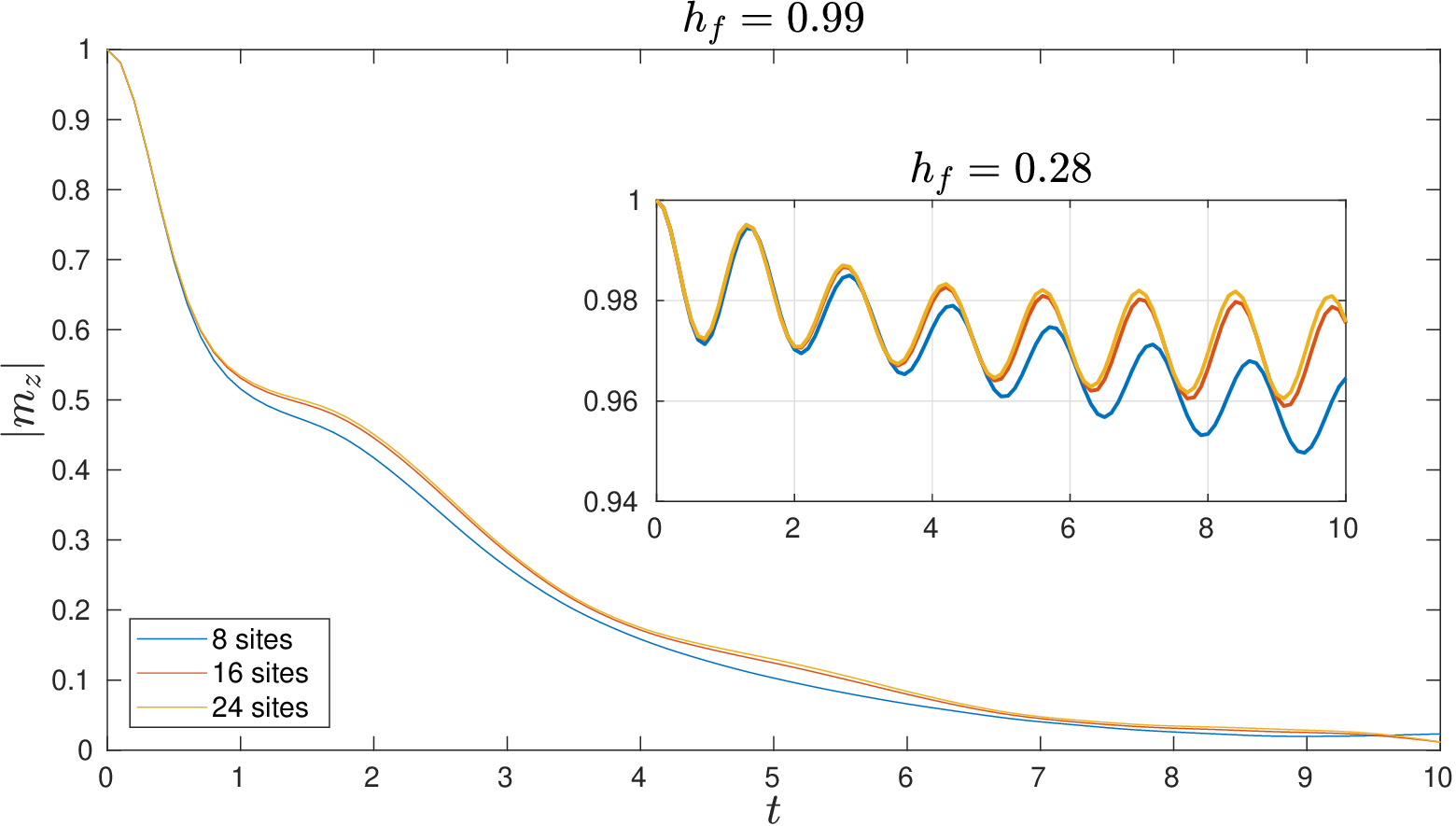}
    \end{minipage}\hfill
    \begin{minipage}{0.48\textwidth}
    \centering
    \caption{Comparison between the time evolution of the magnetisation between TDVP and gTEBD.}
    \label{overlapFigure}
    \includegraphics[width=0.90\textwidth]{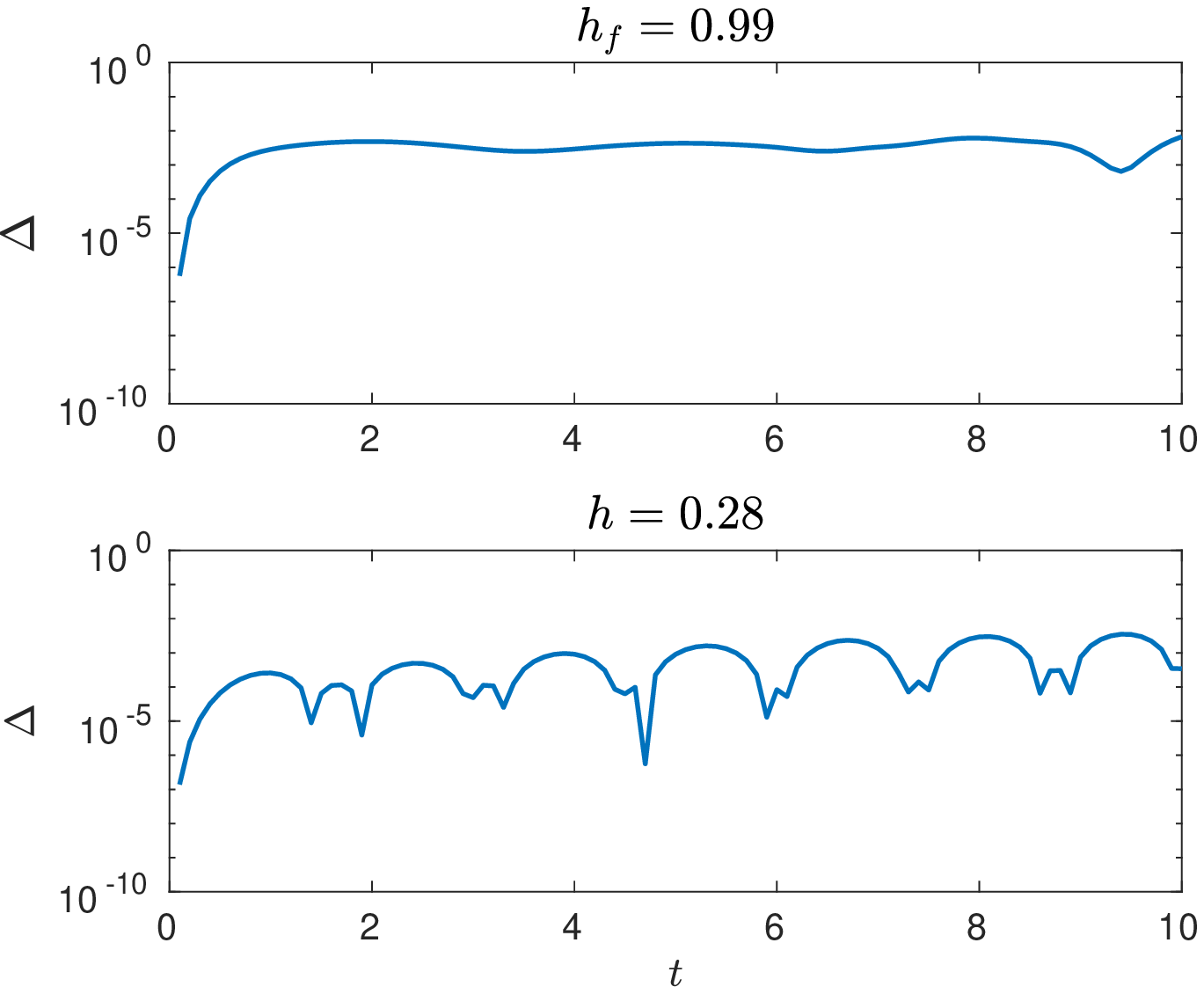}
    \legend{$\Delta$ is an absolute difference between TDVP results and gTEBD result.
    It shows that both of the algorithm agrees with one another stably over the timescale of the thermalisation.}
    \end{minipage}
\end{figure}

The results from the algorithm showed agreement with the TDVP results within $10^{-4}$ (\fig{overlapFigure}).
Both result gave an initial quadratic decay which comes from the time-energy uncertainty principle:
\begin{align}
    \sigma_{\hat{H}}\frac{\sigma_{\hat{B}}}{\left|\deriv{\braket{\hat{B}}}{t}\right|} \geq \frac{\hbar}{2}
\end{align}
where $H$ is a Hamiltonian of a system, $B$ is some operator and $\sigma_{\hat{O}}$ is the uncertainty of an operator $\hat{O}$
(see \App{Proofofenergy} for the proof).
Since the initial state is the eigenstate of $\hat{S}^z$ operator, 
the first derivative of the magnetisation with respect to time must be $0$ when $t=0$. 
Therefore it must decay quadratically or slower. 

For the shallow quench with $h_f=0.28$, increasing the number of sites in the unit cell made the evolution of the state converge towards 
the true no-truncation limit. 
As observed by Halimeh \etal \cite{Halimeh2016},
the result from the gTEBD magnetisation reached a plateau around $0.97$ for $N=24$ 
with the correct oscillation period ($T = 1.4$) and amplitude ($\sim0.2$).
For the deep quench, both methods predicted a small plateauing behaviour around $t=1.25$ 
followed by fast decay of magnetisation with a half time of $t_{1/2}\approx 1.8$. 

\subsection{Differences From Preexisting Algorithms}
To conclude this section, let us discuss the difference between the 
three time-evolution algorithms for MPS: iTEBD (infinite Time-Evolving Block Decimation), TDVP, and gTEBD.
(i)TEBD is an algorithm developed to evolve a state with a Hamiltonian with nearest-neighbour interactions.
It evolves a state by computing the time evolution unitary operator of nearest neighbour interaction terms directly and applies it for even bonds and
odd bonds iteratively \cite{Vidal2004,Schollwock2011}.
When the unitary evolution operator is applied to two sites, the canonical form is recovered by applying a truncation to the evolved sites by performing
SVD and rejecting the smallest singular values.
This truncation forces the time-evolved state to leave the variational space.
As a result, the state at longer time may deviate from the true result.

TDVP, on the other hand, does not perform a truncation. Instead, the maximum dimension of the local matrix is fixed, and by constructing linear
coupled Ordinary Differential Equations (ODE)  of the elements of local matrices from Schr\"{o}dinger's 
Equation, the elements of MPS are evolved directly \cite{Haegeman2011,Milsted}.
By solving the matrix elements, a state time-evolved using TDVP does not leave the variational space. 
This means that TDVP finds an optimal time-evolved state within the given dimensions of the MPS. 
This enables the high precision calculation beyond iTEBD. 
Also TDVP is capable of performing time evolution with long-range interactions. 
However, due to gauge fixing, which occurs during the construction of coupled ODE, the Hamiltonian and the state requires translational invariance.

gTEBD is not restricted to nearest neighbour interactions.
It performs well with both imaginary and real time evolution of Hamiltonians with long-range interactions.
It is also shown that the accuracy is comparable to TDVP, suggesting that the gTEBD-evolved states stay close to the variational 
space spanned by the elements of the local MPS matrices.
This is a side effect from the minimisation steps performed in the Lanczos steps.
Although the truncation step in the canonicalisation steps takes some information away, the updated $L$ matrices has more information 
than the previous ones. This effectively keeps the evolved MPS close to the variational space.
At the end of a \emph{sweep}, the MPS, in principle, is recovering its full variational characteristic by not rejecting $D$ and $R^\dag$ matrices. 
This can be a possible reason for the agreement of gTEBD and TDVP at a very high precision. 

\newpage
\section{Conclusion}
In the first half of the chapter, the Matrix Product State formulation of the quantum mechanics was introduced.
With the help of Tensor Network Diagrams, features of quantum mechanics such as the notions of states, operators, and 
computation of expectation values are recovered.
The key feature of MPS is a truncation and projection which enables MPS with large bond dimensions to be expressed in a compact form. 

In the latter half of the chapter, a new algorithm for evolving MPS with long-range interaction was introduced.
It is an algorithm which combines Krylov subspace expansion and Suzuki-Trotter decomposition to perform the evolution with
long-range interactions.
It is shown that there is a degree of freedom on how the Hamiltonian is decomposed.

For minimising the error at the Suzuki-Trotter steps, a bias function is introduced using the error function.
This function fixes the gauge present in the decomposition.
An optimisation of the function was performed with an XY model. 
The optimisation showed that parameter ranges with a steep slope produces a catastrophic numerical instability during the MPO construction phase. 
Therefore the optimal set of parameters was chosen from the numerically stable region.

To benchmark the algorithm, it was tested on three distinct models:
real time evolution in a nearest neighbour model, imaginary time evolution with long-range interactions, 
and real time evolution with long-range interactions. 
Those evolutions agreed well with the analytical and literature values
and the result from the imaginary time evolution shows the capability to evolve with non-hermitian operators.
It also shown that the algorithm asymptotically approaches toward the dynamics of full long-range interactions in the limit of $N\to\infty$.

It was discussed that gTEBD has few constraints on the form of the Hamiltonian that it can be applied compared to the other existing algorithms.
This makes the algorithm more flexible than the known time-evolution algorithms for MPS.
By taking a large unit cell size it has a potential to perform a time evolution on two-dimensional lattice models.

The invention of gTEBD opens a new door to large-scale precision computational simulations of strongly correlated quantum lattices.
The algorithm, however, has many aspects which can be improved in future work. 
First, the optimal form of the bias function is not known.
In this chapter, a possible form of a bias function was conjectured in some simple cases from the symmetries in the Hamiltonian.
However, because of the different bond profile for different Hamiltonians, 
the set of optimal coefficients must be different.
Establishing a generic scheme for finding out such a set of coefficients is strongly suggested for the future. 

Secondly, in the current implementation of the algorithm, 
when a state is evolved for longer times, the matrix dimensions needed to support the evolved MPS exceeds some threshold.
A direct application of the MPO causes the MPS dimensions to increase from $m$ to $mM$.
As it was shown in the introductory section, most of the components of MPO are $0$,
therefore, up to some point MPS projection provides a very accurate result. 
However as the number of MPO-MPS operation increases, the accuracy of the projection decreases, leading to non-physical behaviour. 
It was not observed in the benchmark cases discussed in this chapter, but as the number of sites and complexity of the Hamiltonian increases,
this may lead to some catastrophic behaviour
(this topic is discussed further in the next chapter).
Therefore an implementation of some time adaptive scheme for determining time dependent threshold of MPS dimensions is needed.

Finally, one can alter the estimation method for matrix exponentials.
For this implementation of gTEBD, the Lanczos steps are used to estimate the matrix exponential, 
but there is no reason not to chose other existing method. 
Such methods include $n$\textsuperscript{th} script Runge–Kutta methods and other crude finite-system time-evolution method discussed by
Garcia-Ripoll\cite{GarcIa-Ripoll2006}. Construction of the unitary time-evolution MPO introduced by Zaletel \etal\cite{Zaletel2015}
may also be a viable option for a possible replacement. 

In short gTEBD is a fully expandable algorithm that can go beyond what previously known algorithms can perform. 
As it is just developed, there are many optimisations that could be done to improve the performance of the algorithm.
Future research on the large-scale simulation using this algorithm as well as implementation of more sophisticated schemes for performance enhancement 
are anticipated.

\graphicspath{{figures/ch3/}}
\chapter{Global Quenching on a Two-Dimensional Triangular Ising Lattice}
\label{dynamicalphasetransitions}
\section{Spin-1/2 Two-Dimensional Triangular Ising Lattice}
\begin{figure}[tb]
        \centering
        \includegraphics[width=1.0\textwidth]{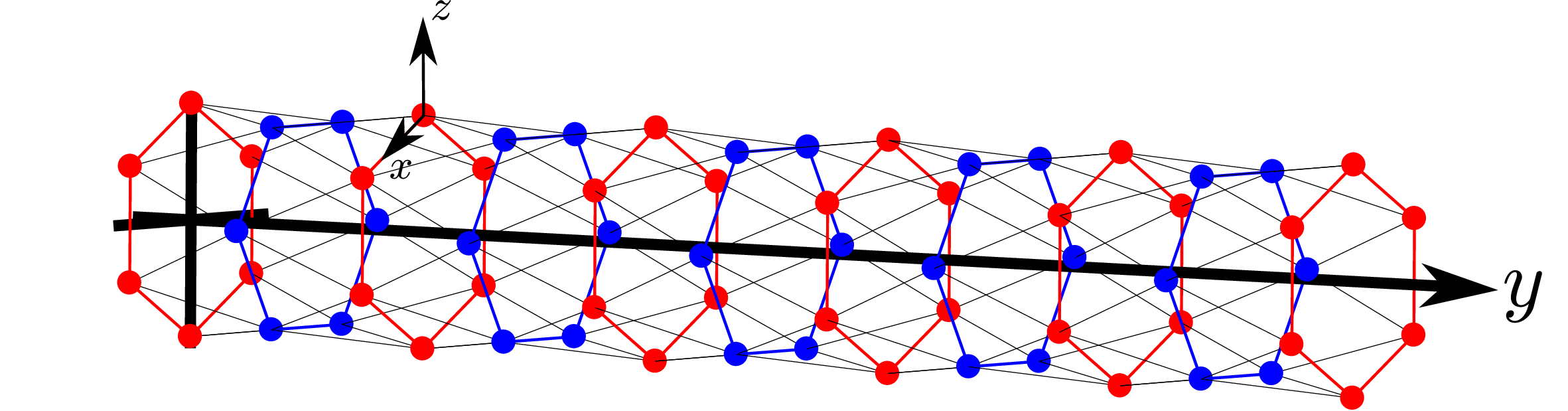}
        \caption{A triangular spin lattice on a surface of a cylinder}
        \label{trilatcyl}
\end{figure}
\begin{figure}
        \centering
        \includegraphics[height=0.25\textheight]{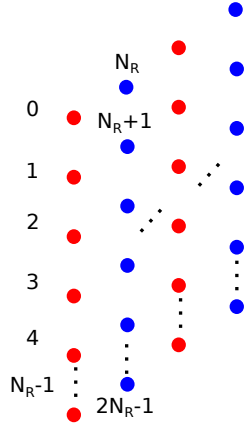}
        \caption{Numbering of the sites upon projection onto the one-dimension}\label{Numbering}
\end{figure}
In this chapter, the full features of a two-dimensional triangular spin-1/2 Ising lattice (2DTIL\index{2DLTI}) described by the Hamiltonian
\begin{align}
    \hat{H}(\alpha,\Gamma) &= \sum_{i}\sum_{ j > i } \frac{\hat{S}^z_i\hat{S}^z_j}{|\bm{r}_i-\bm{r}_j|^{\alpha}} 
    + \Gamma \sum_{i} \hat{S}^x_i &(\alpha,\Gamma > 0)
\end{align}
is investigated.
To simplify the calculation, the model on the surface of an infinite cylinder is considered. 
This can be done by providing periodic boundary conditions to
one of the dimensions of the lattice (\fig{trilatcyl}). 

This simplification also allows a further projection of the surface on to a one-dimensional chain,
where the MPS becomes the most suitable representation. 
The projection is done as follows:
First, $N_R$ sites are evenly placed on a ring of a cylinder.
The radius of the cylinder ($R$) is chosen such that the $\mathbb{R}^3$ Euclidean nearest neighbour distance is $1$.
\begin{align}
    R &= \frac{1}{\sqrt{2\left( 1-\cos\frac{2\pi}{N_r} \right)}}
\end{align}
Then nearest layers of sites in the cylinder are placed with $\pi/N_r$ rotation.
The $\mathbb{R}^3$ distance of the cross sections between the layers ($\Delta y$) are chosen such that all the cross-layer
nearest neighbour (e.g. site $1$ and site $N_R+1$ in \fig{Numbering}. ) distances becomes $1$ as well. 
\begin{align}
    \Delta y &= \sqrt{1 - 2R^2\left(1-\cos\frac{\pi}{N_R}\right)}
\end{align}

In this representation the distance between sites are measured as $\mathbb{R}^3$ Euclidean distance 
of the space which embeds the cylinder. 
The construction allows all the nearest neighbour distances to be $1$. 
The Hamiltonian can be represented in terms of MPO with the unit cell size of $N_R$
\footnote{Notice that this is a Hamiltonian unit cell size, not the unit cell size of MPS. 
    To make the unit cell size of the Hamiltonian MPO to be equivalent to MPS unit cell size, one must seek a least common multiple of the unit cell sizes;
and repeat them until the size becomes equivalent.}
\newpage
\section{Ground State Phase Diagram}
\begin{figure}[tb]
    \begin{minipage}{0.5\textwidth}
        \centering
        \includegraphics[width=0.8\textwidth]{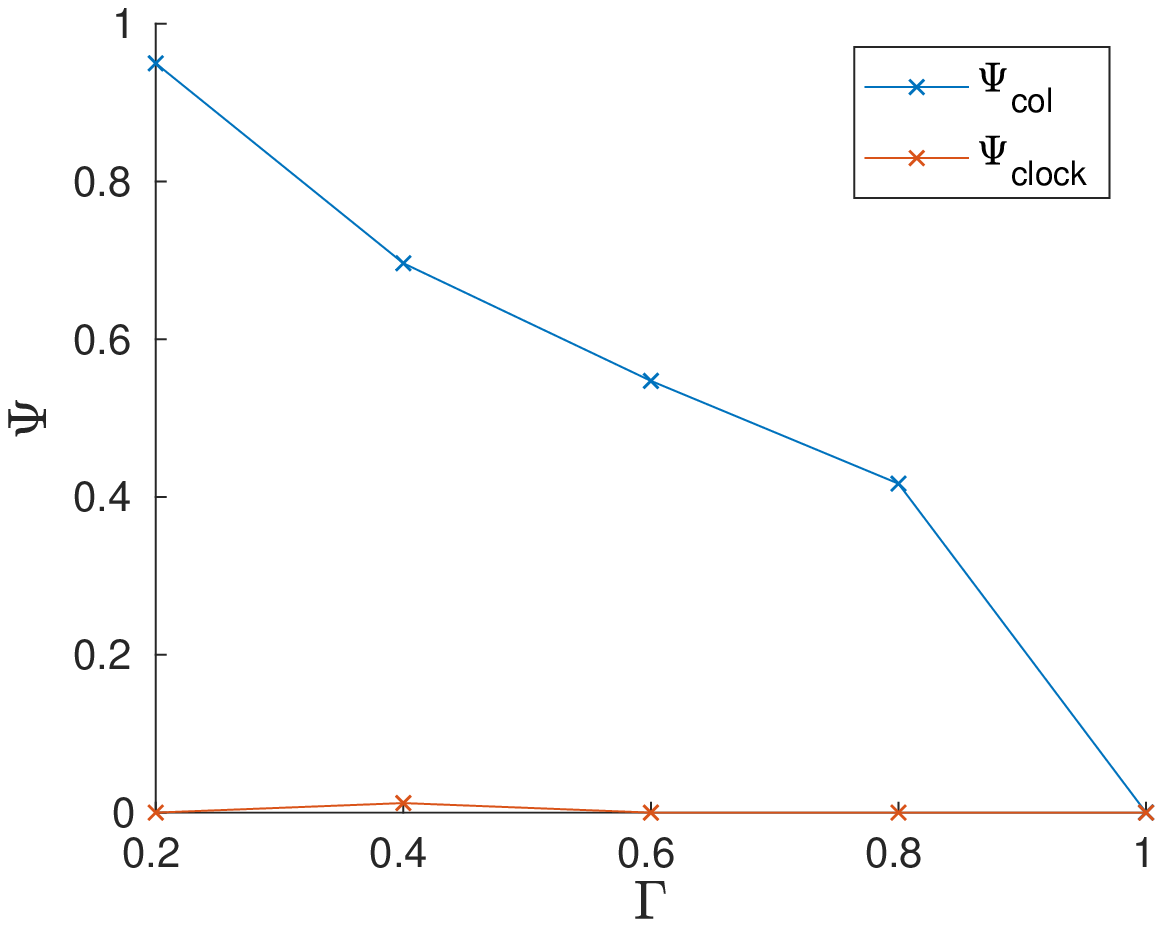}
        \caption{The values of the order parameters along $\alpha=3$.}\label{alpha3op}
    \end{minipage}\hfill
    \begin{minipage}{0.5\textwidth}
        \centering
        \includegraphics[width=0.8\textwidth]{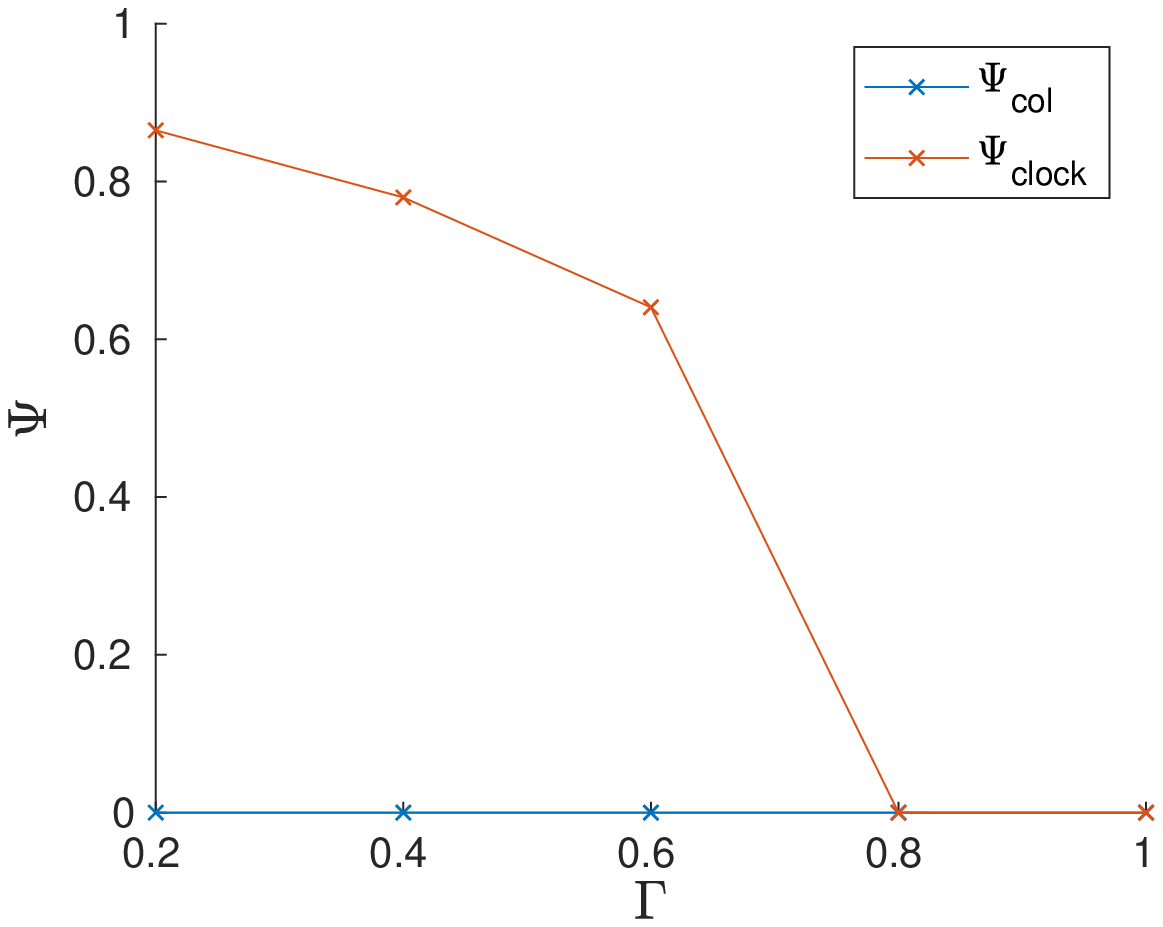}
        \caption{The values of the order parameters along $\alpha=8$.}\label{alpha8op}
    \end{minipage}
\end{figure}
To perform a global quench on a ground state, 
the quantum phase diagram spanned by the parameters $\alpha$ and $\gamma$ is constructed.
From the mean-field and QMC study of the model \cite{Humeniuk2016}, it is known that the ground state phase has three distinct regions (\fig{MFQMC}):
\begin{list}{}{}
    \item[Columner:] Antiferromagnetic order which is analogous to a Neel state in a one-dimensional models. 
        The order parameter is 
        \begin{equation}
            \Psi_{\mathrm{col}} = \frac{2}{N} \left|\bra{\psi} \left(\sum_{j} (-1)^j \hat{S}^{z}_j\right)\ket{\psi}\right|
        \end{equation}
        The phase is four fold degenerate because for each ring, spin up and down can be exchanged and the spin profile of each ring can be flipped.
    \item[Clock:] Antiferromagnetic order in a triangular sub lattice spanned by the three nearest sites. 
        Given a sub lattice (site $<i,j,k>$), if one site have a magnetisation of $m_z$ then other two sites has a magnetisation of $-m_z/2$.
        The order parameter is \cite{Isakov2003}
            \begin{align}
                m_{z;i} &= 2\bra{\psi} S^{z}_{i}\ket{\psi} \\
                \Psi_{\mathrm{clock}} &= \frac{3}{N} \sum_{<i,j,k>}\left| m_{z;i} + m_{z;j}e^{i4\pi/3} + m_{z;k}e^{-i4\pi/3}\right| /\mathcal{C} 
            \end{align}
        with $\mathcal{C} = 3/2$.
        Alternatively a triangular sub lattice with the magnetisation profile of $(m_z,-m_z,0)$ would have $\mathcal{C} = \sqrt{3}/2$.
        The phase is 3 fold degenerate for the (m,-1/2m,-1/2m) configuration because there are three possible sites which have $0$ $z$-magnetisation. 
        The phase is 6 fold degenerate for the (m,-m,0) configuration
        because fixing a site in a sublattice and exchanging the other two sites does not change the order;
        and a magnetisation at a site can be $1$, $-1$ or $0$
    \item[Disordered:] It is characterised by no net magnetisation in the $z-$direction in any of the sites. 
        All the local states are perpendicular to the eigenstates of the local $\hat{S}^z$ operators.
        Although it is technically a disordered state, a pseudo order parameter can be defined as 
            \begin{equation}
                \Psi_{\mathrm{dis}} = \left|\frac{1}{N} \sum_{i} 2\bra{\psi}S^{x}_i\ket{\psi}\right|\label{disoparam}
            \end{equation}
            to quantify how deep a state is in the disordered phase. 
\end{list}

However, the diagram provided in \fig{MFQMC} may not be applicable to our model because it is constructed with a finite two-dimensional lattice
with an open boundary condition.
Therefore to find the true quantum phase diagram of the model on the surface of a cylinder, DMRG calculations are conducted 
(c.f. Nariman \etal \cite{Nariman2017}).
To implement the long-range interaction term, the Hamiltonian is re-defined in terms of an MPO unit cell:
\begin{align}
    \hat{H}(\alpha,\Gamma) &=\sum_{i}\sum_{j=0}^{N_R-1}\sum_{k=0}^{N_R-1}\sum_{l=1}^{\infty}
    \frac{\hat{S}^{z}_{iN_R + j}\hat{S}^{z}_{(i+l)N_R + k}}{|\bm{r}_{iN_R + j}-\bm{r}_{(i+l)N_R + k}|^\alpha}\nonumber\\
    &+\sum_{p}\sum_{q=0}^{N_R-2}\sum_{r=q+1}^{N_R-1} \frac{S^{z}_{pN_R+q}S^{z}_{pN_R+r}}{|\bm{r}_{pN_R+q} - \bm{r}_{pN_R+r}|^\alpha} + 
    \Gamma \sum_{w}\hat{S}^{x}_w
\end{align}
To utilise the MPO representation of exponential decaying correlations discussed in \S\ref{Matrixproductoperators}, the cross layer
long-range interactions term (the first term with four sums) are approximated with exponential series with a method \cite{Kaufmann2003}.

A rough quantum phase diagram is created by scanning the parameter space for $\alpha=3$ and $\alpha=8$ (nearest neighbour limit) with 
$0<\Gamma<1$. 
As it can be seen in \fig{alpha3op}, a slice at $\alpha=3$ does not contain a clock ordered state.
It has a strong columnar order in the $x$-direction at $\Gamma=0.2$ and transition point roughly at $\Gamma \sim 0.9$.
On the other hand, as it is shown in \fig{alpha8op}, a slice at $\alpha=8$ does not contain a columnar ordered phase. 
The slice has a clock order to disorder transition around $\Gamma\sim0.7$.

\begin{figure}
    \begin{minipage}{0.6\textwidth}
        \centering
        \includegraphics[width=0.98\textwidth]{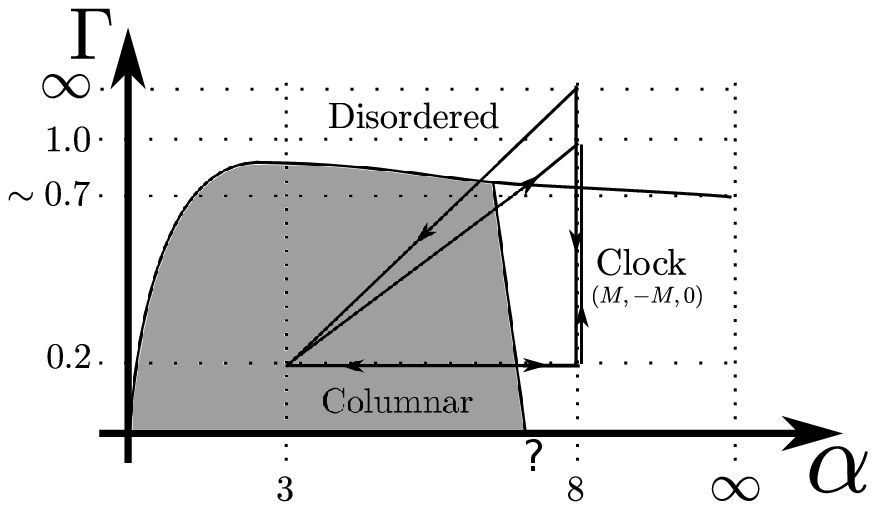}
        \caption{Schematic quantum phase diagram of the 2DTIL}\label{PhaseDiagram}
    \end{minipage}\hfill
    \begin{minipage}{0.40\textwidth}
        \centering
        \includegraphics[width=0.80\textwidth]{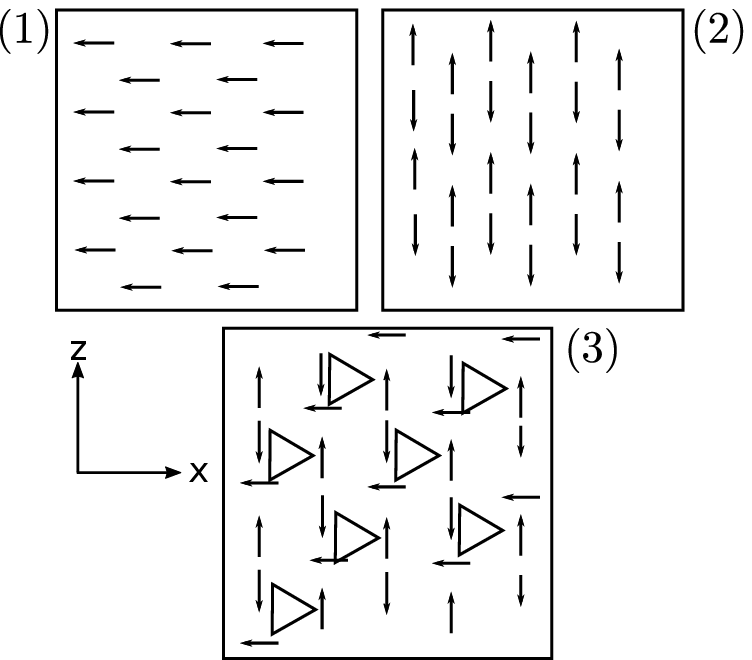}
        \caption{A graphical representation of: (1) disordered, (2) columnar, and  (3) clock ordered phase in 2DTIL}\label{PhaseOrdercyc}
    \end{minipage}
\end{figure}

Also a tomography of the clock ordered phase in the $\alpha=8$ slice revealed that 
the phase does not have the triangular sublattice of ($m_z,-m_z/2,-m_z/2$).
Instead, it is found that the it has triangular sublattices with ($m_z$, $-m_z$, $0$) (\fig{PhaseOrdercyc}, (3)).
Therefore the quantum phase diagram of the model is slightly different from what is expected from the literature. 
The deduced phase diagram is shown in \fig{PhaseDiagram}
\footnote{For the full DMRG study see \cite{Nariman2017}}.

\newpage
\section{Global Quenching of the Ground States}
With the new algorithm introduced in the last chapter, global quenching on the ground states of the model are performed.
They are performed for all the possible cases: columnar to disordered, columnar to clock, clock to disordered, clock to columnar, 
disordered to columnar, and disordered to clock (they are indicated as arrows in \fig{PhaseDiagram}).

As it can be seen from \fig{PhaseOrdercyc} (3), clock ordered phase requires $N_R$ to be a multiple of $6$ and 
the height of a unit cell in a multiple of three layers. Also, the columnar order (\fig{PhaseOrdercyc} (2)) 
requires $N_R$ to be even and the number of layers to be a multiple of $2$. 
Therefore to fully take account of all the possible phases which can be present in the real two-dimensional systems,
$N_R$ must be in multiple of $6$ and the number of layers in a unit cell must also be the multiple of $6$. 
Also, to give an accurate long-range order interaction profile using an exponential series, each site on the ring acquires $N_e$
exponential terms. This makes the dimension of MPO extremely large ($\sim \mathcal{O}(2dN_R^2N_e)$).

Performing numerical simulations with this many sites and dimensions of MPO was extremely difficult.
However it turned out to not be impossible.
Even with an initial state of a product state\index{Product state} (state with $m=1$), the time evolution 
from $t=0$ to $t=10$ with a time step of $dt=0.001$ took roughly a month
For all the other initial states with the initial MPS dimensions much larger than $1$, it took roughly the same 
amount of computational time because the initial product states quickly saturated to the maximum states in hours.

Although one month sounds like a very long time to perform a computation of this type, 
it is much better compared to the estimated computational time of $3$ to $10$ 
years with the original naive implementation of the algorithm.
A great reduction comes from the use of implicit applicator of MPO by projection instead of the 
explicit application of MPO and then performing truncation. 
Although it is not used in the calculations in this thesis, further speed up can be achieved by exploiting $Z(2)$ symmetry for some of the calculations. 

\newpage
\subsection{Entanglement Entropy}
Given an infinite chain (or lattice), the entanglement entropy is defined as the von Neumann entropy of a reduced density matrix of 
a subsystem defined by left (or right) of a reference bond. 
The reduced density matrix of a subsystem can be obtained by tracing out the other half of the infinite chain.
Given a state $\ket{\Psi}= \sum_{L} \sum_{R} A_{L.R} \ket{L}\ket{R}$ with $\ket{L}$ and $\ket{R}$ are the possible configurations of the spins in 
the left and the right of the system, reduced density matrix is:
\begin{align}
    \begin{cases}
        \hat{\rho}^{L} = \mathrm{Tr}_R\left( \ket{\Psi}\bra{\Psi} \right)\\
    \hat{\rho}^{R} = \mathrm{Tr}_L\left( \ket{\Psi}\bra{\Psi} \right)
    \end{cases}
\end{align}
The von Neumann entropy (refer to as entropy if not explicitly stated) of the reduced density matrix\cite{Schollwock2011} is:
\begin{align}
    S_{L|R} &= -\mathrm{Tr} \left(\hat{\rho}^{L} \log \hat{\rho}^{L} \right)
\end{align}

For a pure state, the entropy of the whole state is stationary in time.
However the entropy of a subsystem can exceed the entropy of the system itself. 
In fact, the entanglement entropies of the two subsystems (semi-infinite chains L and R) follows the subadditivity relation:
\begin{align}
    S_L + S_R \geq S_{\text{whole}}
\end{align}

The maximum entropy of a state that can be expressed using MPS with dimension $m$ is trivially expressed as $\log(m)$. 
However, this value is not the true upper limit because the 
truncation of MPS requires the magnitude of the singular values of the reduced density matrix to decrease
in an exponential manner\cite{Schollwock2011}.
The entanglement entropy of a system can be then used to determine the suitable $m$. 
A numerical scheme which adaptively changes $m$ may lead to a better computational efficiency. 

\subsection{Order Parameter Operator}
The definitions of the order parameters from the previous section takes value $0$ when the system consists of a superposition of ordered states.
To resolve it, the order parameter operators are defined:
\begin{align}
    \hat{\Psi}_{\mathrm{col}} &=\frac{2}{N} \left( \sum_{i} \hat{S}^{z}_{A;i} - \hat{S}^{z}_{B;i} \right)\\
    \hat{\Psi}_{\mathrm{clock}}  &= \frac{6}{N}\left( \sum_{i} \hat{S}^{z}_{A;i} + e^{i4\pi/3}\hat{S}^{z}_{B;i} 
   + e^{-i4\pi/3}\hat{S}^{z}_{C;i} \right)
\end{align}
where $A$, $B$ and $C$ are the sites in the sublattices (columnar=two nearest neighbour sites, clock=neighbouring three sites; c.f. \fig{PhaseOrdercyc}),
and $i$ goes over all the sublattices in the lattice.

With these operators, new order parameters are defined as:
\begin{align}
    |\Psi_{\mathrm{col}}|^8 &= \Braket{\left(\hat{\Psi}_{\mathrm{col}}^{\dag}\hat{\Psi}_{\mathrm{col}}\right)^4} \\
    |\Psi_{\mathrm{clock}}|^6 &= \Braket{\left(\hat{\Psi}_{\mathrm{clock}}^{\dag}\hat{\Psi}_{\mathrm{clock}}\right)^3}
\end{align}

With this construction, $4$ and $6$-fold degeneracies in the states will get resolved and 
finite values will be given if any of the eigenstates of an order parameter operator exists in a state. 

\subsection{Results and Discussions}
\subsubsection{Quenching on Initial Disordered states}
\begin{figure}
    \begin{minipage}{0.48\textwidth}
    \centering
    \caption{Change in the order parameter quenched initial disordered state into clock ordered phase}
    \includegraphics[width=0.98\textwidth]{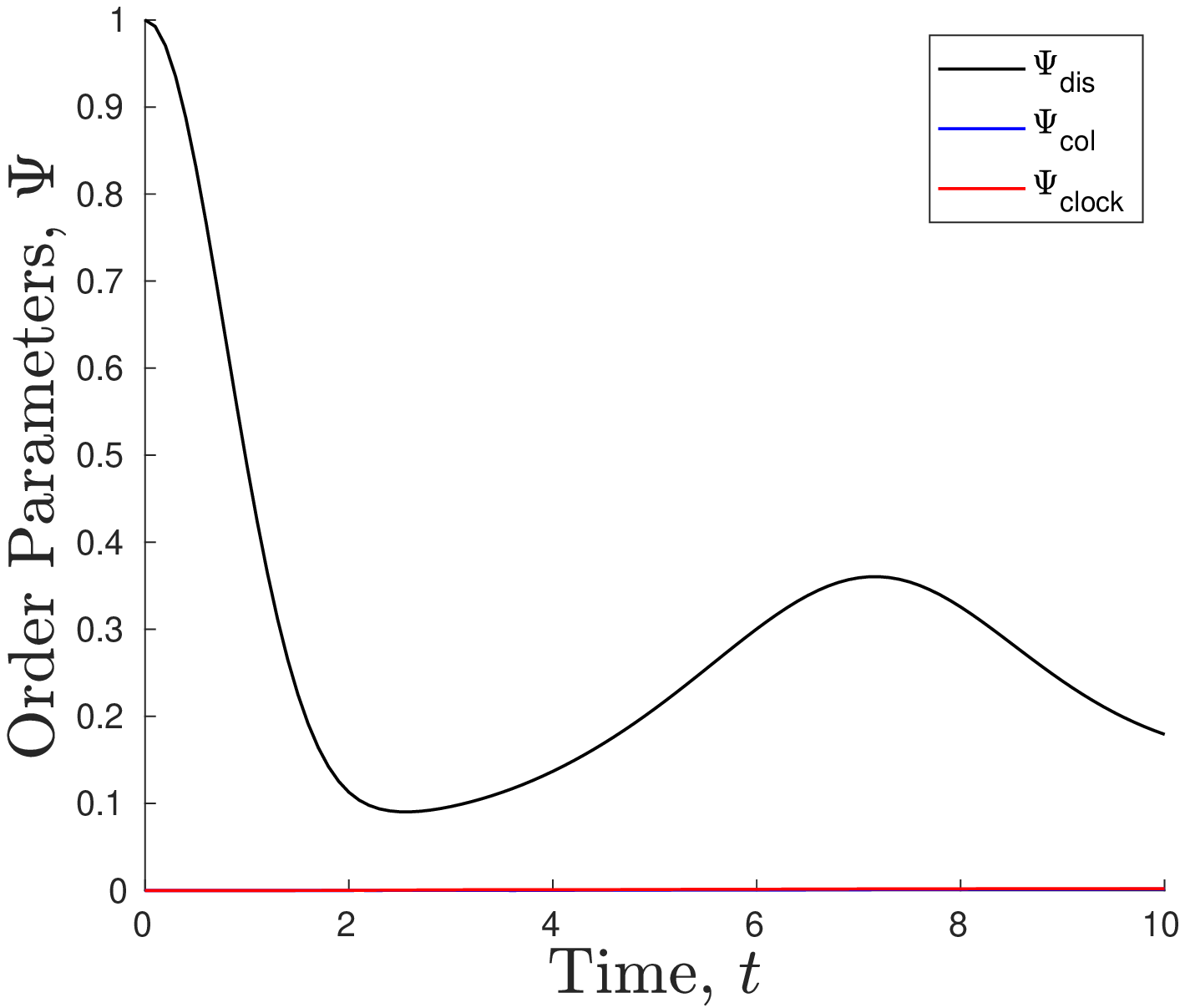}\label{ResFMtoClock}
    \end{minipage}\hfill
    \begin{minipage}{0.48\textwidth}
    \centering
    \caption{Change in the order parameter quenched initial disordered state into columnar ordered phase}
    \includegraphics[width=0.98\textwidth]{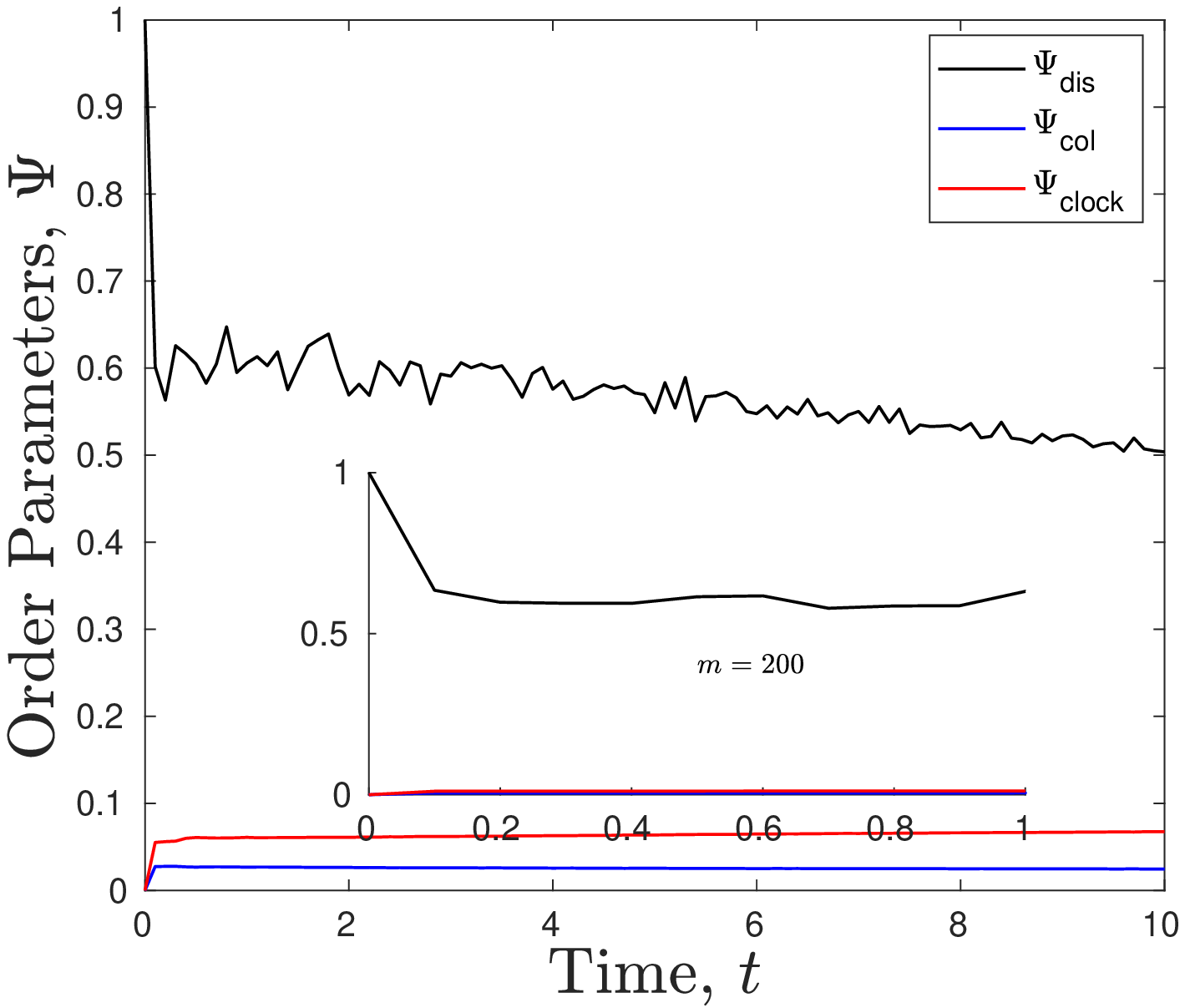}\label{ResFMtoCol}
    \end{minipage}\hfill
\end{figure}
A ground state is prepared in the limit of $\Gamma\to\infty$.
In this limit, the ground state is trivially a ferromagnetic state where all the spins are pointing towards the negative $x$-direction.
The state is evolved in Hamiltonians with clock ordered ground state ($\hat{H}(\alpha=8.0,\Gamma=0.2)$) and columnar ordered ground state
($\hat{H}(\alpha=3.0,\Gamma=0.2)$).
To quantitatively measure the thermalisation behaviour, $\Psi_{\mathrm{dis}}$ (\eqn{disoparam}) is used as a pseudo-order parameter.
For the computational reason, the maximum MPS dimension which can be reached is set to $m=100$.

Shown on \fig{ResFMtoCol} is a change in order parameters over time for quenching into the clock ordered Hamiltonian.
As expected the thermalisation is numerically obtained as a decaying behaviour in $\Psi_{\mathrm{dis}}$.
Clearly, the state does not recover other orders in the time scale of the simulation, 
and the steady state must be a superposition of non-trivial states.

However it seems that there is a hump at $t=2$. Around this time, the order parameter bounces, 
and recovers some orders for a moment and starts to thermalise again.
This can just be a numerical fluke caused by the increasing correlations which MPS failed to accurately represent.
There are two other possible reasons. 
Like the shallow quenching case in the transverse-Ising model, an order parameter can oscillate while it gets dumped by the decoherence.
This behaviour may be telling us that while the ground state of the quenched Hamiltonian is in a deep columnar phase, 
in terms of dynamical phase, this quenching parameter does not give a Hamiltonian with an ordered ground state that is strong enough to 
kill the pseudo-ordered state. 

\begin{figure}[tb]
    \begin{minipage}{0.48\textwidth}
        \centering
        \includegraphics[width=0.8\textwidth]{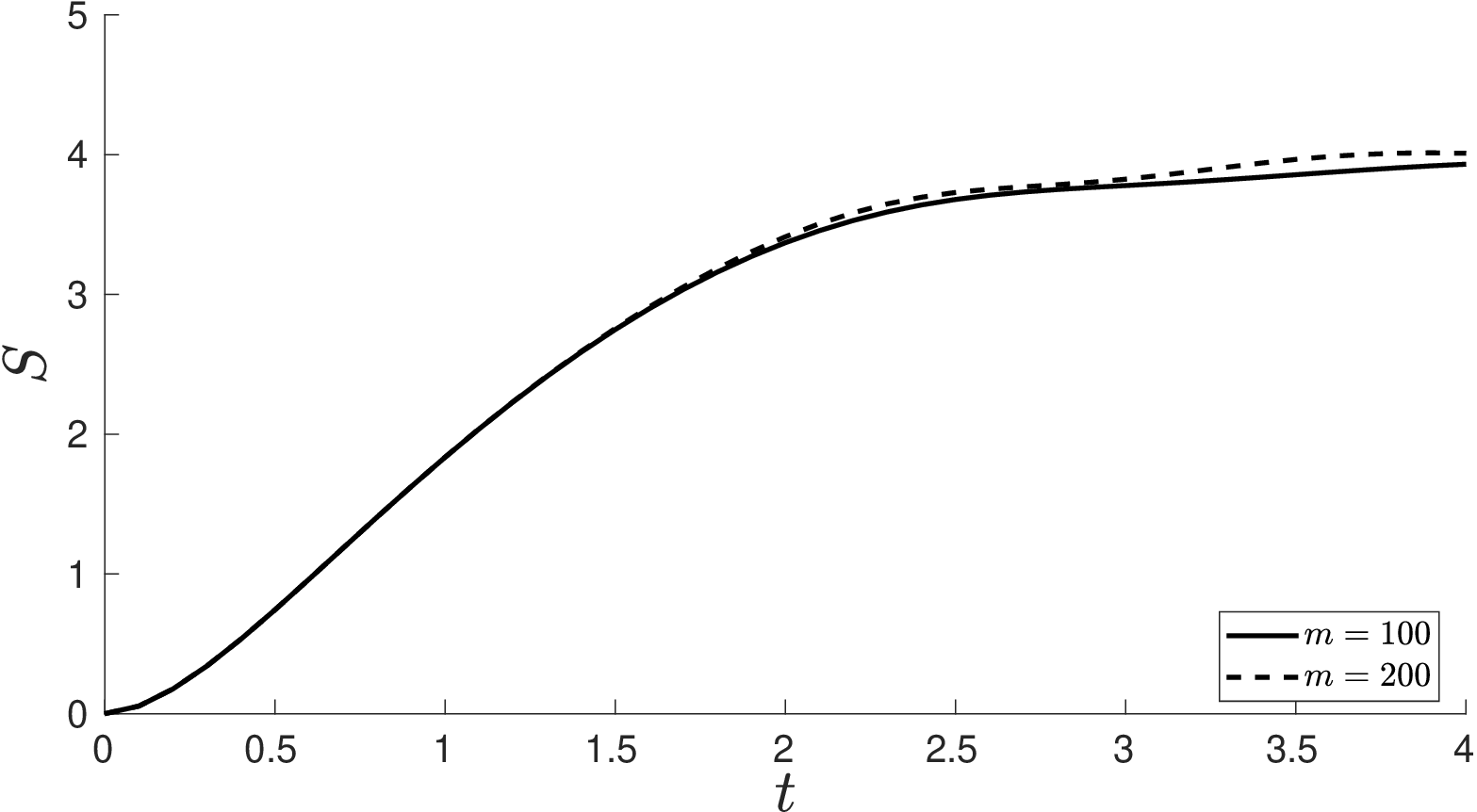}
        \caption{Change in the entanglement entropy (S) over time for disorder $\to$ clock ordered quenching.}
        \label{entropyFMClock}
    \end{minipage}\hfill
    \begin{minipage}{0.48\textwidth}
        \centering
        \includegraphics[width=0.8\textwidth]{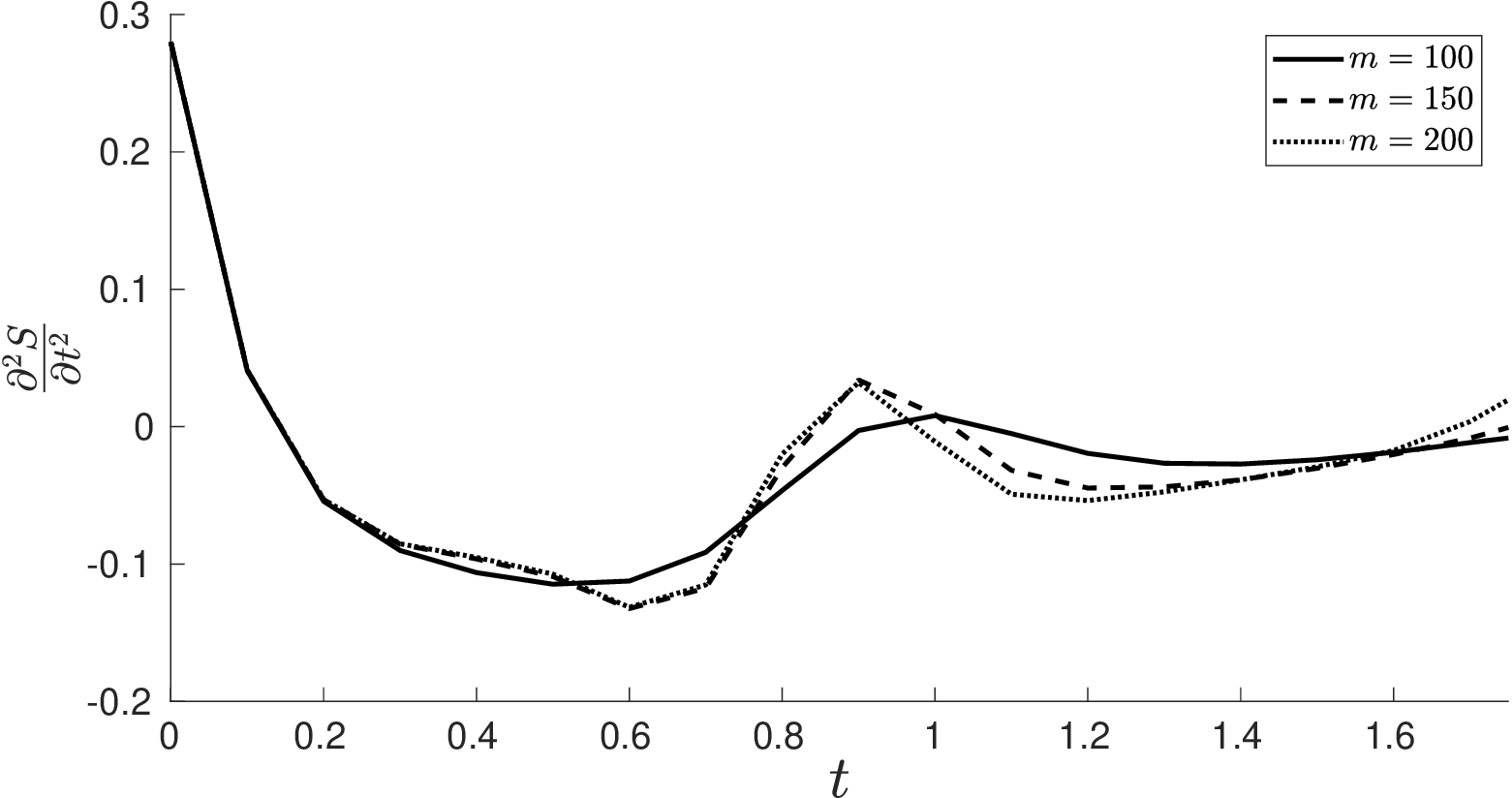}
        \caption{Change in the second derivative of the entanglement entropy (S) over time for disorder $\to$ clock ordered quenching.}
        \label{entropyFMClock2D}
    \end{minipage}
\end{figure}

It is also possible that this may be a result of the saturation of the fluctuations along a ring of the cylinder. 
This possibility can be verified by looking at the values of the entanglement entropy. 
Shown in the \fig{entropyFMClock} is the change in the entanglement entropy over time.
At the time in question, the value of the entropy is well within the limit of $log(100)$.
This tells us that the bounce in order parameter is not happening due to the information lost by the projections and truncations.

This fact can further be confirmed by looking at the second derivative of the entanglement entropy with respect to time.
As it can be seen in \fig{entropyFMClock2D}, there is a number-of-states-independent infliction point. 
This is smoking gun evidence that the entropy does saturate over the ring due to the periodic boundary condition
which caused the initial rapid increase in entropy to a gradual linear increase at the later period. 

Shown in \fig{ResFMtoCol} is a change in order parameter of the $x-$ferromagnetic product state evolved in a Hamiltonian $\hat{H}(\alpha=3,\Gamma=0.2)$.
Strikingly, the state undergoes a sharp drop in its pseudo-order parameter $\Psi_{\text{dis}}$ and a slow but steady linear decrease.
It is predicted that sharp oscillatory behaviours are coming from the output resolution limit 
($dt_{\text{out}} = 0.1$ due to the constraint in the data storage unit).
The transition occurs at $t=0.2$  with the value of entanglement entropy of $1.22$.
However, this transition point shifted greatly towards the entropy sector as $m$ increases. 
Therefore it is likely that the dynamics shown in the figure after $t=0.2$ is unphysical.
Also the raise in order parameter that is happening in the larger plot of \fig{ResFMtoCol} is also a numerical fluke. 
When it is simulated with $m=200$, such behaviour was not observed. 

These two facts imply that a $\Gamma=\infty$ to $(\alpha=3.0,\Gamma=2.0)$ quench allows a state to thermalise magnitudes faster than 
quenching into the Hamiltonian with clock ordered phase.
The faster time scale implies that the $x-$ferromagnet are in a superposition of many non-degenerate eigenstates of the Hamiltonian. 
It is suspected that the large degeneracy in clock ordered ground state, which has a $0$ $z-$magnetisation site in its sublattice,
is naively preserving the order to be protected for a longer time. 

\begin{figure}[bth]
    \centering 
    \includegraphics[width=0.5\textwidth]{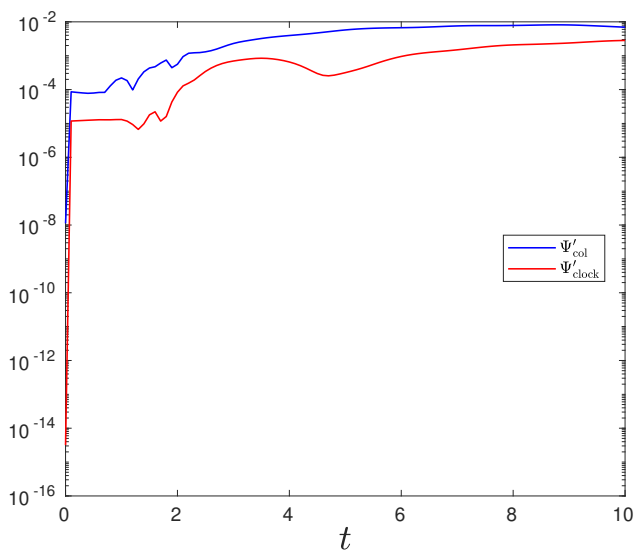}
    \caption{Variance of the order parameters in disordered $\to$ clock order quenching.}\label{varOP}
\end{figure}

Also, as shown in \fig{varOP}, a recovery of columnar and clock order is observed. 
At later times, the initial state recovers the symmetry of the Hamiltonian as a result of a time evolution, 
but a particular realisation of the system may be ordered. 
The initial spike in order parameter is the smoking gun that proves there is indeed a dynamical phase transition from the disordered to a weakly ordered state. 
Although it is very weak, an emergence of long-range correlations was observed,
hinting that there might be a set of parameters which provides an initial state to evolve into a highly entangled states. 

\subsubsection{Quenching on Initially Ordered states to Disordered Hamiltonian}
\begin{figure}
    \begin{minipage}{0.48\textwidth}
    \centering
    \caption{Change in the order parameter quenched initial clock ordered phase into a disordered state.}
    \includegraphics[width=0.98\textwidth]{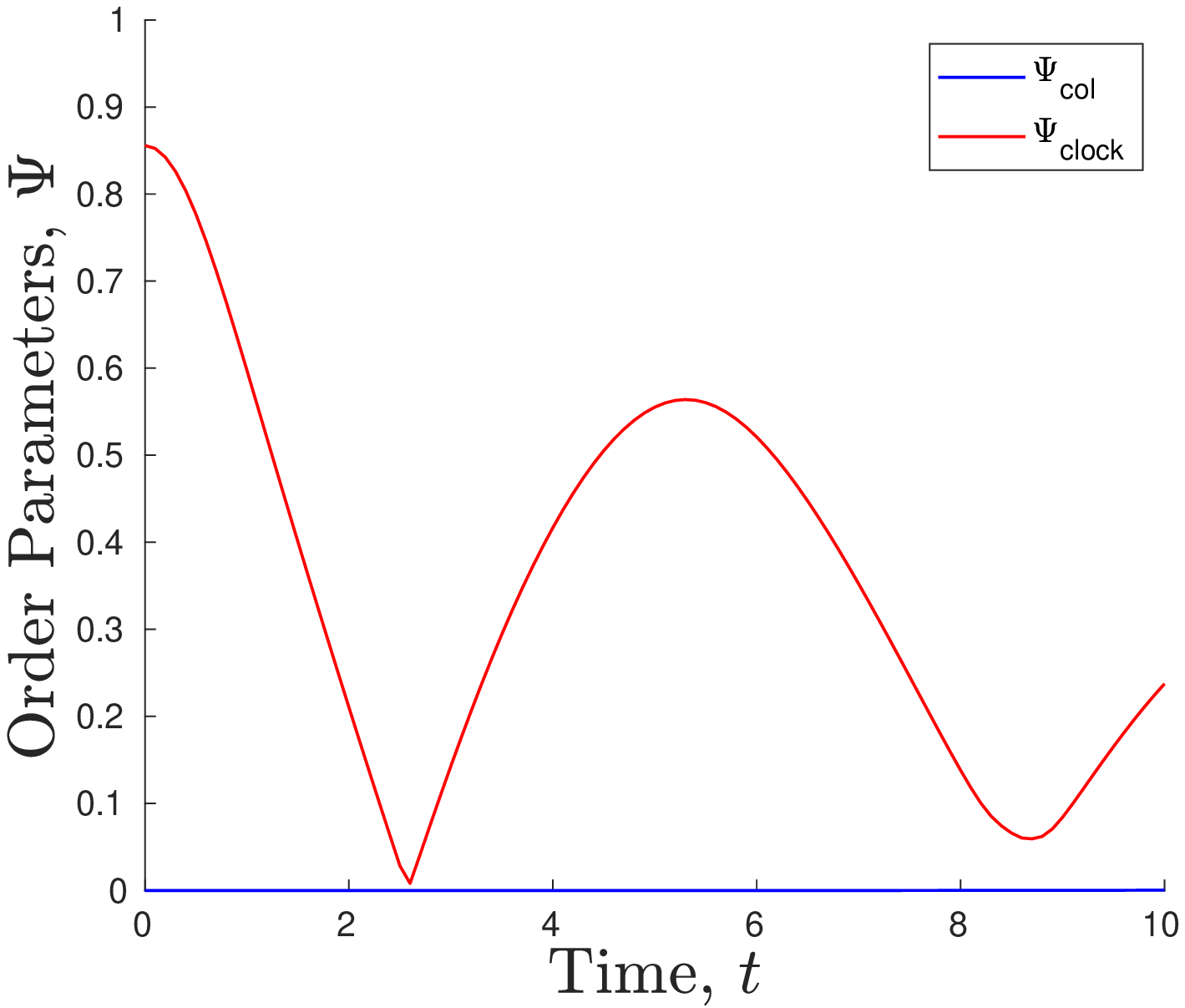}\label{ResClocktoFM}
    \end{minipage}\hfill
    \begin{minipage}{0.48\textwidth}
    \centering
    \caption{Change in the order parameter quenched initial columnar ordered phase into a disordered state.}
    \includegraphics[width=0.98\textwidth]{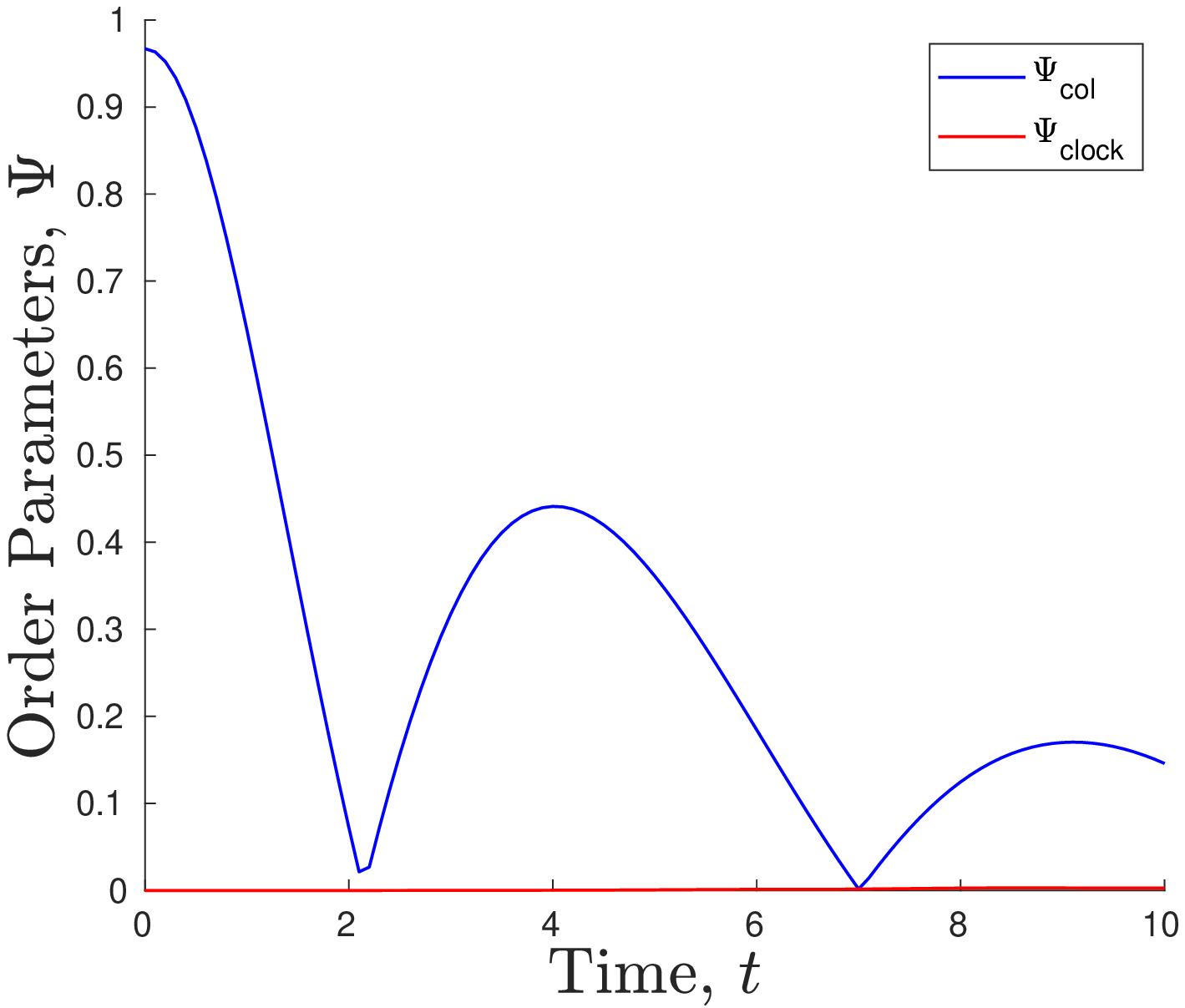}\label{ResColtoFM}
    \end{minipage}\hfill
\end{figure}

Ground states are prepared in Hamiltonians with parameters which have clock phase ($\hat{H}(\alpha=8,\Gamma=0.2)$) and columnar phase
($\hat{H}(\alpha=3,\Gamma=0.2)$). Those states are then evolved independently with a Hamiltonian ($\hat{H}(\alpha=8,\Gamma=1)$), 
which has a disordered ground state. 
The thermalisation are observed in terms of the decaying amplitude of the oscillation in the order parameters of the initial states
(\fig{ResClocktoFM} and \fig{ResColtoFM}).

It seems that despite the difference in the initial long-range interaction strength, 
The time scales of the thermalisation are found to be roughly the same for the both quenches.
The sharp turning points around $t=2$ indicates that due to the oscillations of the local spins, 
the sign of the order parameter changes (the profile of the spins in $z$-direction inverts).
As the number of oscillations increase, the state gets thermalised and all the order disappears at the limit of $t\to\infty$.

Currently, a time evolution up to $t=10$ is a limit of what can be achieved using available computational resources.
The nature of the thermalised state (the steady state of the quench) is unknown.
Unlike disordered to ordered transition, this is a loss of an order due to the time evolution. 
It is not typically interesting but it agrees order to disorder quench in the case of one-dimensional spin chain\cite{Halimeh2016}. 
It is important that through this quench resemblance between one and two-dimensional system are shown. 

\subsubsection{Quenching on Initially Ordered states to Ordered Hamiltonian}
\label{OrderOrderQ}
\begin{figure}
    \begin{minipage}{0.48\textwidth}
    \centering
    \caption{Change in the order parameter quenched initial clock ordered phase into columnar ordered phase.}
    \includegraphics[width=0.98\textwidth]{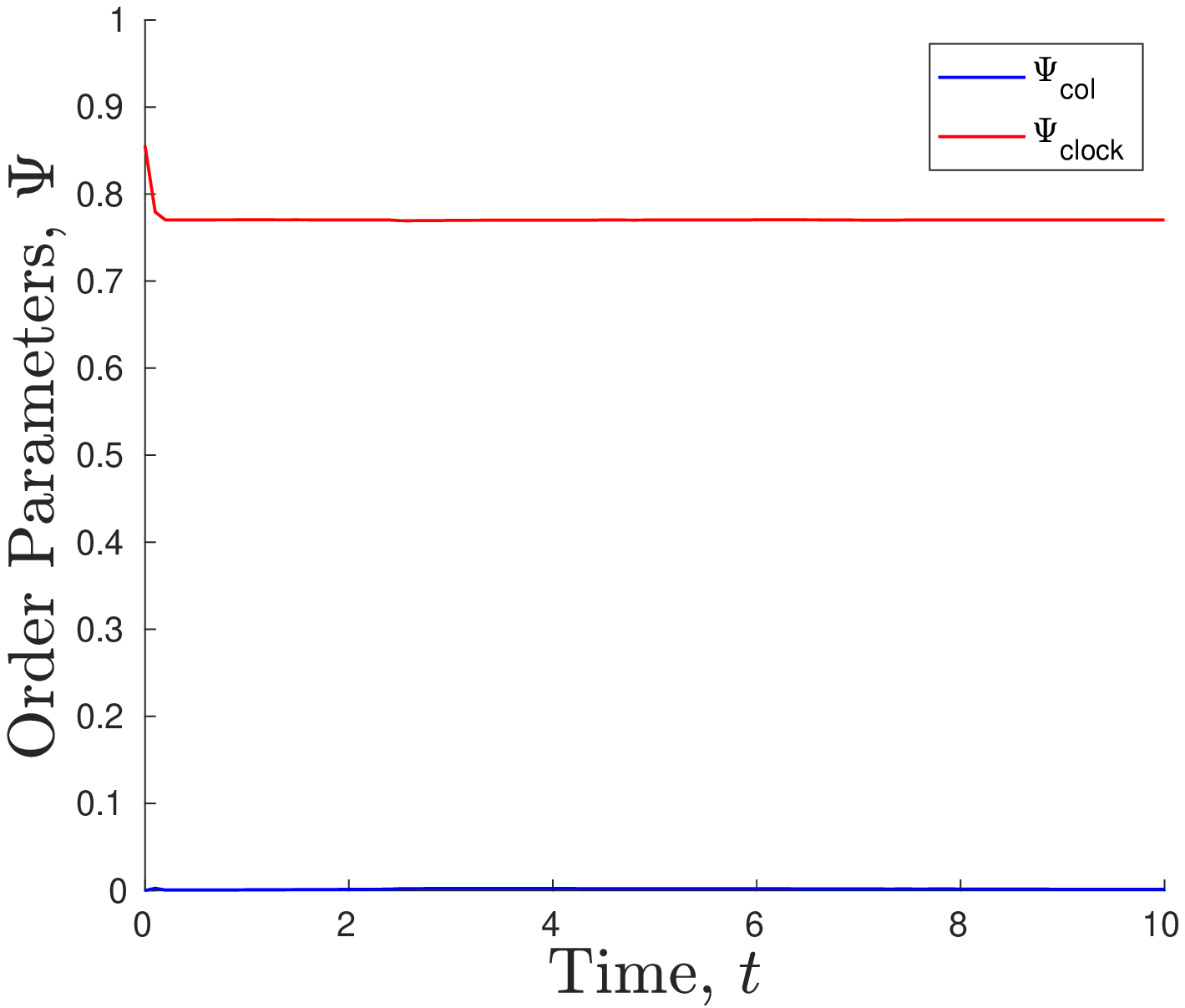}\label{ResClocktoCol}
    \end{minipage}\hfill
    \begin{minipage}{0.48\textwidth}
    \centering
    \caption{Change in the order parameter quenched initial columnar ordered phase into clock ordered phase.}
    \includegraphics[width=0.98\textwidth]{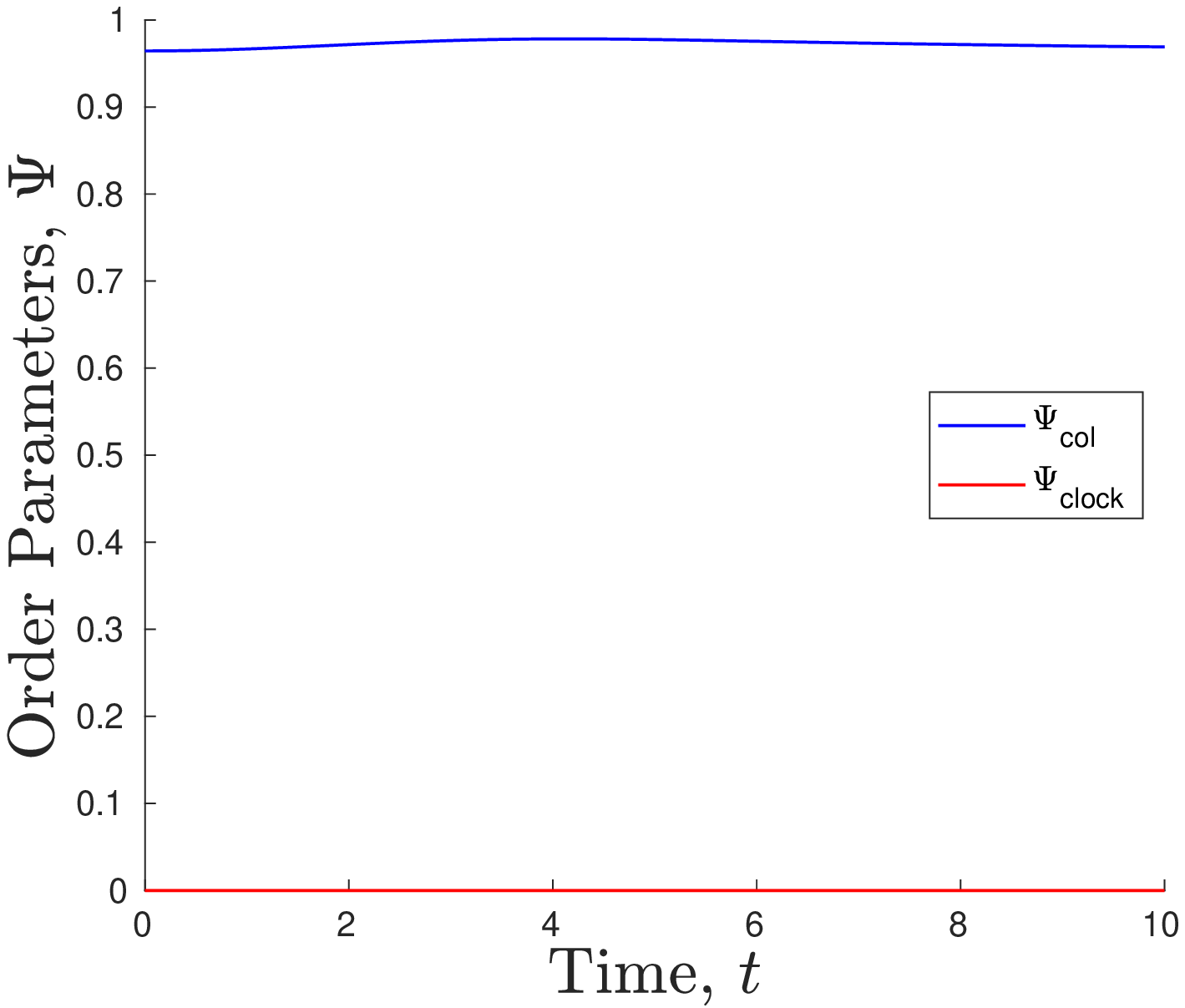}\label{ResColtoClock}
    \end{minipage}\hfill
\end{figure}

To perform an ordered to ordered quench, ground states are prepared in Hamiltonians with parameters which have 
clock phase ($\hat{H}(\alpha=8,\Gamma=0.2)$) and columnar phase ($\hat{H}(\alpha=3,\Gamma=0.2)$). 
Those states are then evolved independently with the Hamiltonian of the other. 
In this quench, the thermalisation has not occurred (\fig{ResClocktoCol} and \fig{ResColtoClock}). 
Instead, the order parameter retained its initial value. 

A constant order parameter over time implies that the phases are stable at least in the time scale of the simulations.
In DMRG simulations that are done to find the ground states of the  Hamiltonians, a pseudo-Convergence into a 
wrong order (if clock then columnar, if columnar then clock) is observed implying that those ordered states are very close in energy spectrum. 
From the energy per site shown in  \tab{GSenergy}, it can be seen that these ordered states are very low energy to each other. 

\begin{table}[h]
\centering
\label{GSenergy}
\begin{tabular}{c|ll}
$\hat{H}$\textbackslash$\ket{\Psi}$ & Columnar         & Clock            \\\hline
Columnar ($\alpha=3,\Gamma=0.2$)               & $-0.310\pm0.001$ & $-0.215\pm0.001$ \\
Clock ($\alpha=8,\Gamma=0.2$)           & $-0.260\pm0.001$ & $-0.286\pm0.001$ 
\end{tabular}
\caption{Energy per site of the states in different orders measured in the different Hamiltonians.}
\end{table}

The interesting thing to be noted is a kink in the clock order parameter in the clock to columnar quench.
A similar effect happened in the disordered to columnar quench. 
As both states are evolved in the Hamiltonian parameterised by 
$\alpha=3,\Gamma=0.2$, it is suspected that this initial kink is 
a dynamical property coming from the strong long-range interactions. 
Unlike in the disordered $\to$ columnar quenching, 
it seems that the clock $\to$ columnar quenching retains 
a low entanglement entropy ($\sim 1.4$), which is much less than $log(m)=4.1$. 
This indicates that this kink is not a numerical fluke but it is a physically realisable phenomenon.
Further investigation with a larger number of $m$ and a shorter time interval is needed. 

\section{Conclusion}
In this chapter, the dynamical properties of the two-dimensional triangular transverse quantum Ising model with long-range interactions on the surface of
an infinitely long cylinder are investigated.
Global quench are performed on the ground states with different orders to see how they evolve in time. 
To find the possible quenching parameters, a quantum phase diagram of the model is constructed. 
Using DMRG, it is found to have three distinctive regions: 
A columnar ordered phase, which is characterised by a four fold degenerate state composed of up and down spins arranged in stripes;
a clock ordered phase, which is composed of triangular sublattices of three neighbouring sites with a $z$-magnetisation 
of positive, negative, and zero; and disordered phase, where its states are typically ferromagnetic in the negative $x$-direction, 
with no magnetic order in the $z$-direction.

Using the new algorithm introduced in the last chapter,
quenching are performed for all the possible combinations of the initial and final parameters of the Hamiltonian with different quantum phases.
In order $\leftrightarrow$ disorder quenching, the thermalisation of the initial states are observed as an oscillating and decaying values of 
the order parameters.
For the quenching Hamiltonian with the parameter of $\alpha=8$, the number-of-state-independent infliction point in the entanglement entropy 
revealed that the numerical simulation is physical. 
In particular, in the disorder $\to$ order quenching, recovery of the symmetry is observed as 
the initial spikes in the order parameters and their saturation towards non-zero values.
They are indications of the emerging long-range correlations from a classical state. 
The order $to$ order quenching, on the other hand, showed that the ordered state stays ordered with the time scale of the simulation. 

For the time evolutions with the Hamiltonian with $\alpha=3$, initial sharp drops in the clock order parameter are observed. 
From the analysis of the entropy, it is assumed to be a physical phenomenon. 
Similar behaviour is also observed in disorder $\to$ order quenching with the Hamiltonian $\alpha=3$. However, in this case 
due to rapid increase in the entropy, the accuracy of the simulation remain uncertain.
It seems that strong long-range interaction in the Hamiltonian increases the time-scale of dissipation by an order of magnitude in comparison 
to when the interaction is only between the nearest neighbour sites. 
Further investigations on shorter time scales with much smaller time steps and larger MPS dimensions are required.

Most importantly, the series of quenching showed that the gTEBD scheme is capable of evolving a Hamiltonian with non-trivial long-range interactions.
Although there are some anomaly behaviours with strong long-range interactions ($\alpha=3$), the algorithm gave a series of predictions that 
are seems to be physically and qualitatively correct. 
the quantitative comparison are not made for four reasons: Firstly, there is no exactly solvable case for two-dimensional long-range interactions. 
Secondly, although there are experimental realisations of the model, there is no experimental probe of its dynamical properties.
Also, there is no other literature which investigates the dynamical properties of the model or its variants. 
Finally, there are no other algorithms that are capable of time-evolving this kind of complex interaction profile.
Therefore, to perform more rigorous tests of the algorithm on two-dimensional quantum lattices, 
tests with other numerical schemes and experimental realisations are needed.

\chapter{Conclusion and Future Prospects}
\label{Conclusion}
In this thesis, a two-dimensional quantum Ising lattice with long-range interactions is investigated. 
The model has an experimental realisation as large number of spins in a Penning ion trap.
Motivated by the experiments, a new numerical scheme is developed to perform an investigation on 
the dynamical properties of the model.

In \Chap{Anewtimeevolution}, a new algorithm which can evolve a system with long-range interactions is introduced.
The new algorithm (generalised TEBD) splits Hamiltonian into two least non-commuting parts,
uses Lanczos method to perform local time evolution, and Suzuki-Trotter decomposition to 
evolve globally. 
With this separation of local and global time evolution , interactions beyond the nearest neighbour was made possible. 
To minimise the error from the non-commuting terms at the Suzuki-Trotter step, a good bias function is introduced and optimised.

Using the optimised bias function, 
benchmarks of the algorithm were performed on exactly solvable short-range interaction model, 
exactly solvable imaginary time evolutions, and long-range real time evolution using other algorithms.
They showed great agreement with the predicted results, revealing that the algorithm is capable of time evolving MPS with
not only translationally invariant Hamiltonians, but also a Hamiltonian with non-trivial long-range interactions.

In \Chap{dynamicalphasetransitions}, properties of a two-dimensional transverse long-range quantum Ising model is investigated.
To make the problem computationally tractable, the model is constructed on to the surface of an infinitely long cylinder.
The quantum phase of the model is found to be different from the mean-field and QMC studies of the finite two-dimensional model. 
Although they all consist of two distinctive ordered phases and one disordered phase, the order of the triangular sublattice in 
clock ordered phase was found to be composed of up, down, and $0$ magnetisation in its triangular sublattice.

Global quenching are performed on all the possible initial and final sets of parameters which goes in and out from the
Hamiltonians with different ground states.
The simulations of the global quenching revealed the presence of thermalisation disorder $\leftrightarrow$ order transitions. 
The saturation of the entropy of a system in the direction of the ring is observed as a number-of-state-independent 
infliction points in entanglement entropy. 
Also, despite the thermalisation, in the disorder $\to$ ordered quenching, 
it is found that weak order gets recovered as the emergence of long-range correlations.
This suggests that weak long-range order may be recovered as a time evolution of a pure quantum system.
On the other hand, ordered $\to$ ordered transition had no sign of thermalisation in a computationally reachable time scale.

Most importantly, the numerical calculations of the time evolutions of two-dimensional lattices have produced results 
that are qualitatively agrees with phenomena which can be deduced from the dynamics of nearest-neighbour one-dimensional models.
The calculation that are performed in this thesis can be a stepping stone starting point of investigations of 
more complex two-dimensional quantum lattice systems. 

As of October of 2017, there is no other calculations
that are performed on time evolving two-dimensional quantum lattice systems with long-range interactions.
Therefore the computational efficiency of the algorithm and its predictability could not be verified
on such models.
Since the model that is investigated has experimental realisations, it is possible to conduct an experiments as a bench mark of this algorithm. 
Also, verification of the experimental realisation can be done by predicting important features using the algorithm.
Large-scale experiments and larger scale calculations are needed for further confirmation of both numerical schemes and 
their physical applicabilities. 

The dynamical properties of the strongly correlated systems are not known in detail. 
Especially effects from the non-trivial order and frustration in the energy eigenspectrum and the existence of long-range correlations
in a complex many-body quantum system such  as on the dynamics of strongly correlated quantum systems are still largely unknown. 
All of these are fundamental questions that must be asked when one deals with a physical realisation of high-temperature superconductivities and 
quantum computers. 
Therefore there are many research opportunities that have rich theoretical and real-world implications in this field.

For the algorithm, future research can be done on the improvement of errors and the computational time at each step. 
The method of the local time evolution can be altered to a more efficient method, such as Unitary Matrix Product Operator \cite{Zaletel2015}. 
This may give great reduction in computational time because it only needs to do MPO-MPS contraction once.
An investigation on the optimal form of a generic bias function for suppressing the overlapping terms of the two parts of the 
Hamiltonian can be another possible research topic. 
It may give deeper insight into a large scale commutation relation, and further improvements in the errors from the Suzuki-Trotter step. 
The independent study of the two-dimensional quantum lattice models using different algorithms are also needed to test the accuracy of the algorithm
beyond spin-chain models.

For the dynamical properties of strongly correlated quantum lattices, there are so many future research opportunities. 
First of all, the future research on the experimental verification of the algorithm. 
This should not be impossible, but as $\alpha=3$ is an upper limit of the experimental setup it may be extremely difficult. 
The further investigation into the quenching within the low-$\alpha$ regime is needed.
Further investigations into the transition point in the parameter space, 
such as a point where the order conserving dynamics breaks into the order destroying dynamics, is another viable research topic.
In particular a search for a transition which strongly recovers the orders.
The long-range correlation which can be realised through this process may provide interesting applications in quantum information. 

In this thesis, the focus was on global quenching with no explicit symmetry breaking scheme. 
The extension of the algorithm to perform local quenching should not be difficult.
Also, adding a random field to give pseudo interactions with the environment to cause spontaneous symmetry breaking 
may not be difficult either. 
Such extensions should give richer dynamical properties compared to the crude evolution of a pure system; 
and it should be closer to what will actually happen in the real experiments. 

To conclude, the dynamical nature of two-dimensional quantum lattice system is poorly known.
Most of the questions introduced in \Chap{Introduction} remain unanswered. 
The numerical simulations provided in this thesis shed some lights into key features that must be investigated in the future. 
The future theoretical, numerical, and experimental investigation may reveal richer 
dynamical properties beyond semi-classical, mean-field, and weakly interacting studies.
\footnote{After the submission of this thesis, an experimental result on one dimensional transverse Ising model with long range interactions is published
    (J. Zhang, G. Pagano, P. W. Hess, A. Kyprianidis, P. Becker, H. Kaplan, A. V. Gorshkov,
    Z. X. Gong, and C. Monroe. \textit{Observation of a Many-Body Dynamical Phase Transition
    with a 53-Qubit Quantum Simulator} pp. 1–8 (2017). URL \url{http://arxiv.org/abs/1708.01044}.).}

\appendix

\input{back/app4_time_energy_uncertainty_relation.ptex}
\backmatter

\bibliography{thesis}
\printindex
\end{document}